\documentclass[aps,10pt,prd,twocolumn,superscriptaddress,preprintnumbers,eqsecnum,%
showkeys,nofootinbib,showpacs]{revtex4-2}
\usepackage{bm}
\usepackage{graphicx}
\usepackage[T1]{fontenc} 
\usepackage{slashed}
\usepackage{mathrsfs}  
\usepackage{amsmath}
\usepackage{enumerate}
\usepackage{dcolumn,booktabs}
\newcolumntype{d}[1]{D{.}{.}{#1}}
\allowdisplaybreaks
\newcommand{\deut}{D}
\newcommand{\lambdadp}{{\lambda'}_{\!\!\deut}}
\newcommand{\lambdanp}{{\lambda'}_{\!\! n}}
\newcommand{\lambdapp}{{\lambda'}_{\!\! p}}
\usepackage{xr-hyper}
\usepackage{hyperref}
\hypersetup{colorlinks=true,citecolor=blue,linkcolor=blue,urlcolor=blue}
\newcommand{\firstpart}{{Part I~\cite{CW_kin_spin1}}}
\usepackage{orcidlink}
\begin{document}
\title{Semi-inclusive deep-inelastic scattering on a polarized spin-1 target. \\
II.\ Deuteron and spectator nucleon tagging}
\author{W.~Cosyn\,\orcidlink{0000-0002-9312-8569}}
\email[ E-mail: ]{wcosyn@fiu.edu}
\affiliation{Department of Physics, Florida International University, Miami, Florida 33199, USA}
\author{C.~Weiss\,\orcidlink{0000-0003-0296-5802}}
\email[ E-mail: ]{weiss@jlab.org}
\affiliation{Theory Center, Jefferson Lab, Newport News, VA 23606, USA}
\begin{abstract}
We develop the theoretical framework for semi-inclusive deep-inelastic scattering on a polarized spin-1 target
and apply it to scattering on the polarized deuteron with spectator nucleon tagging.

In Part I (previous article) we present the general form of the semi-inclusive cross section
and polarization observables for the spin-1 target.

In Part II (this article) we consider deep-inelastic scattering on the polarized deuteron with spectator
nucleon tagging as a special case of target fragmentation.
Methods of light-front quantization are employed to separate nuclear and hadronic structure
in the high-energy process and achieve a composite description.
The light-front wave function of the polarized deuteron is obtained from a rotationally covariant 3-dimensional
wave function in the center-of-mass frame of the proton-neutron system.
The tagged structure functions are computed in the impulse approximation.
The momentum and spin distribution of the active nucleon are controlled
by the deuteron polarization and the detected spectator momentum ($D/S$ wave ratio).
The cross section and spin asymmetries are evaluated for general deuteron polarization
(vector and tensor, longitudinal and transverse) as functions of the spectator momentum.
Tensor-polarized spin asymmetries of order unity are achieved for spectator momenta
$\sim$ 300 MeV, which select configurations with large $D$-wave.
Sum rules for the tagged spin structure functions are derived.
The results can be used for simulations of spectator tagging in future polarized
fixed-target experiments (Jefferson Lab) or at the Electron-Ion Collider.
\end{abstract}
\keywords{}
\preprint{JLAB-THY-26-4661}
\maketitle
\tableofcontents
\section{Introduction}
\label{sec:intro}
This is Part II of a study of semi-inclusive deep-inelastic scattering on a polarized spin-1 target.
\firstpart{} presents the general form of the semi-inclusive cross section and polarization
observables for the spin-1 target. Part II considers deep-inelastic scattering on the
polarized deuteron with spectator nucleon tagging.

The deuteron nucleus plays an important role as a spin-1 target for inclusive and semi-inclusive
deep-inelastic lepton scattering (DIS and SIDIS).
A particular case of SIDIS is the detection of a ``slow'' nucleon (proton or neutron)
with momentum $|\bm{p}_N| \sim$ few 100 MeV in the target rest frame.
It is an example of target fragmentation, or hadron production in the
target rapidity region caused by soft interactions with rapidity range $\sim 1$.
In QCD the target fragmentation cross section can be computed in a factorization
scheme with so-called fracture functions, which combine aspects of parton distribution and
fragmentation functions \cite{Trentadue:1993ka,Collins:1997sr,Anselmino:2011ss}.
In the case of slow nucleon production in DIS on nuclei, the dominant
production mechanism is the nuclear breakup, understood as the liberation of a
bound nucleon in the initial nucleus by the DIS process, and the cross section
can be computed in an approximation scheme on this basis \cite{Frankfurt:1983qs,Frankfurt:1981mk}.
In this context the measurement is referred to as ``spectator nucleon tagging.''
It presents a unique situation where the SIDIS cross section and polarization observables
can be predicted on the basis of well-known nuclear structure and dynamics.

The deuteron in the initial state of the DIS process is described in terms of nucleon degrees
of freedom \cite{Frankfurt:1983qs,Frankfurt:1981mk} (regarding the role of non-nucleonic
degrees of freedom in high-energy processes, see the references). 
It represents a quantum-mechanical system that exists as a superposition
of $pn$ configurations in the space of relative momentum (or coordinate) and spin variables.
The detection of the spectator nucleon and measurement of its momentum ``selects''
certain configurations from the superposition.
This allows one to control the nuclear configurations during the DIS process
and use that information for theoretical analysis.

The typical nucleon momenta in the deuteron are of the order of the so-called binding momentum
$\sqrt{\epsilon_\deut m} = 45$ MeV, where $\epsilon_\deut$ is the deuteron binding energy and $m$ the nucleon mass.
Tagging of nucleons with momenta $|\bm{p}_N| \sim$ few 10 MeV (in the deuteron rest frame)
selects average configurations in the deuteron, where interaction effects are moderate.
Extrapolation to the nucleon pole at unphysical momenta $|\bm{p}_N|^2 = -\epsilon_\deut m$ selects
configurations of infinite spatial size, where interactions are absent and the nucleons are free;
this technique can be used to extract free neutron structure in a model-independent
manner \cite{Sargsian:2005rm,Cosyn:2015mha,Strikman:2017koc,Jentsch:2021qdp}.
Tagging of nucleons with $|\bm{p}_N| \sim$ few 100 MeV selects small-size configurations,
where interactions are strong; this method can be used to study the interaction dependence of
partonic structure modifications \cite{Melnitchouk:1996vp,Sargsian:2002wc} and their
connection with short-range $NN$ correlations \cite{Hen:2013oha,Segarra:2020plg,Ratliff:2023two};
see Ref.~\cite{Hen:2016kwk} for a review.

The orbital motion of the nucleons in the deuteron includes an $S$ and $D$ wave component,
in which the spins are coupled parallel and antiparallel to the deuteron spin.
The $S$-wave is dominant at average nucleon momenta; the $D$ wave becomes prominent at momenta
$\gtrsim$ 200 MeV.
By selecting configurations with a certain momentum using spectator tagging,
one effectively controls the $D/S$ ratio and thus the spin state of the nucleons
during the DIS process.
In DIS on a vector-polarized deuteron, proton tagging controls
the effective polarization of the active neutron \cite{Frankfurt:1983qs,Cosyn:2019hem,Cosyn:2020kwu}.
This can be used to perform measurements
of neutron spin structure functions in defined spin states and eliminate the $D$-wave depolarization.
In DIS on the tensor-polarized deuteron, tagging controls the $D$-wave sustaining the
tensor-polarized spin asymmetries. By selecting configurations with a large $D/S$ ratio,
tensor-polarized asymmetries of order unity can be achieved, even saturating the mathematical
bounds $A^T \in [-2, 1]$ (see \firstpart) \cite{Frankfurt:1983qs}.
This is very different from inclusive DIS without spectator tagging,
where scattering predominantly happens in average configurations with small $D$-wave
and the tensor-polarized asymmetries are $\ll 1$ \cite{Frankfurt:1983qs,HERMES:2005pon,Poudel:2025nof}.
It offers a striking example of the entanglement of spin and orbital degrees of freedom
in the deuteron and the idea of selection of configurations by spectator tagging.

Measurements of DIS on the deuteron with spectator tagging are performed in fixed-target scattering
experiments at Jefferson Lab, using specialized detectors for detecting slow protons or neutrons
emerging from the target, covering momenta $|\bm{p}_N| \lesssim$ 100 MeV
(BONuS, ALERT) \cite{CLAS:2011qvj,CLAS:2014jvt,Armstrong:2017wfw, Albayrak:2024vcy}
and $\sim$ few 100 MeV (Deeps, BAND, LAD) \cite{CLAS:2005ekq,EMCSRC:2011,Segarra:2020txy}.
Measurements of spectator tagging are also planned at the Electron-Ion Collider (EIC), using the
far-forward detectors for charged and neutral beam fragments \cite{Jentsch:2021qdp,AbdulKhalek:2021gbh}.
The collider setup offers several advantages for nuclear breakup detection: no target material
preventing the detection of slow protons or neutrons;
excellent coverage and momentum resolution for far-forward protons (magnetic spectrometer);
possibility of detecting far-forward neutrons (zero-degree calorimeter). Measurements of DIS
with spectator tagging could also be performed on the polarized deuteron if polarized deuteron
beams become available at EIC; see Refs.~\cite{Huang:2020uui,Huang:2021gbi,Huang:2025gqx}
for the technical prospects. An important advantage of the collider in this regard is that
the deuteron polarization does not interfere with the spectator detection (no holding magnets
as in polarized fixed targets). This would make it possible to explore the spin effects of
polarized spin-1 SIDIS using spectator tagging.

In \firstpart\ we have derived the general form of the cross section of
semi-inclusive scattering from a polarized spin-1 target.
In Part II here we consider the specific process of DIS on the polarized deuteron
with spectator nucleon tagging. We focus on the description of nuclear structure in the
DIS process and present quantitative predictions for the polarized tagged structure functions.

Methods of light-front (LF)
quantization \cite{Dirac:1949cp,Weinberg:1966jm,Kogut:1969xa,Leutwyler:1977vy,Brodsky:1997de}
are employed to separate nuclear and hadronic structure
in the high-energy process and achieve a composite description \cite{Frankfurt:1981mk}.
The deuteron is described by a LF wave function in nucleon degrees of freedom.
The LF wave function is matched with the rotationally covariant 3-dimensional wave function
in the center-of-mass frame of the $NN$ system \cite{Kondratyuk:1983kq,Terentev:1976jk}; this formulation
ensures rotational invariance and is particularly useful for the LF description of spin degrees of freedom.

The polarized tagged DIS cross section is computed in the impulse approximation (IA).
Two different formulations are employed: LF quantum mechanics \cite{Frankfurt:1981mk}
and the virtual nucleon scheme \cite{Frankfurt:1996xx,Sargsian:2001ax};
the comparison of the results illustrates the theoretical uncertainties of the method.
The cross section is expressed in terms of the LF momentum distributions of the active nucleon
in the deuteron, which depend on the tagged spectator momentum and the deuteron polarization.
These distributions have the same structural properties as the transverse-momentum-dependent (TMD)
parton distributions in QCD \cite{Boussarie:2023izj}
(spin-orbit effects, $T$-even and odd structures) and can be discussed in these terms.
They exemplify the idea of the selection of configurations through spectator tagging.

The tagged deuteron structure functions are computed for general deuteron polarization
(longitudinal vector, transverse vector, tensor) as functions of the spectator momentum.
Predictions for the tagged polarization observables (differential cross section,
azimuthal harmonics, spin asymmetries, see {\firstpart}) are presented and studied as functions
of the tagged spectator momentum. The role of the $D/S$ wave ratio in the vector and
tensor-polarized spin asymmetries is explained. The dependence of the observables
on the nuclear structure model is investigated.

The results of this study are interesting in two respects: (i) They can be used for the
simulation and analysis of polarized spectator tagging experiments at EIC or future
fixed-target facilities, focusing on measurements of neutron spin structure or nuclear
structure effects in high-energy scattering.
(ii) They illustrate the unique structures appearing in SIDIS on a polarized spin-1 target,
in a realistic example where the dynamics is well understood and quantitative predictions can be made.
The spin asymmetries in nuclear breakup in the target fragmentation region are generally
much larger than those predicted or measured for quark fragmentation in the current fragmentation
region SIDIS, presenting favorable conditions for experiments.

In the present study we describe the nuclear DIS process in the IA,
where it is assumed that the DIS final state and the spectator nucleon evolve independently
after the DIS process. It is known that final-state interactions (FSI) have a sizable effect on the
observed spectator momentum distributions at momenta $|\bm{p}_N| \gtrsim$ 100 MeV and need to be
included in complete treatment \cite{CiofidegliAtti:2000xj,CiofidegliAtti:2003pb,%
Cosyn:2011jnm,Cosyn:2015mha,Cosyn:2017ekf,Strikman:2017koc}.
The IA results presented here are useful
as a baseline for the experimental analysis and for further theoretical treatment including FSI.
In the spin-dependent tagged structure functions only the $T$-even structures
are non-zero in the IA; the $T$-odd structures require explicit FSI
(the situation is the same as with $T$-odd structures in the current fragmentation
such as the Sivers function \cite{Sivers:1989cc,Boussarie:2023izj}).

The production of a slow nucleon in DIS on the deuteron is a special case of
target fragmentation. In QCD the cross section of SIDIS in the
target fragmentation region can be computed using a generalized factorization theorem, as the product
of the cross section for electromagnetic scattering on a quark/gluon and the fracture
function of the target \cite{Trentadue:1993ka,Collins:1997sr}.
The fracture function describes both the target structure in the initial state and the
fragmentation process producing the identified hadron in the final state; as such it
combines aspects of the parton densities and the parton fragmentation functions.
In the present study we do not use QCD factorization but compute the cross section for
``nucleon production'' directly in terms of the nuclear wave function and the structure
functions of the active nucleon (without factorizing them into quark/gluon coefficient
functions and distributions in the nucleon). This approximation is justified in the present kinematics,
where the nuclear breakup process is the dominant source of nucleon production.
The results obtained in this approximation are compatible with the QCD factorization theorem
and respect the underlying scale separation \cite{Strikman:2017koc}.
Our results may be regarded as a particular nonperturbative model of the fracture function
for nucleon production in DIS on the deuteron (this interpretation appears literally
when the active nucleon structure functions in our expressions are factorized into
quark/gluon coefficient functions and distributions in the nucleon).

Polarization effects in electron-deuteron scattering have been studied extensively in
quasi-elastic scattering (deuteron breakup into $NN$ final state), including
general structural analysis \cite{Dmitrasinovic:1989bf,Arenhovel:1993cs,%
Arenhovel:2002sg,Arenhovel:2004bc,Flores:2023pgp}
and dynamical calculations using nonrelativistic nuclear theory \cite{Leidemann:1991qs,Arenhoevel:1992xu},
relativistic bound state equations \cite{Jeschonnek:2009tq,Jeschonnek:2016jyn},
and relativistic amplitude methods \cite{Rekalo:1987ym,Gakh:2004zq,Grassi:2022flz};
see also references in the quoted works.
In the present study we focus on the deep-inelastic regime and use methods of LF quantization
to separate nuclear and hadronic structure. The necessity of using LF quantization in
high-energy scattering processes and tests of the consistency of the approximations
(momentum sum rule, spin sum rules) are discussed in the text.

The article is organized as follows.
Section~\ref{sec:process} summarizes the specific setup of the tagged DIS process,
including the kinematic variables, collinear-frame coordinates, and deuteron polarization
parameters, as a special case of the general framework developed in \firstpart{}.
Section~\ref{sec:lightfront} covers the treatment of nuclear structure in the DIS process
including LF quantization, spin structure, and the IA
in both the 3-dimensional quantum-mechanical and the 4-dimensional virtual nucleon formulation.
Section~\ref{sec:effective} explores in detail the nucleon LF momentum
distributions in the deuteron sampled in tagged DIS, including their
definition and properties, dependence on the nucleon and deuteron spin, probabilistic formulation,
and numerical estimates.
Section~\ref{sec:IA_strucfunc} presents the tagged DIS structure functions in the
IA, for unpolarized and polarized electron and vector- and tensor-polarized
deuteron, and the sum rules for the tagged spin structure functions.
Section~\ref{sec:polarization_observables} presents examples of polarization observables
in tagged DIS with the vector- and tensor-polarized deuteron, illustrating the idea of
``selection of configurations'' using the spectator momentum, and estimates of
the nuclear structure model dependence.
Section~\ref{sec:conclusions} summarizes the conclusions and possible applications
and extensions of the results.
Appendix~\ref{app:spin_sum_rules} gives the derivation of the spin sum rules
for the polarized neutron distributions in spectator tagging, for both longitudinal
and transverse spin.

\section{Tagged DIS on polarized deuteron}
\label{sec:process}
\subsection{Kinematic variables}
\begin{figure}[t]
\begin{center}
\includegraphics[width=0.35\columnwidth]{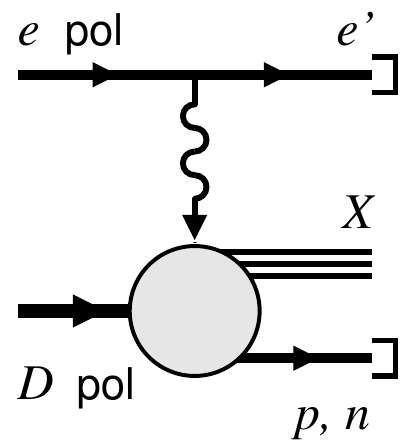}
\end{center}
\caption{
\label{fig:deuteron_tagged}
DIS of polarized electrons on the polarized deuteron with detection of a proton (or neutron)
in the nuclear fragmentation region (``tagged DIS''), Eq.~(\ref{eq:tagged_reaction}).}
\end{figure}

The semi-inclusive scattering process considered in this study is DIS of polarized electrons
on a polarized deuteron with detection of a nucleon (proton or neutron) in the final state
(see Fig.~\ref{fig:deuteron_tagged}),
\begin{align}
e(l| \lambda_e) + \deut (p_\deut| S_\deut, T_\deut) \rightarrow e'(l') + X + N (p_N),
\label{eq:tagged_reaction}
\end{align}
where $l, l'$ are the initial/final electron 4-momenta, $p_\deut$ is the initial deuteron 4-momentum,
$p_N$ is the final nucleon 4-momentum. The electron polarization is specified by the helicity $\lambda_e$,
the deuteron polarization by the vector and tensor parameters $S_\deut$ and $T_\deut$ (described in the
following). 
We focus on the kinematic region where the detected nucleon momentum in the deuteron
rest frame remains finite in the DIS limit (target fragmentation region) and has values
$|\bm{p}_N| \lesssim$ few 100 MeV.
In this region the nucleon originates predominantly from the breakup of the deuteron nucleus,
and the measurement is commonly referred to as ``spectator nucleon tagging'' or ``tagged DIS'';
this interpretation is not imposed a priori here but emerges from the theoretical analysis
of Sec.~\ref{sec:lightfront}. For definiteness we consider the case that the detected nucleon
is a proton and refer to it as such in the following,
\begin{align}
N \equiv p, \hspace{2em} p_N \equiv p_p;
\end{align}
equivalent formulas can be written for the case of a detected neutron.

The process Eq.~(\ref{eq:tagged_reaction}) is characterized by the invariant kinematic variables described
in \firstpart. The original scaling variable for scattering on the deuteron is defined as
\begin{align}
x_\deut \equiv \frac{Q^2}{2 p_\deut q}, \hspace{1em} 0 < x_\deut < 1 .
\end{align}
We use here the rescaled scaling variable
\begin{align}
x \equiv 2x_\deut = \frac{Q^2}{p_\deut q}, \hspace{2em} 0 < x < 2 ,
\label{x_rescaled}
\end{align}
which can be regarded as the scaling variable for scattering on a nucleon in the
absence of nuclear binding (this interpretation is optional; the variable is
rigorously defined). We define the average nucleon mass as $m \equiv (m_n + m_p)/2$,
so that the deuteron and nucleon masses are related as $M_\deut = 2 m - \epsilon_\deut$,
where $\epsilon_\deut =$ 2.2 MeV is the binding energy. With these definitions
\begin{align}
x_\deut M_\deut = x m + \mathcal{O}(\epsilon_\deut),
\end{align}
where the $\mathcal{O}(\epsilon_\deut)$ terms are numerically irrelevant and can be neglected.
Kinematic quantities depending on the combination $x_\deut M_\deut$ are therefore practically
the same for scattering on the deuteron and on the unbound nucleon with the
equivalent $x$. The $\gamma$ parameter defined in \firstpart, Eq.~(\ref{P1:gamma_def}), is
\begin{align}
\gamma \equiv \frac{2 x_\deut M_\deut}{Q} = \frac{2 x m}{Q} + \mathcal{O}(\epsilon_\deut).
\label{gamma_approx}
\end{align}
%
%
\subsection{Collinear frames}
\label{subsec:collinear_frames}
The theoretical analysis of tagged DIS is performed in the class of collinear frames introduced
in \firstpart.
The deuteron momentum $\bm{p}_\deut$ and the momentum transfer $\bm{q}$ are along the $z$-axis, with
$\bm{q}$ pointing in the negative $z$ direction. We describe 4-vectors in the collinear frames by
their LF components 
\begin{subequations}
\label{eq:lf_coords}
\begin{align}
a^\pm &\equiv a^0 \pm a^3, \hspace{2em} \bm{a}_T \equiv (a^1, a^2),
\\
a\cdot b &= \tfrac{1}{2} (a^+ b^- + a^- b^+) - \bm{a}_T \cdot \bm{b}_T.
\end{align}
\end{subequations}
The components of $p_\deut$ and $q$ in the collinear frames are fixed by the invariant kinematic
variables as 
\begin{subequations}
\label{collinear_frame}
\begin{align}
&p_\deut^+ > 0 \; \textrm{arbitrary}, \hspace{2em} p_\deut^- = \frac{M_\deut^2}{p_\deut^+},
\\[-1ex]
&q^+ = - \frac{\xi p_\deut^+}{2}, \hspace{2em} q^- = \frac{2 Q^2}{\xi p_\deut^+},
\\
&\bm p_{\deut T} = \bm q_T = 0,
\end{align}
\end{subequations}
where
\begin{align}
\xi \equiv \frac{2 x}{1 + \sqrt{1 + \gamma^2}} .
\end{align}
In the DIS limit, $Q \to \infty$ and $x$ fixed, one has $\xi \approx x$ up to power corrections.

The collinear frames are an equivalence class of frames connected by boosts along the
$z$-axis (longitudinal boosts). A particular frame in selected by specifying the value of $p_\deut^+$.
Longitudinal boosts between frames with different $p_\deut^+$ are performed by rescaling the LF
components of the 4-vector as
\begin{align}
a^{+\prime} = \lambda a^{+},
\hspace{2em}
a^{-\prime} = \lambda^{-1} a^{-}, 
\hspace{2em}
\lambda \equiv p_\deut^{+\prime} / p_\deut^+ .
\label{boost_collinear}
\end{align}
If the $+$ and $-$ LF components of the 4-vector are expressed as fractions of
$p_\deut^+$ and $1/p_\deut^+$, as in Eq.~(\ref{collinear_frame}), the boost is performed simply
by changing the value of $p_\deut^+$ in the expression. In particular, the class of collinear
frames contains the deuteron rest frame, which is selected by setting
\begin{align}
p_\deut^+ = M_\deut ;
\end{align}
for other examples of collinear frames, see Ref.~\cite{Strikman:2017koc}.

The final-state proton in the process Eq.~(\ref{eq:tagged_reaction}) is characterized
by its LF plus and transverse momentum [see \firstpart, Sec.~\ref{P1:subsec:final_state}]
\begin{align}
p_p^+ = \frac{\alpha_p p_\deut^+}{2}, \hspace{2em} \bm{p}_{pT}.
\label{eq:alpha_proton_def}
\end{align}
The LF momentum fraction $\alpha_p$ is
invariant under longitudinal boosts and can be expressed 
in terms of the proton momentum in the deuteron rest frame as
\begin{align}
\alpha_p = \frac{2(E_N(\bm{p}_p) + p_p^z)}{M_\deut}, \hspace{2em} E_N (\bm{p}_p) \equiv 
{\textstyle\sqrt{|\bm{p}_p|^2 + m^2}} .
\label{eq:alpha_proton_restframe}
\end{align}
For small rest-frame momenta $|\bm{p}_p| \ll m$,
\begin{align}
\alpha_p \approx 1 + \frac{p_p^z}{m} + \mathcal{O}\left(\frac{|\bm{p}_p|^2}{m^2}\right),
\label{eq:alpha_proton_small}
\end{align}
and the values of $\alpha_p$ are close to unity in the kinematic region under consideration.
The invariant phase space element in the final-state proton momentum is given by
[see \firstpart, Eq.~(\ref{P1:eq:dG_h})]
\begin{subequations}
\label{eq:recoil_phasespace}
\begin{align}
&d\Gamma_p = \frac{d^3 p_p}{(2\pi)^3 2 E_N(\bm{p}_p)}
= \frac{1}{[2 (2\pi)^3]} \times \frac{d\alpha_p}{\alpha_p}
d^2 p_{pT},
\\[1ex]
&d^2 p_{pT} = |\bm{p}_{pT}| \, d|\bm{p}_{pT}| \, d\phi_p ;
\end{align}
\end{subequations}
for other equivalent forms, see Ref.~\cite{Strikman:2017koc}. The same factor $[2 (2\pi)^3]$ will
appear in the results for the tagged structure functions in Sec.~\ref{sec:IA_strucfunc}
and will cancel in the final result for the cross section, see Eq.~(\ref{cross_section_deuteron});
we always write the factor in brackets to distinguish it.

The domain of the kinematic variables is restricted by the condition that the LF plus momentum
of the unobserved hadronic final state $X$ must be positive,
\begin{align}
p_X^+ = p_\deut^+ + q^+ - p_p^+ > 0,
\end{align}
which puts an upper limit on the proton plus momentum. In the DIS limit, neglecting power corrections
$\mathcal{O}(x^2 m^2/Q^2)$, the condition becomes
\begin{align}
\alpha_p < 2 - x .
\label{alphap_limit}
\end{align}

\subsection{Cross section}
The general structure of the cross section for semi-inclusive scattering on a spin-1
target and its decomposition in structure functions are described in \firstpart,
see master formula Eq.~(\ref{P1:cross_section_final}).
In the application to tagged DIS on the deuteron ($A \equiv D$), we put $dx_\deut = dx/2$,
Eq.~(\ref{x_rescaled}), and consider the cross section differentially
in $dx$. The expression takes the form
\begin{align}
d\sigma &= \frac{2\pi \alpha_{\rm em}^2 y^2}{Q^4 (1 - \epsilon)} \; \frac{dx}{2} dQ^2
\frac{d\psi_{l'}}{2\pi}
\nonumber \\
&\times
\left( \mathcal{F}_U + \mathcal{F}_S + \mathcal{F}_T \right) \; d\Gamma_p ,
\end{align}
where $d\Gamma_p$ is defined in Eq.~(\ref{eq:recoil_phasespace}), and $\mathcal{F}_{U, S, T}$
are the expressions containing the unpolarized, vector-polarized and tensor polarized structures
as defined in \firstpart, Eqs.~(\ref{P1:cross_section_unpol}), (\ref{P1:cross_section_vector}),
and (\ref{P1:cross_section_tensor}),
\begin{align}
\mathcal{F}_U = F_{[UU,T]\deut} + \epsilon F_{[UU,L]\deut} + ... \;\; \textrm{etc.}
\label{cross_section_deuteron}
\end{align}
The azimuthal angle in the expressions is now the final-state proton angle, $\phi_h \equiv \phi_p$,
defined as in \firstpart, Fig.~\ref{P1:fig:kin}.
The structure functions depend on the DIS variables $x$ and $Q^2$
and the final-state proton variables $\alpha_p$ and $|\bm{p}_{pT}|$,
\begin{align}
F_{[UU,T]\deut}(x, Q^2; \alpha_p, |\bm{p}_{pT}|) \;\; \textrm{etc.}
\end{align}

The semi-inclusive DIS process of Eq.~(\ref{eq:tagged_reaction}) can be measured and analyzed
without making any assumption about the composite structure of deuteron. A composite description
in terms of deuteron structure in nucleon degrees of freedom and a DIS process on the nucleon
will be constructed in Sec.~\ref{sec:lightfront}, using methods of LF quantization.

For reference we note that the differential cross section in Eq.~(\ref{cross_section_deuteron})
can also be represented in the form \cite{Jentsch:2021qdp}
\begin{subequations}
\label{cross_section_reduced}
\begin{align}
&d\sigma = \frac{2\pi \alpha_{\rm em}^2 y^2}{Q^4 (1 - \epsilon) x} \; dx\, dQ^2
\frac{d\psi_{l'}}{2\pi}
\; \times \; \sigma_{{\rm red}, \deut} \; d\Gamma_p ,
\\
&\sigma_{{\rm red}, \deut} \equiv \frac{x}{2}
\left( \mathcal{F}_U + \mathcal{F}_S + \mathcal{F}_T \right) ,
\end{align}
\end{subequations}
where the flux factor is the same as in DIS on the nucleon, and $\sigma_{{\rm red}, \deut}$ is the
reduced cross section for tagged DIS on the deuteron. It is defined such that, up to nuclear binding effects,
\begin{align}
\int d\Gamma_p  \, \sigma_{{\rm red}, \deut} \approx \sigma_{{\rm red}, n} = F_{2n} - (1 - \epsilon) F_{Ln},
\end{align}
which is the reduced cross section of DIS on the neutron in the convention used in high-energy
scattering experiments; see e.g.\ Ref.~\cite{H1:2009jxj}. This representation is useful for validating
the normalization of the theoretical expressions including nuclear structure, and for estimating
experimental cross sections and rates.

\subsection{Deuteron polarization}
\label{subsec:deuteron_polarization}
The deuteron in the initial state of the tagged DIS process Eq.~(\ref{eq:tagged_reaction}) is assumed
to be in a mixed polarization state with vector and tensor polarization.
The polarization is described by the spin density matrix as introduced in \firstpart,
Sec.~\ref{P1:sec:densitymatrix}. The covariant spin density matrix is parametrized by the deuteron polarization
4-vector and 4-tensor
\begin{align}
S_\deut^\alpha, \hspace{1em} T_\deut^{\alpha\beta},
\end{align}
whose components are specified by the invariant
polarization parameters\footnote{For brevity we do not put a
label $\deut$ on the invariant polarization parameters; it is clear that they refer to the deuteron.}
\begin{align}
\{ S_L, S_T, \phi_S \},
\hspace{1em}
\{ T_{LL}, T_{LT}, T_{TT}, \phi_{T_L} , \phi_{T_T} \}.
\label{polarization_parameters}
\end{align}
In the class of collinear frames the polarization vector and tensor are described by their
LF components and can be converted between frames in a simple manner using
Eq.~(\ref{boost_collinear}) \cite{Cosyn:2020kwu}.

In the nuclear structure calculations in Sec.~\ref{sec:lightfront} we will
start from the polarization vector and tensor in the deuteron rest frame and transport
that information to the other collinear frames through boosts. The rest-frame polarization 3-vector
and 3-tensor are expanded in the 3-dimensional basis vectors $\{ \bm{e}_{x'}, \bm{e}_{y'}, \bm{e}_{z}\}$
in the deuteron rest frame defined in \firstpart, Sec.~\ref{P1:subsec:collinear_frames}
and Fig.~\ref{P1:fig:kin}. The unit vector $\bm{e}_{z}$ is along the
collinear axis in the direction opposite to $\bm{q}$. The $x'$ direction is along
the final-state proton transverse momentum; the $y'$ direction is the right-handed normal to it;
the transverse unit vectors are thus given by
\begin{align}
\bm{e}_{x'} = \bm{p}_{pT} / |\bm{p}_{pT}|,
\hspace{1em}
\bm{e}_{y'} = \bm{e}_{z} \times \bm{e}_{x'}.
\label{basis_vectors_prime}
\end{align}

The rest-frame polarization vector is expanded in terms of the basis vectors
as in \firstpart, Eqs.~(\ref{P1:vector_restframe})
\begin{align}
\bm{S}_\deut &= S_L \bm{e}_{z}
\nonumber \\
&+ S_T [\cos (\phi_p - \phi_S) \bm{e}_{x'}
+ \sin (\phi_p - \phi_S) \bm{e}_{y'}],
\label{vector_restframe}
\end{align}
where $\phi_p$ is the azimuthal angle of $\bm{p}_{pT}$ and $\phi_S$
is the spin polarization angle, measured relative to the lepton scattering plane
as specified in \firstpart, Fig.~\ref{P1:fig:kin} ($\phi_p \equiv \phi_h$ in the figure).
The rest-frame polarization tensor is expanded in the 3-dimensional spherical tensors
formed from the basis vectors as in \firstpart, Eq.~(\ref{P1:tensor_restframe}),
\begin{align}
T^{ij}_\deut &=
{\textstyle\sqrt{\textstyle\frac{3}{2}}} \;
T_{LL} e_{LL}^{ij}
\nonumber \\[-.5ex]
&+ \sqrt{2} \; T_{LT} \cos(\phi_p-\phi_{T_L}) e_{LT}^{ij}
\nonumber \\
&+ \sqrt{2} \; T_{LT} \sin(\phi_p-\phi_{T_L}) e_{LT'}^{ij}
\nonumber \\
&+ \tfrac{1}{\sqrt{2}} \; T_{TT} \cos(2\phi_p-2\phi_{T_T}) e_{TT}^{ij}
\nonumber \\
&+ \tfrac{1}{\sqrt{2}} \; T_{TT} \sin(2\phi_p-2\phi_{T_T}) e_{TT'}^{ij} ;
\label{tensor_restframe}
\end{align}
the expressions of the 3-dimensional spherical tensors are given
in \firstpart, Eq.~(\ref{P1:basis_tensors}).

The values of the invariant polarization parameters Eq.~(\ref{polarization_parameters})
are determined by the experimental setup. The values for standard setups
(deuteron polarization relative to the electron beam axis in the deuteron rest frame)
can be computed using the formulas of \firstpart, Sec.~\ref{P1:subsec:preparation}.
The nuclear structure calculations here are aimed at the structure functions and
do not require the experimental values of the invariant polarization parameters.

In the nuclear structure calculations it will be useful to have the polarization
parameters corresponding to pure deuteron spin states in the rest frame.
In a pure spin state, with spin projection $\Lambda = (-1, 0, 1)$ along
an axis defined by the unit vector $\bm{N}$ in the deuteron rest frame, the
rest-frame polarization vector and tensor are given by
\begin{subequations}
\label{pure_state}
\begin{align}
\bm{S}_\deut &= \Lambda \bm{N},
\\[1ex]
T_\deut &= W(\Lambda) \, (-\tfrac{1}{6} + \tfrac{1}{2} \bm{N} \otimes \bm{N}),
\\[1ex]
W(\Lambda) &\equiv \textrm{$(1,-2,1)$ for $\Lambda = (+1,0,-1)$},
\end{align}
\end{subequations}
see \firstpart, Eq.~(\ref{P1:pure_state_density}).
Note that $\bm{S}_\deut \neq 0$ only in the states with spin projections $\Lambda = \pm 1$,
while $T_\deut \neq 0$ in all spin projections. The polarization vector and tensor
of Eq.~(\ref{pure_state}) can be expanded in the basis vectors and tensors
as in Eq.~(\ref{vector_restframe}) and (\ref{tensor_restframe}).
In particular, if the axis is along the longitudinal direction, $\bm{N} = \bm{e}_z$,
Eq.~(\ref{pure_state}) gives\footnote{The following expressions are
special cases of the general formula \firstpart, Eq.~(\ref{P1:pure_state_invariant}).}
\begin{subequations}
\label{pure_state_longitudinal}
\begin{align}
\bm{S}_\deut &= \Lambda \bm{e}_z,
\\[1ex]
T_\deut
&= W(\Lambda) \left[ \tfrac{1}{\sqrt{6}} e_{LL} \right] ,
\end{align}
\end{subequations}
and the invariant polarization parameters are
\begin{subequations}
\label{pure_state_longitudinal_invariant}
\begin{align}
&S_L = \Lambda, \hspace{1em} S_T = 0,
\\[1ex]
& \{ T_{LL}, T_{LT}, T_{TT} \} =
W(\Lambda) \times 
\{ \tfrac{1}{3}, 0, 0 \} .
\end{align}
\end{subequations}
If the axis is along a transverse direction, $\bm{N} = \bm{N}_T$ with $|\bm{N}_T| = 1$ and
azimuthal angle $\phi_N$ relative to the lepton plane (see \firstpart, Fig.~\ref{P1:fig:kin}),
\begin{align}
\bm{N}_T &= \cos (\phi_p - \phi_N) \bm{e}_{x'} + \sin (\phi_p - \phi_N) \bm{e}_{y'},
\end{align}
the rest-frame polarization parameters are
\begin{subequations}
\label{pure_state_transverse}
\begin{align}
\bm{S}_\deut &=
\Lambda \left[ \cos (\phi_p - \phi_N) \bm{e}_{x'} + \sin (\phi_p - \phi_N) \bm{e}_{y'} \right] ,
\\[1ex]
T_\deut
&= W(\Lambda) \left[ -\tfrac{1}{2\sqrt{6}} e_{LL}
+ \tfrac{1}{2\sqrt{2}} \cos (2 \phi_p - 2 \phi_N) e_{TT}
\right.
\nonumber \\
&\left. + \tfrac{1}{2\sqrt{2}} \sin (2 \phi_p - 2 \phi_N) e_{TT'}
\right] ,
\end{align}
\end{subequations}
and the invariant polarization parameters are
\begin{subequations}
\label{pure_state_transverse_invariant}
\begin{align}
&S_L = 0, \hspace{1em}  S_T = |\Lambda|,
\\[1ex]
& \{ T_{LL}, T_{LT}, T_{TT} \} =  W(\Lambda) \times \{ -\tfrac{1}{6}, 0, \tfrac{1}{2} \} ,
\\[1ex]
&\phi_S = \phi_N, \hspace{1em} \phi_{T_T} = \phi_N .
\end{align}
\end{subequations}
\section{Nuclear structure and DIS process}
\label{sec:lightfront}
\subsection{Light-front quantization}
The tagged DIS process involves strong interaction dynamics at very different scales:
the DIS process at energy and momentum transfers $\gg$ 1 GeV, and the nuclear binding
and breakup process at momenta $\sim$ few 10-100 MeV. The theoretical analysis aims to
separate the dynamics at the two scales and construct a composite description of the
nuclear scattering process in terms of nucleon degrees of freedom.

The description of high-energy scattering processes on nuclei (projectile energy
$\gg 1$ GeV in the nuclear rest frame) in terms of nucleon degrees of freedom presents
some specific challenges, resulting from the combination of quantum mechanics and relativity.
(i) Because of nuclear binding the projectile-nucleon scattering amplitudes are generally
``off-shell'', meaning off the energy shell in the time-ordered description of the process,
or off the mass shell in the covariant description. One must ensure that the off-shellness
does not lead to artifacts that grow proportionally to the projectile energy and become
large in the high-energy limit.
(ii) Because of relativity, non-nucleonic degrees of freedom of the nucleus (mesons,
excited baryons) can contribute as initial states of the high-energy scattering process.
One must justify why a truncation to nucleonic degrees of freedom is possible,
and what corrections to this approximation are expected.

LF quantization is an appropriate method for describing high-energy scattering
on relativistic composite systems \cite{Frankfurt:1981mk}. In this approach the structure
of the target is described at fixed LF time $x^+ = x^0 + x^3 = 0$, as ``seen'' by the high-energy
projectile propagating through the system \cite{Weinberg:1966jm,Brodsky:1997de}.
It has several unique features: (i) The off-shellness of the projectile-nucleon scattering
amplitudes remains finite in the high-energy limit \cite{Frankfurt:1981mk}. This permits the
matching of the off-shell amplitudes in scattering on the nucleus with on-shell amplitudes
measured in scattering of the free nucleon. (ii) The contributions of non-nucleonic degrees
of freedom as initial states of the scattering process remain finite in the high-energy
limit and can be controlled \cite{Frankfurt:1981mk}.
Using a dispersion-theoretical representation of the LF
nuclear scattering amplitudes, one can quantify the contributions of non-nucleonic degrees
of freedom in the initial state and study the convergence of the nuclear scattering
amplitudes as functions of the invariant mass of the nuclear initial state.
In high-energy scattering on the deuteron, the truncation to the $NN$ initial state
(neglecting $NN\pi$ and $\Delta\Delta$ configurations) is a good approximation if the
final state is restricted to breakup momenta $|\bm{p}_N| \lesssim$ few 100 MeV \cite{Frankfurt:1981mk}.
Altogether, LF quantization permits the factorization
of nuclear and nucleonic structure in high-energy scattering, with finite effects due to nuclear binding.

In the LF description of nuclear high-energy scattering the nucleus in the initial state
is described by a LF wave function in nucleon degrees of freedom. The LF wave function encodes
the low-energy nuclear structure as relevant for the high-energy process and is boost-invariant
(independent of the reference frame) \cite{Brodsky:1997de}. The frame independence allows
one to consider the LF wave function in the nuclear rest frame, where it can be matched with
the nonrelativistic wave function. In this way the extensive theoretical and empirical knowledge
of nonrelativistic nuclear few-body systems can be recruited for describing high-energy processes.

The LF description of nuclear high-energy scattering processes can be implemented using
two formulations:
\begin{enumerate}[a)]
\item a three-dimensional formulation based on LF quantum mechanics (LFQM);
\item a four-dimensional formulation based on Feynman diagrams, referred to as
the virtual nucleon approximation (VNA) \cite{Frankfurt:1996xx,Sargsian:2001ax}.
\end{enumerate}
When used as approximations with on-shell nucleon structure as input, the two formulations
give equivalent results up to instantaneous terms in $x^+$ (LFQM) or off-shell
contributions (VNA), which represent the systematic uncertainty arising from the
lack of complete knowledge of the dynamics. In the following we calculate the tagged
DIS cross section in the IA using first the LFQM
formulation (Secs.~\ref{subsec:quantum_mechanical}--Secs.~\ref{subsec:impulse_approximation}).
We then repeat the calculation in the VNA formulation and show that the results
are identical up to instantaneous or off-shell terms (Sec.~\ref{subsec:VNA}).

\subsection{Quantum-mechanical formulation}
\label{subsec:quantum_mechanical}
In LFQM the nucleon states are characterized by their LF momenta, consisting of the
plus component $p_N^+$ and the transverse momentum $\bm{p}_{NT}$, defined as in Eq.~(\ref{eq:lf_coords})
(here $N = p$ or $n$). The LF energy or minus component is fixed by the mass-shell condition $p_N^2 = m^2$
as 
\begin{align}
p_N^- \; = \; \frac{|\bm{p}_{NT}|^2 + m^2}{p_N^+} .
\label{pminus_onshell}
\end{align}
The spin degrees of freedom are described by the LF helicity $\lambda_N$ (details are
discussed in Sec.~\ref{subsec:spin_structure}). The nucleon states are denoted as
\begin{align}
| N (p_N, \lambda_N) \rangle \equiv | N (p_N^+, \bm{p}_{NT}, \lambda_N) \rangle;
\end{align}
for brevity we label them by the 4-momentum $p_N$; it is understood that the independent
variables are $p_N^+$ and $\bm{p}_{NT}$, and that $p_N^-$ is always fixed by
Eq.~(\ref{pminus_onshell}). The states are normalized as
\begin{align}
&\langle N (p_N', \lambda_N') | N (p_N, \lambda_N) \rangle
\; = \; (2\pi)^3 \, 2 p_N^+ \, \delta(p_N^{+\prime} - p_N^+) \;
\nonumber \\
&\hspace{3em} \times \delta( \bm{p}_{NT}^{\prime} - \bm{p}_{NT}) \delta (\lambda_N', \lambda_N).
\label{normalization_n}
\end{align}

The deuteron state is expanded in $pn$ states. We assume that the deuteron state has
arbitrary LF plus momentum $p_\deut^+ > 0$ and transverse momentum $\bm{p}_{\deut T} = 0$;
this is sufficient for describing DIS processes in the class of collinear frames
(see Sec.~\ref{subsec:collinear_frames}).
Because LF momentum is conserved in LFQM, the LF momenta of the
$p$ and $n$ in the $pn$ configurations in the deuteron satisfy
\begin{subequations}
\label{pn_lf_momenta}
\begin{align}
&p_p^+ + p_n^+ = p_\deut^+,\\
&\bm p_{pT} + \bm p_{nT} = \bm p_{\deut T} = 0.
\label{pn_transverse_momentum}
\end{align}
\end{subequations}
The plus components are parametrized as fractions of the deuteron plus momentum
[see Eq.~(\ref{eq:alpha_proton_def})]
\begin{align}
p_p^+ = \alpha_p \frac{p_\deut^+}{2},
\hspace{1em}
p_n^+ = \alpha_n \frac{p_\deut^+}{2},
\hspace{1em}
\alpha_p + \alpha_n = 2.
\end{align}
The expansion of the deuteron state in $pn$ states is written in the form
\begin{align}
&\langle p(p_p, \lambda_p), n(p_n, \lambda_n) | \deut (p_\deut, \lambda_\deut) \rangle
\nonumber \\[.5ex]
&= (2\pi)^3 \, 2 p_\deut^+ \, \delta (p_p^+ + p_n^+ - p_\deut^+) \,
\delta^{(2)}(\bm{p}_{pT} + \bm{p}_{nT}) \;
\nonumber \\
&\times \; (2\pi)^{3/2} \Psi_\deut (\alpha_p , \bm{p}_{pT}; \lambda_p, \lambda_n|\lambda_\deut),
\label{eq:lfqm_wf}
\end{align}
where the LF wave function $\Psi_\deut$ depends on the proton LF momentum variables
$\alpha_p$ and $\bm{p}_{pT}$; the corresponding values for the neutron in the configuration are
\begin{align}
\alpha_n = 2 - \alpha_p, \hspace{2em} \bm{p}_{nT} = -\bm{p}_{pT} .
\label{momentum_neutron}
\end{align}
The wave function is normalized such that
\begin{align}
& \sum_{\lambda_p, \lambda_n}
\int \frac{d\alpha_p \; d^2 p_{pT}}{\alpha_p (2 - \alpha_p)} \;
\Psi_\deut^\ast (\alpha_p, \bm{p}_{pT}; \lambda_p, \lambda_n | \lambdadp)
\nonumber \\
& \times \Psi_\deut (\alpha_p, \bm{p}_{pT}; \lambda_p, \lambda_n | \lambda_\deut)
\;\; = \;\; \delta (\lambdadp, \lambda_\deut) .
\label{wf_general_normalization}
\end{align}
The LF wave function is invariant under longitudinal boosts; it does not depend on $p_\deut^+$
but only on the boost-invariant fraction $\alpha_p$. Under transverse boosts it transforms
kinematically \cite{Brodsky:1997de};
only states with $\bm{p}_{\deut T} = 0$ are needed in the present calculation.

In the expansion Eq.~(\ref{eq:lfqm_wf}) the LF energy of the $pn$ states is different from
that of the deuteron state,
\begin{align}
p_p^- + p_n^- \neq p_\deut^- .
\label{lf_energy_difference}
\end{align}
It implies that the 4-momentum of the $pn$ states is different from that of the deuteron state
\begin{align}
p_p + p_n \neq p_\deut .
\label{lfqm_four_momenta}
\end{align}
The transition is therefore accompanied by a difference of the invariant masses of the states,
defined as the square of their 4-momenta
\begin{align}
&M_{pn}^2 \equiv (p_p + p_n)^2,
\hspace{2em}
M_\deut^2 = p_\deut^2.
\end{align}
The invariant mass of the $pn$ state is a function of the LF momentum variables
\begin{align}
M_{pn}^2 \equiv
M_{pn}^2 (\alpha_p , \bm{p}_{pT})
= \frac{4(|\bm{p}_{pT}|^2 + m^2)}{\alpha_p (2 - \alpha_p)}
\label{invariant_mass_def}
\end{align}
(this form applies to the states with zero overall transverse momentum).
The LF energy difference Eq.~(\ref{lf_energy_difference}) is proportional to the invariant
mass difference between the $pn$ state and the deuteron state,
\begin{align}
&p_p^- + p_n^- - p_\deut^- = \frac{M_{pn}^2 - M_\deut^2}{p_\deut^+} .
\label{lf_energy_difference_from_invariant_mass}
\end{align}
The invariant mass difference thus acts as the invariant representation of the
LF ``energy denominator'' and plays a central role in LFQM.
In the deuteron rest frame ($p_\deut^+ = M_\deut$), expressing the LF variables
$\alpha_p$ and $\bm{p}_{pT}$ in terms of the proton 3-momentum $\bm{p}_p$,
and expanding in $|\bm{p}_p|/m$, one obtains
\begin{align}
M^2_{pn} - M^2_\deut = 4 m \left( \frac{|\bm{p}_p|^2}{m} + \epsilon_\deut \right)
\left[ 1 + \mathcal{O}\left(\frac{p_p^z}{m} \right) \right] ,
\label{invariant_mass_nonrelativistic}
\end{align}
and the invariant mass difference becomes proportional to the energy difference between the $pn$
and deuteron states in the nonrelativistic theory.

The deuteron LF wave function satisfies a two-body bound-state equation with an $NN$ potential
or effective interaction at fixed LF time \cite{Frankfurt:1981mk,Cooke:2001kz}.
The equation determines in particular the analytic structure of the wave function.
On general grounds the wave function can be written in the form
\begin{align}
&\Psi_\deut (\alpha_p , \bm{p}_{pT}; \lambda_p, \lambda_n| \lambda_\deut)
\nonumber \\
& = (2\pi)^{-3/2} \; 
\frac{\Gamma_{\deut pn} (\alpha_p , \bm{p}_{pT}; \lambda_p, \lambda_n| \lambda_\deut)}
{M^2_{pn} (\alpha_p , \bm{p}_{pT}) - M_\deut^2}.
\label{wf_invariant_mass}
\end{align}
The invariant mass difference in the denominator of Eq.~(\ref{wf_invariant_mass})
is independent of the $NN$ interaction.
It describes the propagation of the nucleons in intermediate states between interactions.
The form Eq.~(\ref{wf_invariant_mass}) contains the free nucleon pole (Bethe-Peierls pole),
which occurs at $|\bm{p}_p|^2 + m \epsilon_\deut = 0$ in the nonrelativistic approximation
of Eq.~(\ref{invariant_mass_nonrelativistic}) and is a universal feature of a two-body
system bound by a short-range interaction. The vertex function in the numerator
of Eq.~(\ref{wf_invariant_mass}) depends on the $NN$ interaction. It describes the binding
effects and the spin structure of the system. It can be determined by solving the LF
bound-state equation with specific models of the $NN$ interaction. Alternatively, it can
be obtained by matching the LF and non-relativistic deuteron wave functions in the rest frame.

\subsection{Spin structure}
\label{subsec:spin_structure}
In LFQM the spin states of particles are chosen as LF helicity states \cite{Soper:1972xc,Brodsky:1997de}.
The LF helicity states are obtained from the rest-frame spin states through
a sequence of LF boosts along the longitudinal and transverse momentum directions.
They differ from the canonical spin states by the so-called Melosh
rotation \cite{Melosh:1974cu} if the transverse momentum of the state is nonzero, $\bm{p}_T \neq 0$.
In the deuteron bound state the nucleons have transverse momenta $\bm{p}_{pT} = -\bm{p}_{nT} \neq 0$,
and their LF helicity states are affected by the rotation. The deuteron state as a whole has
$\bm{p}_{\deut T}=0$ in the collinear frames, and its LF helicity states are the same as canonical spin states.

In formulating the LF helicity structure of the deuteron one faces the challenge that
3-dimensional rotational invariance is not manifest in LF quantization.
In equal-time quantization (e.g. in nonrelativistic theory), where rotational invariance is
manifest, the deuteron has two internal spin states, corresponding to the $S$ and $D$ waves
of the orbital motion. The LF version of this structure can be obtained
by expressing the LF wave function in a form that exhibits rotational invariance
and can be matched with the spin structure in equal-time quantization.
A practical way to do this is to express the deuteron LF wave function as a function of
the 3-momentum of the nucleons in the c.m.\ frame of the $pn$
system \cite{Frankfurt:1981mk,Kondratyuk:1983kq,Frankfurt:1983qs,Cosyn:2020kwu}.

The c.m.\ frame is the special collinear frame in which the LF
components of the total 4-momentum of the $pn$ system satisfy
\begin{align}
(p_p + p_n)^+ &= (p_p + p_n)^-
\end{align}
such that
\begin{align}
(p_p + p_n)^z &= \frac{1}{2} \left[ (p_p + p_n)^+ - (p_p + p_n)^- \right] = 0. 
\label{pz_at_rest}
\end{align}
From Eqs.~(\ref{pminus_onshell}) and (\ref{invariant_mass_def}) one can see that this
is realized in the collinear frame with $p_\deut^+ = M_{pn}$. Note that this frame is
not the deuteron rest frame, but a frame that differs from it by a boost with
parameter $\lambda = M_{pn}/M_\deut$, see Eq.~(\ref{boost_collinear}).
The condition Eq.~(\ref{pz_at_rest}), together with $\bm{p}_{pT} + \bm{p}_{nT} = 0$ valid
in any collinear frame, implies that the ordinary 3-momentum of the $pn$ system is zero,
and the 4-momentum is [here $(p^0, \bm{p})$ denote the ordinary 4-vector components]
\begin{align}
p_p + p_n &= (M_{pn}, \bm{0}).
\end{align}
In this frame the $p$ and $n$ have equal and opposite ordinary 3-momenta,
and their 4-momenta can be written as
\begin{subequations}
\begin{align}
& p_p = (E, \bm{k}), \hspace{2em}
p_n = (E, -\bm{k}),
\\[2ex]
&E \equiv E(\bm{k}) \equiv \sqrt{|\bm{k}|^2 + m^2}.
\end{align}
\end{subequations}
The relation of the c.m.\ momentum $\bm{k}$ to the LF variables $\alpha_p$ and $\bm{p}_{pT}$ is
\begin{subequations}
\label{CM_LF_variables}
\begin{align}
k^z &= \frac{M_{pn}}{2} (\alpha_p - 1), 
\hspace{2em}
\bm{k}_T = \bm{p}_{pT} ,
\\[1ex]
\alpha_p &=  1 + \frac{k^z}{E},
\hspace{2em}
M_{pn} = 2 E
\end{align}
\end{subequations}
The relation between the integration measures is
\begin{align}
\int\frac{d\alpha_p \; d^2 p_{pT}}{\alpha_p (2 - \alpha_p)}  
\; = \; \int \frac{d^3 k}{E(k)} .
\label{integration_k}
\end{align}
The c.m.\ momentum thus serves as an alternative variable to the LF momenta and can be
used as argument of the LF wave function. This representation allows one to formulate
the LF helicity structure of the wave functions such that it respects
3-dimensional rotational covariance.

The explicit form of the deuteron LF wave function in terms of the c.m.\ momentum variable
has been derived in Refs.~\cite{Kondratyuk:1983kq,Frankfurt:1983qs,Cosyn:2020kwu};
we follow the notation of Ref.~\cite{Cosyn:2020kwu}.
\begin{subequations}
\label{wf_3d}
\begin{align}
&\Psi_\deut (\alpha_p , \bm{p}_{pT}; \lambda_p, \lambda_n | \lambda_\deut)
= \sum_{\lambdapp, \lambdanp}
\widetilde \Psi_\deut (\bm{k}, \lambdanp, \lambdapp | \lambda_\deut )
\nonumber \\
& \times U^\ast (\bm{k}, \lambdapp, \lambda_p ) \; U^\ast(-\bm{k}, \lambdanp, \lambda_n ) ,
\label{wf_k_3d_lf}
\\[2ex]
&\widetilde \Psi_\deut (\bm{k}, \lambdanp, \lambdapp | \lambda_\deut )
\nonumber \\
&= \frac{1}{\sqrt{2}} \epsilon_\deut^i (\lambda_\deut) \; 
\left[ 
\delta^{ij} f_0(k) \; + \; \frac{1}{\sqrt{2}} \left( \frac{3 k^i k^j}{k^2} - \delta^{ij} 
\right) f_2(k)
\right]
\nonumber \\
& \times \chi^\dagger (\lambdanp) \left[ \sigma^j (i \sigma^2) \right] \, \chi^\ast (\lambdapp) ,
\label{wf_k_3d}
\\[2ex]
&U (\bm{k}, \lambdapp, \lambda_p )
\nonumber \\
&= \; \chi^\dagger (\lambdapp) \left[
\frac{E + k^z + m + \bm{k}_T \bm{\sigma}_T \sigma^3}{\sqrt{2 (E + k^z) (E + m)}} \right]
\chi(\lambda_p) ,
\label{U_proton}
\\[2ex]
&U (-\bm{k}, \lambdanp, \lambda_n )
\nonumber \\
&= \; \chi^\dagger (\lambdanp) \left[
\frac{E - k^z + m - \bm{k}_T \bm{\sigma}_T \sigma^3}{\sqrt{2 (E - k^z) (E + m)}} 
\right] \chi(\lambda_n) ,
\label{U_neutron}
\end{align}
\end{subequations}
where $\bm{k}$ is related to $\alpha_p$ and $\bm{p}_{pT}$ by Eq.~(\ref{CM_LF_variables}).
$\widetilde \Psi_\deut$ in Eq.~(\ref{wf_k_3d}) is the 3-dimensional relativistic wave function in the
c.m.\ momentum $\bm{k}$ and the canonical spin variables $\lambdapp$ and $\lambdanp$. The 3-vector
$\epsilon^i_\deut (\lambda_\deut)$ is the polarization 3-vector of the $pn$ configuration in the c.m.\
frame  and is identical to the deuteron polarization 3-vector in the deuteron rest frame.
The 2-spinors $\chi (\lambda_p')$ and $\chi (\lambda_n')$ describe the
spin degrees of freedom of the proton and neutron at rest, quantized along the $z$-axis.
The form of $\widetilde\Psi_\deut$ is constrained by 3-dimensional rotational
invariance in $\bm{k}$ and includes the $S$- and $D$-wave of the orbital motion,
similar to the nonrelativistic deuteron wave function. The radial wave functions
are normalized as ($k \equiv |\bm{k}|$)
\begin{align}
& 4\pi \int \frac{dk \, k^2}{E(k)} [f_0^2(k) + f_2^2(k)] \; = \; 1 .
\label{normalization_radial}
\end{align}
The functions $U (\bm{k}, \lambdapp, \lambda_p )$ and $U (-\bm{k}, \lambdanp, \lambda_n )$
in Eqs.~(\ref{U_proton}) and (\ref{U_neutron}) are the Melosh rotations of the proton
and neutron spins. They connect the canonical spin variables $\lambdapp$ and $\lambdanp$
with the LF helicities $\lambda_p$ and $\lambda_n$ and depend on the proton/neutron
momentum in the c.m. frame, with $U(\bm{k}_T = 0) = 1$. The star '$^\ast$' in Eq.~(\ref{wf_3d})
denotes complex conjugation, which appears because the wave function describes the
matrix element $\langle p n| D\rangle$, Eq.~(\ref{eq:lfqm_wf}), where the proton/neutron
appear in the complex conjugate state.

Equation~(\ref{wf_3d}) serves as a general parametrization of the deuteron LF wave function
and is used in all applications in the present study. The dynamical content is in the
radial wave functions in the c.m.\ momentum variable, $f_L(k) (L = 0, 2)$.
They can be obtained by solving the dynamical equation
for the 2-body bound state in LFQM with specific models of the $NN$ interaction
\cite{Frankfurt:1981mk,Frankfurt:1992ny,Cooke:2001kz}. Alternatively, the radial
wave functions can be approximated by the nonrelativistic radial wave functions as
\begin{align}
f_L(k) \; = \; \sqrt{E(k)} \, f_{L, {\rm nr}}(k) \hspace{2em} (L = 0, 2),
\label{nonrel_approx}
\end{align}
where the factor $\sqrt{E}$ accounts for difference between the normalization
Eq.~(\ref{normalization_radial}) and the standard nonrelativistic normalization
\begin{align}
& 4\pi \int dk \, k^2 [f_{0, {\rm nr}}^2(k) + f_{2, {\rm nr}}^2(k)] \; = \; 1 .
\label{normalization_radial_nonrel}
\end{align}
The approximation Eq.~(\ref{nonrel_approx}) has been shown to be accurate for
$|\bm{k}| \lesssim$ 300 MeV \cite{Frankfurt:1981mk,Frankfurt:1992ny}
and is used in the present study.

LFQM in principle permits ``nonspherical'' components of the deuteron wave function,
beyond the $S$- and $D$-waves in the c.m.\ frame included in the parametrization Eq.~(\ref{wf_3d})
\cite{Carbonell:1995yi,Sargsian:2022rmq}. The incomplete P-wave described in
Ref.~\cite{Sargsian:2022rmq} is connected with the presence of non-nucleonic degrees of
freedom in the deuteron at very large c.m.\ momenta $|\bm{k}| \gtrsim$ 800 MeV,
which causes a departure from the logic based on the requirement of rotational invariance
in the $NN$ sector. These structures can be neglected at the nucleon momenta
$\sim$ few 100 MeV considered in the present study.

The deuteron LF wave function can also be represented in 4-dimensional form, as a sum of
bilinear forms in the nucleon 4-spinors describing the LF helicity states, contracted with the
4-dimensional polarization vector of the $pn$ system
\cite{Kondratyuk:1983kq,Frankfurt:1983qs,Cosyn:2020kwu}. The invariant functions accompanying the
bilinear forms can be connected with the radial functions in the c.m.\ frame parametrization
Eq.~(\ref{wf_3d}). The 4-dimensional form of the deuteron LF wave function permits the
evaluation of spin sums using gamma matrix algebra. It will not be used explicitly in the
present study, but results obtained with it will be quoted in Sec.~\ref{sec:effective}.
\subsection{Impulse approximation}
\label{subsec:impulse_approximation}
\begin{figure}[t]
\begin{center}
\includegraphics[width=0.85\columnwidth]{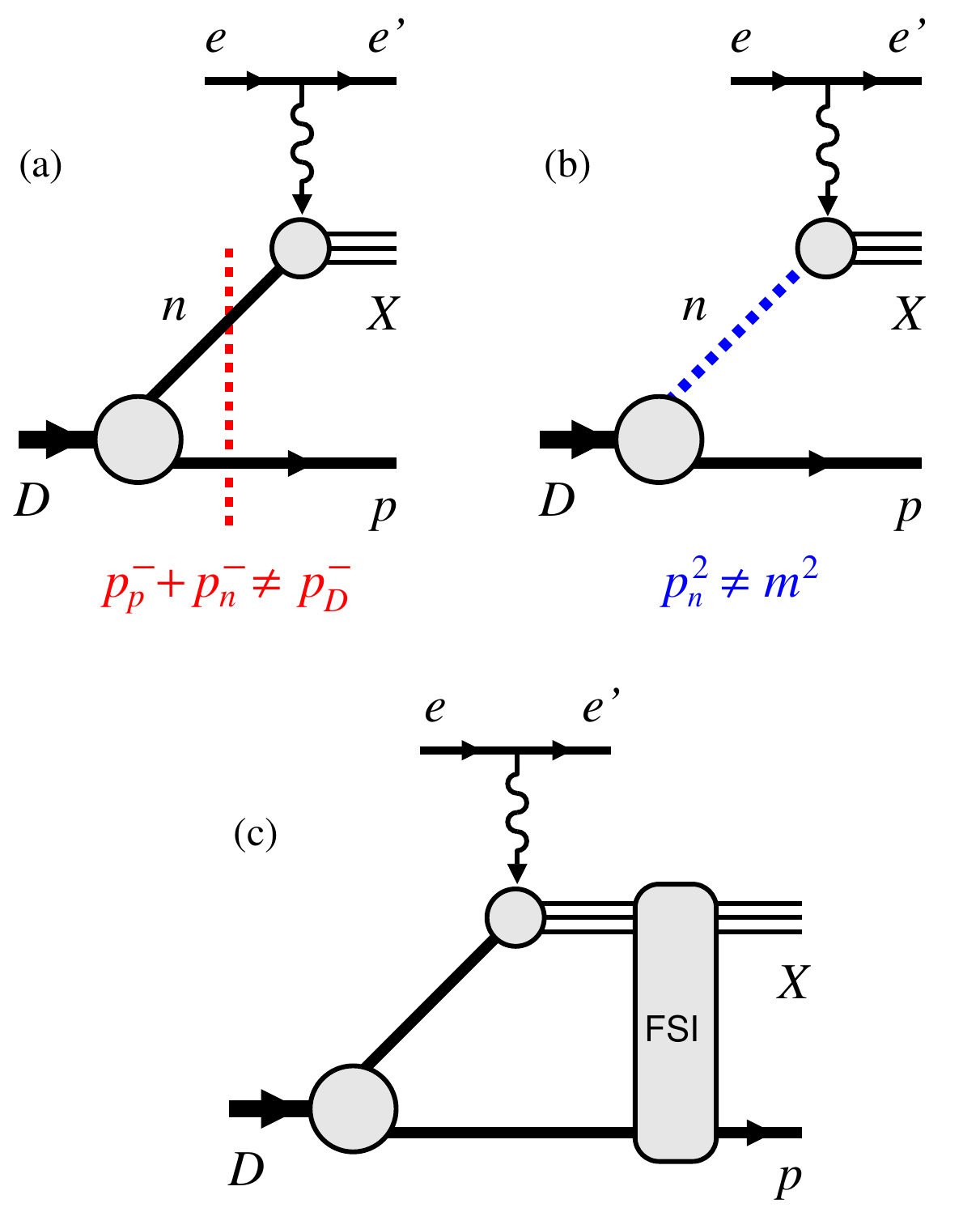}
\end{center}
\caption{Composite description of tagged DIS process in LF quantization.
(a)~IA in quantum-mechanical formulation (LFQM).
(b)~IA in virtual nucleon formulation (VNA).
(c)~FSI.}
\label{fig:IA_FSI}
\end{figure}

LF quantization permits a composite description of tagged DIS on the deuteron as
a DIS process on the nucleons and a distribution of the nucleons in the deuteron.
One distinguishes two contributions (see Fig.~\ref{fig:IA_FSI}): 
\begin{enumerate}[a)]
\item
The impulse approximation (IA), where the electromagnetic current interacts with a
single nucleon, and the final state produced in the DIS process
evolves independently from the other nucleon (``spectator'') (see Fig.~\ref{fig:IA_FSI}a).
The tagged nucleon in the final state is assumed to be the spectator nucleon
(production of slow nucleons in the DIS process is kinematically restricted and
can be neglected at $x \gtrsim 0.1$ \cite{Strikman:2017koc}).
Measurement of the the tagged nucleon LF momentum fixes the relative momentum of
the nucleons in $pn$ configuration, and therefore the LF momentum of active nucleon.
\item Final-state interactions (FSI), where certain components of the DIS final state
interact with the spectator (see Fig.~\ref{fig:IA_FSI}c).
The interactions transfer momentum to the spectator nucleon and require integration
over all relevant $pn$ configurations in the initial state \cite{Strikman:2017koc}
The $pn$ configurations in the initial state are therefore only indirectly related
to the LF momentum of the tagged nucleon. The phase space for the integration strongly
grows with the tagged nucleon momentum, leading to a strong kinematic dependence
of the FSI effects. The FSI process can also transfer quantum numbers (charge, spin)
to the spectator nucleon.
\end{enumerate}
%
%

The IA dominates the tagged DIS cross section at low momenta of the tagged nucleon,
$|\bm{p}_p| \lesssim$ 300 MeV in the deueteron rest frame, where phase space for
FSI is suppressed \cite{Strikman:2017koc}. The IA becomes exact
when extrapolating to the free nucleon pole, which occurs at a slightly
unphysical value of the tagged proton nucleon momentum; this feature is related
to the analytic properties of the deuteron wave function and can be used to
extract free neutron structure from tagged
DIS \cite{Sargsian:2005rm,Strikman:2017koc,Cosyn:2019hem,Jentsch:2021qdp}.
In the present study we use the IA to compute the cross section of tagged DIS
on the deuteron with general vector and tensor polarization.
The IA provides the baseline for the theoretical and experimental analysis
and permits simple study of the various polarization effects.
Dynamical models of the FSI in unpolarized tagged DIS were developed in
Refs.~\cite{CiofidegliAtti:2000xj,CiofidegliAtti:2003pb,Cosyn:2011jnm,Cosyn:2015mha,%
Cosyn:2017ekf,Strikman:2017koc}.
The inclusion of FSI in polarized tagged DIS requires further theoretical
development and will be the object of a further study.

We now compute the polarized tagged DIS cross section in the IA using LFQM \cite{Strikman:2017koc}.
In this formulation the scattering process is viewed as a transition between
quantum-mechanical states induced by the electromagnetic current operator.
The nuclear initial state is expanded in free nucleon states using the LF wave function,
and the electromagnetic current couples to a single free nucleon with definite LF momentum.
The single-nucleon states are on the mass shell, but the LF energy of the two-nucleon
intermediate states is different from that of the initial/final state.
This formulation is in close correspondence to the IA in nonrelativistic
quantum mechanics \cite{Benhar:2005dj}.

The calculation is performed in the frame where the deuteron and virtual photon momenta
are collinear and along the $z$-axis, see Eq.~(\ref{collinear_frame}) and \firstpart.
We consider the current matrix element for the
$\deut \rightarrow X + p$ transition (see Fig.~\ref{fig:IA_FSI}),
\begin{align}
\langle X, p(p_p, \lambda_p) | J^\mu| \deut (p_\deut, \lambda_\deut) \rangle ,
\end{align}
and insert a complete set of intermediate proton and neutron states, using
\begin{align}
1_{pn} &= \sum_{\lambda_p', \lambda_n} 
\int \frac{dp_n^+ d\bm p_{nT}}{(2\pi)^3 2p_n^+}
\int \frac{dp_p^{+\prime} d\bm p_{pT}'}{(2\pi)^3 2p_p^{+\prime}}
\nonumber \\
&\times | p(p_p', \lambda_p'), n(p_n, \lambda_n) \rangle
\langle p(p_p', \lambda_p'), p(p_n, \lambda_n)| 
\end{align}
(we put a prime on the intermediate-state proton variables to distinguish them from those of
the final-state proton). Assuming that the current couples to the neutron,
and expressing the $\deut \rightarrow pn$ matrix element in terms of the deuteron LF wave function 
Eq.~(\ref{eq:lfqm_wf}), we obtain
\begin{align}
&\lefteqn{\langle X, p(p_p, \lambda_p)| J^\mu| \deut (p_\deut, \lambda_\deut) \rangle}
\nonumber \\[1ex]
&= \sum_{\lambda_p', \lambda_n} 
\int \frac{dp_n^+ d\bm p_{nT}}{(2\pi)^3 2p_n^+}
\int \frac{dp_p^{+\prime} d\bm p_{pT}'}{(2\pi)^3 2p_p^{+\prime}}
\nonumber \\[1ex]
&\times \langle X |J^\mu| n(p_n, \lambda_n) \rangle \langle p(p_p, \lambda_p) | p(p_p', \lambda_p') \rangle
\nonumber \\[2ex]
&\times \langle p(p_p', \lambda_p'), n(p_n, \lambda_n) | \deut (p_\deut, \lambda_\deut) \rangle 
\nonumber
\\[2ex]
&= \sum_{\lambda_n} \langle X|J^\mu|n(p_n, \lambda_n) \rangle
\nonumber \\
&\times  \frac{2 (2\pi)^{3/2}}{2 - \alpha_p}
\Psi_\deut (\alpha_p, \bm{p}_{pT}; \lambda_p, \lambda_n| \lambda_\deut).
\label{eq:IA_lfqm}
\end{align}
The neutron LF momentum variables are fixed in terms of the final-state proton variables
by Eq.~(\ref{momentum_neutron}),
\begin{align}
\alpha_n = 2 - \alpha_p, \hspace{2em} \bm{p}_{nT} = -\bm{p}_{pT} .
\label{momentum_neutron_alt}
\end{align}
The factor
\begin{align}
p_\deut^+/p_n^+ = 2/\alpha_n = 2/(2 - \alpha_p) 
\end{align}
in Eq.~(\ref{eq:IA_lfqm})
arises from the normalization of the nucleon and deuteron states, Eq.~(\ref{eq:lfqm_wf}),
and represents a relativistic flux factor specific to LF quantization.
The hadronic tensor for tagged DIS, see Eq.~(\ref{P1:eq:hadronictensor}) in {\firstpart}, is
then obtained as
\begin{align}
&W^{\mu\nu}_\deut (p_\deut, q, p_p | \lambdadp, \lambda_\deut)
\nonumber \\[2ex]
&= (4\pi )^{-1} \sum_{\lambda_p} \sum_X \; (2\pi)^4\delta^4(p_\deut + q - p_p - p_X)
\nonumber\\
&\times 
\langle \deut (p_\deut, \lambdadp) | J^\mu| X, p(p_p, \lambda_p) \rangle
\nonumber\\[2ex]
&\times 
\langle X, p(p_p, \lambda_p) | J^\nu| \deut (p_\deut, \lambda_\deut) \rangle
\nonumber \\[2ex]
&= \frac{2 [2(2\pi)^3]}{(2 - \alpha_p)^2} \sum_{\lambda_p} \sum_{\lambda_n, \lambda_n'} 
\nonumber \\[1ex]
&\times
\Psi_\deut^\ast (\alpha_p, \bm{p}_{pT}; \lambda_p, \lambda_n'| \lambda_\deut')
\Psi_\deut      (\alpha_p, \bm{p}_{pT}; \lambda_p, \lambda_n|  \lambda_\deut)
\nonumber
\\[2ex]
&\times (4\pi)^{-1} \sum_{X} (2\pi)^4\delta^4(q + p_\deut - p_p - p_X)
\nonumber
\\
&\times
\langle n(p_n, \lambda_n') |J^{\dagger\mu}(0)|X \rangle \langle X|J^\nu(0)| n(p_n, \lambda_n) \rangle .
\label{eq:htensor2}
\end{align}
The particular arrangement of the prefactors is explained in the following.

The hadronic tensor Eq.~(\ref{eq:htensor2}) is a matrix in the LF helicities of the
initial-state deuteron appearing in the current matrix element and its complex-conjugate,
$\lambda_\deut$ and $\lambda_\deut'$. As such it is to be averaged with the deuteron
spin density matrix in the LF helicity representation, see
Eq.~(\ref{P1:eq:hadronictensor_contracted}) in {\firstpart},
\begin{align}
&\langle W^{\mu\nu}_\deut \rangle (p_\deut, q, p_p)
\nonumber \\
&= \sum_{\lambda_\deut, \lambdadp} \rho_\deut(\lambda_\deut, \lambdadp) \;
W^{\mu\nu}_\deut (p_\deut, q, p_p | \lambdadp, \lambda_\deut) .
\label{hadronic_tensor_deuteron_averaged}
\end{align}
The final-state proton helicity $\lambda_p$ is summed over in Eq.~(\ref{eq:htensor2}),
since the proton polarization is not observed in measurements considered here.

The deuteron hadronic tensor Eq.~(\ref{eq:htensor2}) can be expressed in terms of the hadronic tensor
of the active neutron. This allows one to connect the cross section of tagged scattering on the
deuteron with that of scattering on the free neutron.
In this step must take into account the energy off-shellness in LFQM,
which causes non-conservation of 4-momentum in the intermediate of the scattering process,
see Eq.~(\ref{lfqm_four_momenta}). In the IA hadronic tensor this implies
\begin{align}
p_n & \neq p_\deut - p_p  \hspace{2em} \textrm{(4-vectors)},
\end{align}
so that the ``internal'' 4-momentum of neutron is not equal to the difference
of ``external'' 4-momenta $p_\deut$ and $p_p$.
The neutron hadronic tensor in Eq.~(\ref{eq:htensor2}) therefore cannot be identified
with the free neutron hadronic tensor with 4-momenta $q$ and $p_n$, as this would lead to
\begin{align}
q + p_n &\neq q + p_\deut - p_p = p_X ,
\end{align}
so that 4-momentum is not nonconserved between the initial and final state of the virtual
photon-neutron scattering process. However, the neutron hadronic tensor in Eq.~(\ref{eq:htensor2})
can be identified with the free neutron tensor at the effective 4-momentum transfer
\begin{align}
\widetilde{q} &\equiv q + p_\deut - p_n - p_p \neq q,
\label{qtilde_def}
\end{align}
which takes into account that the virtual photon-neutron interaction occurs in an
intermediate state where the 4-momentum of the nucleons is shifted. With Eq.~(\ref{qtilde_def})
\begin{align}
\widetilde q + p_n = q + p_\deut - p_p = p_X,
\end{align}
and 4-momentum is conserved in the virtual photon-neutron scattering process.
The neutron hadronic tensor at the effective momentum transfer is defined as
\begin{align}
&W_n^{\mu\nu} (p_n, \widetilde{q}| \lambda_n', \lambda_n)
\nonumber \\[1ex]
&\equiv \; (4\pi)^{-1} \sum_{X} 
\; (2\pi)^4 \; \delta^{(4)} (\widetilde{q} + p_n - p_{X})
\nonumber \\
&\times \langle n(p_n, \lambda_n') | J^{\dagger\mu} (0) | X \rangle \;
\langle X | J^\nu (0) | n(p_n, \lambda_n) \rangle .
\end{align}
Substituting this expression in Eq.~(\ref{eq:htensor2}), the spin-averaged
deuteron hadronic tensor Eq.~(\ref{hadronic_tensor_deuteron_averaged})
is finally obtained as
\begin{align}
&\langle W^{\mu\nu}_\deut \rangle (p_\deut, q, p_p)
\nonumber \\[1ex]
& = \frac{2 [2 (2\pi)^3]}{(2 - \alpha_p)^2} \; \sum_{\lambda_\deut, \lambdadp}
\rho_\deut(\lambda_\deut, \lambdadp) \; 
\sum_{\lambda_p} \sum_{\lambda_n, \lambda_n'}
\nonumber \\
& \times \Psi_\deut^\ast  (\alpha_p , \bm{p}_{pT}; \lambda_p, \lambda_n'| \lambdadp)
\Psi_\deut       (\alpha_p , \bm{p}_{pT}; \lambda_p, \lambda_n|  \lambda_\deut) \; 
\nonumber \\[2ex]
&\times W_n^{\mu\nu} (p_n, \widetilde{q}| \lambda_n', \lambda_n ),
\label{w_impulse}
\end{align}
where the neutron LF momentum variables are again given by Eq.~(\ref{momentum_neutron_alt}).
Equation~(\ref{w_impulse}) is the IA result for the tagged DIS cross section in LFQM
and exhibits the factorization of deuteron and nucleon structure, in the form of
a product of a density in the deuteron LF wave function and the effective neutron hadron tensor.
In Sec.~\ref{sec:IA_strucfunc} we use Eq.~(\ref{w_impulse}) to derive the expressions of the
various structure functions appearing in the tagged DIS cross section.

The arrangement of the prefactors in Eq.~(\ref{w_impulse}) is explained as follows.
The factor $[2 (2\pi)^3]$ ensures that the tagged cross section is differential in the
physical phase space element Eq.~(\ref{eq:recoil_phasespace}) with the normalization
of the LF wave function given by Eq.~(\ref{wf_general_normalization}).
The factor $1/(2 - \alpha_p)^2$ is a relativistic flux factor arising from LF quantization.
The factor 2 provides the connection between the deuteron and nucleon hadronic tensors
and ensures the proper relation between the structure functions (see Sec.~\ref{sec:IA_strucfunc}).

Some  comments  on  the  accuracy  of  the  LFQM  calculation  are  in order.
First, the 4-momentum nonconservation arising from the off-shellness of the LF energy does
not produce large invariants in the DIS limit,
\begin{align}
p_\deut q \sim Q^2 \rightarrow  \infty, \hspace{2em} Q^2/(p_\deut q) = x \;\; \textrm{fixed.}
\label{dis_limit}
\end{align}
Using the LF components of $p_n + p_p - p_\deut$ given by Eqs.~(\ref{pn_lf_momenta})
and (\ref{lf_energy_difference_from_invariant_mass}), and
the LF components of $q$ in the collinear frame given by Eq.~(\ref{collinear_frame}b),
one obtains \cite{Frankfurt:1981mk,Strikman:2017koc}
\begin{align}
&(p_p + p_n - p_\deut) q = \frac{1}{2} (p_p^- + p_n^- - p_\deut^-) q^+
\nonumber \\
&= - \frac{\xi}{4} (M_{pn}^2 - M_\deut^2) \approx - \frac{x}{4} (M_{pn}^2 - M_\deut^2),
\label{offshell_power_suppressed}
\end{align}
which remains finite in the DIS limit. Note that the individual invariants $p_p q$, $p_n q$ and
$p_\deut q$ all become large $\sim Q^2$, because the momenta $p_p$, $p_n$ and $p_\deut$ have
nonzero plus components, but the difference in Eq.~(\ref{offshell_power_suppressed}) remains
finite because the difference of the 4-momenta has zero plus component.
The effects caused by the off-shellness of the LF energy are therefore power-suppressed as
\begin{align}
\frac{(p_p + p_n - p_\deut)q}{p_\deut q} \sim \frac{x^2 (M_{pn}^2 - M_\deut^2)}{4 Q^2}
\end{align}
in the DIS limit. This feature is unique to LF quantization and arises because the
quantization axis is aligned with the direction of the high-energy process. In other quantization schemes
(e.g. equal-time quantization, where the 0-component of 4-momentum is off-shell) the invariants
produced by the energy off-shellness generally grow $\sim Q^2$. This is the principal
reason for the use of LF quantization in high-energy scattering.
In particular, Eq.~(\ref{offshell_power_suppressed}) implies that
\begin{align}
\widetilde{q}^{\, 2} = q^2 \left[ 1 - \frac{x (M_{pn}^2 - M_\deut^2)}{2 Q^2} \right] ,
\label{qtilde_squared}
\end{align}
so that the invariant momentum transfer in the electron-nucleon scattering process is equal to
that in the electron-deuteron process (which is fixed by the kinematics),
up to power corrections governed by the nuclear motion.

Second, in LF quantization the components of the current operator differ in how they involve
the interactions of the system and are affected by instantaneous terms in $x^+$;
see Ref.~\cite{Strikman:2017koc} for a summary. The ``good'' current $J^+$ is free of interactions
and instantaneous terms; the ``bad'' current $J^T$ involves the interactions and is 
affected by instantaneous terms; the ``worst'' current $J^-$ is entirely dependent on
the interactions. The matrix elements of $J^+$ can be computed reliably in the
LFQM approach leading to Eq.~(\ref{w_impulse}). The matrix elements of $J^-$ can be
reconstructed from the transversality condition in the collinear frame (current conservation),
\begin{align}
q^\mu \langle ...| J_\mu | ... \rangle
&= \frac{q^+}{2} \langle ... | J^- |... \rangle \; + \; 
\frac{q^-}{2} \langle ... | J^+ | ... \rangle
\nonumber \\
&= 0 ,
\end{align}
and one does not need to know the form of the operator $J^-$. The matrix elements of $J_T$ 
involve unknown instantaneous terms, which are not included in Eq.~(\ref{w_impulse}).
Studies comparing LFQM calculations using different current components in different
frames (good and bad currents in a collinear frame vs.\ good currents in a non-collinear frame)
indicate that the effect of the instantaneous terms is power-suppressed in the 
DIS limit as $(\textrm{mass})^2/Q^2 \rightarrow 0$ \cite{Frankfurt:1988nt}.
The use of Eq.~(\ref{w_impulse}) without instantaneous terms is therefore a
reasonable approximation for the leading-twist structure functions in tagged DIS.
Its accuracy for higher-twist structure functions cannot be established at present,
absent a dynamical theory that would allow one to consistently generate instantaneous terms.
Our results for higher-twist structure functions quoted below should therefore be
regarded as a rough estimate.
\subsection{Virtual nucleon formulation}
\label{subsec:VNA}
We now compute the tagged DIS cross section in the equivalent four-dimensional formulation
of LF nuclear structure based on Feynman diagrams (virtual nucleon approximation,
VNA) \cite{Sargsian:2001ax,Sargsian:2009hf}. In this formulation the 4-momenta
are conserved in the intermediate states of the scattering process,
but the active nucleon momentum is off the mass-shell $p^2 \neq m^2$.
The VNA formulation provides an alternative definition of deuteron LF wave function in terms
of the 4-dimensional deuteron-proton-neutron vertex function. It also offers a different perspective
on the instantaneous terms in the bad-current matrix element in the LFQM, by relating them
to off-mass-shell terms in the VNA.

In the VNA the $\deut \rightarrow p + X$ electromagnetic transition
amplitude is regarded as an effective Feynman diagram
(see Fig.~\ref{fig:IA_FSI}b).  As with Feynman diagrams for point-particle interactions,
4-momenta are conserved at all the vertices. Applying 4-momentum conservation to the
$\deut \rightarrow p + n$ effective vertex, the 4-momentum of the active neutron
is now given by
\begin{align}
\widetilde{p}_n \equiv p_\deut - p_p \hspace{1em} \textrm{(4-momenta)},
\label{vna_neutron_momentum}
\end{align}
and is off the nucleon mass-shell,
\begin{align}
\widetilde{p}_n^{\, 2} - m^2 \neq 0.
\label{virtuality}
\end{align}
Applying the effective Feynman rules of Ref.~\cite{Sargsian:2001ax}, the IA current matrix element
is obtained as
\begin{align}
&\langle X(p_X), p (p_p, \lambda_p) |
J^\mu |\deut (p_\deut, \lambda_\deut ) \rangle
\nonumber \\
& = -\Gamma^\mu_{\gamma^\ast n X}
\frac {\widetilde{\slashed{p}}_n + m}{\widetilde{p}_n^{\, 2} - m^2} \Gamma_{\deut pn}^\nu
\epsilon_{\deut\nu}(\lambda_\deut) v(p_p, \lambda_p)\, .
\label{eq:PWIA_virtual}
\end{align}
Here $\Gamma^\mu_{\gamma^\ast n X}$ denotes the abstract vertex of the electromagnetic transition
between the active neutron and the DIS final state $X$; 
its relation to the neutron structure function will be explained below. 
The following factor is the Feynman 
propagator of a pointlike nucleon with 4-momentum $\widetilde{p}_n$. $\Gamma_{\deut pn}^\nu$ is
the deuteron-proton-neutron vertex, which is a 4-vector and a matrix in 
bispinor space. It is contracted with the deuteron polarization 4-vector $\epsilon_\deut (\lambda_\deut)$
and the bispinor of the final proton, $v(p_p, \lambda_p)$. [In the 4-dimensional formulation of the
deuteron-proton-neutron vertex the proton is described by a charge-conjugated bispinor $v \equiv C \bar u^T$,
the neutron by a regular bispinor $u$, so that the vertex can be represented as the bilinear form
$\bar u \Gamma_{\deut pn}^\nu v$; see Ref.~\cite{Cosyn:2019hem} for details. This is the same
assignment as in the 3-dimensional formulation of the wave function in Eq.~(\ref{wf_3d}),
where the proton is described by the complex-conjugate 2-spinor, and the
neutron by the regular 2-spinor.]

The neutron virtuality $\widetilde{p}_n^2 - m^2$ in Eq.~(\ref{eq:PWIA_virtual}) is a function
of the LF momentum variables of the final-state proton, $\alpha_p$ and $\bm{p}_{pT}$, which
determine the neutron 4-momentum through Eq.~(\ref{vna_neutron_momentum}). The correspondence
between the VNA with the LFQM formulations is established by noting that the virtuality in
the VNA amplitude is proportional to the invariant mass difference Eq.~(\ref{wf_invariant_mass})
in LFQM,
\begin{align}
\widetilde{p}_n^2 - m^2 = - \frac{2 - \alpha_p}{2} \, (M^2_{pn} - M^2_\deut).
\label{invariant_mass_virtuality}
\end{align}
This allows one to express the VNA amplitude in terms of an effective LF wave function.
To do so one needs to connect also the numerators in the two approaches.
Here one encounters the problem that a representation in terms of nucleon spin states 
is defined a priori only for 4-momenta on the mass-shell, $p_n^2 = m^2$.
In order to extend it to off-mass-shell momenta one has to adopt a prescription
relating the off-shell nucleon momentum to an on-shell momentum,
\begin{align}
\widetilde p_n \rightarrow p_n \hspace{1em} \textrm{with} \hspace{1em}
p_n^2 = m^2.
\label{onshell_projection}
\end{align}
This prescription is not unique; geometrically, it means projecting the vector $\widetilde p_n$
on the manifold defined by the mass-shell condition. The freedom is related to the choice of
quantization axis in the noncovariant formulation and determines what kind of wave function
(LF, equal-time, etc.) we want to recover from the VNA expression. In our case we identify
the effective on-shell momentum through a projection along a light-like direction.
Introducing a light-like 4-vector $V$, $V^2 = 0$, we define an effective
on-shell momentum as
\begin{align}
p_n = \widetilde{p}_n - \frac{(\widetilde{p}_n^{\, 2} - m^2) V}{2 (V \widetilde{p}_n)} .
\label{onshell_momentum}
\end{align}
Notice that the modification of $\widetilde{p}_n$ along the light-like direction $V$ is proportional 
to the virtuality $\widetilde{p}_n^2 - m^2$. Specifically, if $V$ is chosen along the $z$-axis
in the LF minus direction, $V^\mu \propto (1, 0, 0, -1), V^+ = 0, V^- \neq 0$, the
LF 4-momentum components of the effective on-shell momentum Eq.~(\ref{onshell_momentum}) are
\begin{subequations}
\begin{align}
& p_n^+ = \widetilde{p}_n^{\, +} = p_\deut^+ - p_p^+,
\\[1.5ex]
& \bm p_{nT} = \widetilde{\bm{p}}_{nT} = - \bm p_{pT},
\\
& p_n^- = \frac{|\bm{p}_{nT}|^2 + m^2}{p_n^+},
\end{align}
\end{subequations}
which coincides with the 4-momentum components of the active neutron in the 
LFQM formulation, Eqs.~(\ref{pn_lf_momenta}) and (\ref{pminus_onshell}).

At the effective on-shell momentum, we then introduce neutron spin states as LF helicity states,
described by the LF helicity bispinors $u_{\rm LF}(p_n, \lambda_n)$. The final-state proton
spin states we describe by $v_{\rm LF}(p_p, \lambda_p)$; the explicit expressions of the
LF helicity spinors are given in Ref.~\cite{Cosyn:2019hem}. With these spinors the numerator
of the Feynman propagator in Eq.~(\ref{eq:PWIA_virtual}) can be expressed as
\begin{subequations}
\label{eq:numerator_onshell}
\begin{align}
\widetilde{\slashed{p}}{}_n + m 
&= 
\sum_{\lambda_n} u_{\rm LF}(p_n, \lambda_n) \bar{u}_{\rm LF}(p_n, \lambda_n) 
\nonumber \\
&+ \; \frac{(\widetilde{p}_n^2 - m^2)}{2(V \widetilde{p}_n)} (V\gamma )
\\[1ex]
&= \sum_{\lambda_n} u_{\rm LF}(p_n, \lambda_n) \bar{u}_{\rm LF}(p_n, \lambda_n) 
\nonumber \\
&+ \; \tfrac{1}{2}(\widetilde{p}_n^{\,-} - p_n^{\,-})\gamma^+ .
\end{align}
\end{subequations}
The first term in the expressions on the right-hand side is the on-shell
spin projector for nucleon states with momentum $p_n$;
the second term is proportional to the nucleon virtuality and accounts for the difference 
between the off-shell and on-shell projectors.

The deuteron spin state in Eq.~(\ref{eq:PWIA_virtual}) is described by the
polarization 4-vector $\epsilon_\deut (\lambda_\deut )$, which is defined relative to the deuteron
4-momentum $p_\deut$ and satisfies
\begin{align}
p_\deut\epsilon_\deut = (p_p + \tilde{p}_n) \epsilon_\deut = 0, \hspace{2em} \epsilon_\deut^2 = -1.
\end{align}
When the neutron 4-momentum is changed by the on-shell projection Eq.~(\ref{onshell_projection}),
the polarization vector is no longer aligned with the sum of the proton and neutron 4-momenta
because $p_p + p_n \neq p_\deut$. We define a polarization vector $\epsilon_{pn} (\lambda_\deut )$
aligned with $p_p + p_n$ as
\begin{align}
\epsilon_{pn} &\equiv \epsilon_\deut
+ \frac{(\widetilde{p}_n^{\, 2} - m^2)(V\epsilon_\deut) p_\deut}{M_\deut^2 (V\tilde{p}_n)}
+ \mathcal{O}[(\widetilde{p}_n^{\, 2} - m^2)^2]
\end{align}
(same argument $\lambda_\deut$ in $\epsilon_{pn}$ and $\epsilon_\deut$), which satisfies
\begin{subequations}
\begin{align}
(p_p + p_n) \epsilon_{pn} &= 0 + \mathcal{O}[(\widetilde{p}_n^{\, 2} - m^2)^2],
\\[2ex]
\epsilon_{pn}^2 &= -1 + \mathcal{O}[(\widetilde{p}_n^{\, 2} - m^2)^2],
\end{align}
\end{subequations}
where we have used Eq.~(\ref{onshell_projection}) and neglected terms of higher order in the
virtuality. The new vector $\epsilon_{pn}(\lambda_\deut)$ describes the spin wave function of the
$pn$ system after the on-shell projection, and the vertex formed with it satisfies
LF helicity selection rules with the nucleon spinors.

We now define the deuteron LF wave function in the context of the VNA as
\begin{align}
&\Psi_\deut(p_p; \lambda_p, \lambda_n| \lambda_\deut)[\textrm{VNA}]
\nonumber \\[1ex]
&\equiv - \frac{\bar{u}_{\rm LF} (p_n, \lambda_n)
\Gamma_{\deut pn}^\mu \epsilon_{pn, \mu} (\lambda_\deut) v_{\rm LF} (p_p,\lambda_p)}
{2(2\pi)^{3/2} \, (\widetilde{p}_n^{\, 2} - m^2)} .
\label{eq:d_lfwf}
\end{align}
The covariant expression on the right-hand side is a function of the proton
4-momentum $p_p$ and  depends on invariants $(V \tilde p_n)$ and $(p_\deut p_p)$,
which can be related to the LF variables $\alpha_p$ and $|\bm{p}_{pT}|$.
By virtue of Eq.~(\ref{invariant_mass_virtuality}), the Feynman denominator is proportional
to the invariant mass denominator of Eq.~(\ref{wf_invariant_mass}). The contraction of
the deuteron vertex with the proton and on-shell neutron spinors is equal to the 
deuteron-proton-neutron vertex function in Eq.~(\ref{wf_invariant_mass}). The wave
function defined by Eq.~(\ref{eq:d_lfwf}) therefore coincides with the one of the
LFQM formulation,
\begin{align}
&\Psi_\deut (p_p; \lambda_p, \lambda_n| \lambda_\deut) [\textrm{VNA}]
\nonumber \\[1ex]
&=
\frac{\Psi_\deut (\alpha_p, \bm{p}_{pT}; \lambda_p, \lambda_n| \lambda_\deut) [\textrm{LF}]}{2 - \alpha_p}.
\label{wave_function_equivalence}
\end{align}
The exact connection between the spin structure of the two wave functions is described in
Ref.~\cite{Cosyn:2019hem}. The covariant decomposition of the bilinear form in the numerator of the
VNA wave function can be matched with the spin decomposition of the LFQM wave function
in the rotationally symmetric representation in the c.m.\ frame (see Sec.~\ref{subsec:spin_structure}).

The electromagnetic $\gamma^\ast n X$ vertex in Eq.~(\ref{eq:PWIA_virtual}) is evaluated at
the off-shell neutron momentum $\widetilde{p}_n$. It can be expanded around the on-shell momentum,
resulting in corrections proportional to the virtuality,
\begin{align}
\Gamma^\mu_{\gamma^\ast n X}(\widetilde{p}_n) 
= 
\Gamma^\mu_{\gamma^\ast n X}(p_n) \; + \; \textrm{terms} \propto (\widetilde{p}_n^{\, 2} - m^2).
\label{vertex_offshell}
\end{align}
The contraction of the on-shell vertex $\Gamma^\mu_{\gamma^\ast n X}(p_n)$ with the on-shell
neutron spinor $u(p_n, \lambda_n)$ in the projector Eq.~(\ref{eq:numerator_onshell})
of the propagator reproduces the current matrix element between a physical neutron state
with momentum $p_n$ and the state $X$,
\begin{align}
\Gamma^\mu_{\gamma^\ast n X}(p_n) u(p_n, \lambda_n)
= \langle X | J^\mu | n (p_n, \lambda_n) \rangle .
\end{align}
The momentum transfer in the matrix element is 
\begin{align}
p_X - p_n = q + p_\deut - p_p - p_n = \tilde q \neq q,
\end{align}
which agrees with the effective momentum transfer in the LFQM calculation, Eq.~(\ref{qtilde_def}).
Altogether, neglecting the off-shell terms in the projector Eq.~(\ref{eq:numerator_onshell})
and in the vertex Eq.~(\ref{vertex_offshell}), the IA current matrix element for deuteron
breakup in the VNA, Eq.~(\ref{eq:PWIA_virtual}), becomes
\begin{align}
&\langle X, p(p_p, \lambda_p) | J^\mu | \deut (p_\deut, \lambda_\deut)\rangle
\nonumber \\[2ex]
&= 2 (2\pi)^{3/2} \sum_{\lambda_n}
\langle X | J^\mu | n(p_n, \lambda_n) \rangle 
\nonumber \\
&\times \Psi_\deut(p_p; \lambda_p, \lambda_n| \lambda )[\textrm{VNA}]
\nonumber
\\[2ex]
&= 2 (2\pi)^{3/2} \sum_{\lambda_n}
\langle X | J^\mu | n(p_n, \lambda_n) \rangle
\nonumber \\
&\times \frac{\Psi_\deut(\alpha_p, \bm{p}_{pT}; \lambda_p, \lambda_n| \lambda)[\textrm{LF}]}{2 - \alpha_p} 
\nonumber
\\[1ex]
& [\textrm{up to terms} \propto (\widetilde{p}_n^{\, 2} - m^2)].
\label{eq:PWIAredux}
\end{align}
This agrees with the result obtained in LFQM, Eq.~(\ref{eq:IA_lfqm}). After squaring the matrix element
one again obtains the hadronic tensor, which can be expressed through the effective neutron
hadronic tensor, see Eq.~(\ref{w_impulse}).

Some comments are in order regarding the correspondence between the VNA and LFQM formulations
and the role of the off-shell terms.

(a) The off-shell terms in the VNA correspond to ``instantaneous'' terms in the LFQM formulation --
contributions to the transition amplitude that do not involve physical intermediate states propagating
over LF finite times. The neutron virtuality in the VNA is proportional to the LF energy offshellness
of the intermediate state in LFQM,
\begin{align}
\widetilde{p}_n^{\, 2} - m^2 = - \frac{2 - \alpha_p}{2} p_\deut^+ (p_p^- + p_n^- - p_\deut^-) .
\end{align}
In the terms where the LF energy denominator is cancelled by off-shell terms in the numerator,
no propagation takes place.

(b) The off-shell terms arising from the numerator of the Feynman propagator
Eq.~(\ref{eq:numerator_onshell}) are proportional to $\gamma^+$. Because $(\gamma^+)^2 = 0$
these terms do not affect structures in the neutron electromagnetic vertex that are $\propto \gamma^+$.
These are precisely the ``good'' components $J^+$ of the neutron current matrix element.
In contrast, structures $\propto \gamma_T$ or $\gamma^-$ in the neutron vertex correspond
to bad or worst current components and have nonzero contraction with the off-shell terms
in the propagator.

(c) The off-shellness is caused by nuclear binding. In the nonrelativistic limit
\begin{align}
\widetilde{p}_n^{\, 2} - m^2 &= - 2 (|\bm{p}_p|^2 + m \epsilon_\deut )
\nonumber\\[1ex]
&= -4 m (E_{{\rm kin}, p} + \epsilon_\deut/2),
\end{align}
where $E_{{\rm kin}, p} \equiv |\bm{p}_p|^2/2 m$ is the proton (or neutron) kinetic energy and
$\epsilon_\deut/2$ is the binding energy per nucleon. The typical values of the virtuality are
therefore determined by the nucleon energy in the nucleus.

(d) In tagged DIS the neutron virtuality in the Feynman diagram is equal to the
invariant momentum transfer between the deuteron and the tagged proton,
\begin{align}
\widetilde{p}_n^{\, 2} -m^2 = (p_\deut-p_p)^2 -m^2 \equiv t' < 0,
\end{align}
which is a kinematic variable computed from the proton momentum. In tagged DIS experiments
one can therefore control the virtuality of the active nucleon in the IA
and measure the dependence of the
structure functions on the virtuality in the physical region $|t'| < 2m\epsilon_\deut$ .
The extrapolation to $t' \rightarrow 0$ (on-shell point, or free nucleon point) eliminates
off-shell terms in the cross section and allows one to access the structure function of
the free neutron with tagged DIS \cite{Sargsian:2005rm}. The method also eliminates FSI effects
through the analytic structure of the cross section \cite{Sargsian:2005rm,Cosyn:2015mha,Strikman:2017koc}.
On-shell extrapolation in polarized tagged DIS is discussed in Ref.~\cite{Cosyn:2019hem}.
\section{Neutron distribution in deuteron}
\label{sec:effective}
\subsection{Neutron spin density matrix}
\label{subsec:nucleon_spin_density_matrix}
The IA expresses the hadronic tensor of tagged DIS in terms of the deuteron LF wave function
and the hadronic tensor of the active neutron. The LF momentum of the active neutron is
fixed by the tagged proton momentum. Because of the entanglement of momentum and
spin degrees of freedom in the deuteron wave function, also the LF helicity of the
active neutron is influenced by the spectator momentum. It is useful to present this dependence
in the form of a momentum and spin density of the active neutron that depends on the
tagged proton momentum.
This form allows one to present the expressions for the tagged DIS structure function in a
compact and transparent form. It also allows one to study the LF structure of the deuteron
in nucleon degrees of freedom and exhibit the analogies with the parton picture of hadrons.
For the spin-1/2 nucleus of $^3$He, similar nucleon LF momentum distributions
have been introduced in Ref.~\cite{Alessandro:2021cbg}.

We write the IA result for the deuteron hadronic tensor with proton tagging, Eq.~(\ref{w_impulse}),
in the form 
\begin{subequations}
\label{IA_density_matrix}
\begin{align}
&\langle W^{\mu\nu}_\deut \rangle (p_\deut, q, p_p)
= \frac{2 \, [2(2\pi)^3]}{(2-\alpha_p)^2} \sum_{\lambda_n, \lambdanp}
\rho_n(p_p| \lambda_n, \lambdanp)
\nonumber \\
&\times W_n^{\mu\nu} (p_n, \widetilde{q}| \lambdanp, \lambda_n ),
\\[2ex]
&\rho_n(p_p | \lambda_n, \lambdanp)
\equiv
\sum_{\lambda_\deut,\lambda_\deut'} \rho_\deut(\lambda_\deut, \lambdadp) \sum_{\lambda_p}
\nonumber \\
&\times \Psi_\deut^\ast  (\alpha_p , \bm{p}_{pT}; \lambda_p, \lambdanp | \lambdadp)
\Psi_\deut       (\alpha_p , \bm{p}_{pT}; \lambda_p, \lambda_n|  \lambda_\deut) \,.
\label{rho_n_def}
\end{align}
\end{subequations}
The effective neutron density $\rho_n$ is a density matrix in the neutron LF helicities.
It depends on the tagged proton LF momentum and the deuteron polarization parameters
contained in the deuteron spin density matrix.

It is convenient to introduce a covariant representation of the neutron density matrix \cite{Cosyn:2020kwu},
in analogy to covariant representation of the deuteron density matrix, see \firstpart,
Sec.~\ref{P1:subsec:covariant_density_matrix}. The matrix element of a neutron operator
between neutron LF helicity states can be represented
as a bilinear form in LF bispinors
\begin{align}
&\langle n(p_n, \lambdanp) | \mathcal{O} | n(p_n, \lambda_n) \rangle
\nonumber \\
& = \bar u_{\rm LF} (p_n, \lambdanp) \, \Gamma \,  u_{\rm LF}(p_n, \lambda_n),
\label{neutron_operator_bilinear}
\end{align}
where $\Gamma$ is a matrix in bispinor indices whose form depends on the operator.
The covariant neutron spin density matrix in the deuteron is defined as
\begin{align}
\Pi_n \equiv \sum_{\lambda_n, \lambda_n'}
\rho_n(\lambda_n,\lambda'_n) u_{\rm LF}(p_n, \lambda_n) \bar u_{\rm LF}(p_n, \lambdanp) .
\label{eq:effective_neutron_rho_cov}
\end{align}
The average over the neutron LF helicities can then be computed as
\begin{align}
& \sum_{\lambda_n, \lambda_n'} \rho_n(p_p| \lambda_n,\lambda'_n) \;
\langle n(p_n, \lambdanp) | \mathcal{O} | n(p_n, \lambda_n) \rangle
\nonumber \\
&= \textrm{tr}[ \Pi_n \Gamma ].
\label{neutron_operator_trace}
\end{align}

The neutron density matrix $\rho_n$, Eq.~(\ref{rho_n_def}),
arises as an average over the deuteron spin ensemble
and depends linearly on the vector and tensor polarization parameters. Its covariant representation
$\Pi_n$, Eq.~(\ref{eq:effective_neutron_rho_cov}), can be expressed as the sum of an unpolarized,
vector polarized, and tensor polarized term,
\begin{align}
\Pi_n = \frac{1}{2} (\slashed{p}_n + m)(u_n + \slashed{s}_n \gamma^5 + t_n),
\label{eq:Pi_n}
\end{align}
where $\gamma^5 \equiv -i\gamma^0\gamma^1\gamma^2\gamma^3$ in our convention. The construction
is described in Ref.~\cite{Cosyn:2020kwu}; here we only state the results.
The expressions take a simple form when written in terms of the deuteron wave function
in the $pn$ c.m.\ frame (see Sec.~\ref{subsec:spin_structure}).
The scalar functions $u_n$ and $t_n$ are obtained as [here $f_{0, 2} \equiv f_{0,2}(k)$]
\begin{subequations}
\begin{align}
u_n \; &= f_0^2 + f_2^2,
\label{eq:un}
\\[2ex]
t_n \; &= -\frac{3 (\bm{k} T_\deut \bm{k})}{|\bm{k}|^2}
\left( 2 f_0 + \frac{f_2}{\sqrt{2}} \right) \frac{f_2}{\sqrt{2}},
\label{eq:Pi_n_tensor}
\end{align}
\end{subequations}
where $T_\deut$ is the 3-tensor describing the tensor polarization in the
deuteron rest frame, Eq.~(\ref{tensor_restframe}).
These functions have the same value in any collinear frame; the relation between
the c.m.\ momentum $\bm{k}$ and the LF momentum variables $\alpha_p$ and $\bm{p}_{pT}$
is given in Eq.~(\ref{CM_LF_variables}). Of the 4-vector function $s_n$ in Eq.~(\ref{eq:Pi_n}),
the ordinary vector components in the c.m.\ frame are obtained as
\begin{subequations}
\label{eq:sn}
\begin{align}
s_n[\textrm{c.m.}] \; &= \; (s_n^0, \bm{s}_n),
\\[2ex] 
s_n^0 \; &= \; -\frac{(\bm{S}_\deut\bm{k})}{m} \left( f_0 - \frac{f_2}{\sqrt{2}} \right)^2 ,\label{eq:sn0}
\\[2ex]
\bm{s}_n \; &= \; \left( f_0 - \frac{f_2}{\sqrt{2}} \right)\left[ 
\frac{(\bm{S}_\deut\bm{k}) \bm{k} }{|\bm{k}|^2}\,\frac{E}{m}\,\left(f_0-\frac{f_2}{\sqrt{2}}\right)\right.
\nonumber\\
&+\; \left. \left( \bm{S}_\deut - \frac{(\bm{S}_\deut\bm{k}) \bm{k} }{|\bm{k}|^2}\right)
\left( f_0+\sqrt{2}f_2\right) 
\right] ,
\label{eq:snvec}
\end{align}
\end{subequations}
where $\bm S_\deut$ is the 3-vector describing the vector polarization in the
deuteron rest frame, Eq.~(\ref{vector_restframe}). The LF components of $s_n$
in an arbitrary collinear frame are then obtained by forming the LF plus and minus components
in the c.m. frame, $(s^0 \pm s^3)[\textrm{c.m.}]$, and performing a boost to the desired value of
$p_\deut^+$ , see Eq.~(\ref{boost_collinear}), with the boost parameter given by
$p_\deut^+ / p_\deut^+[\textrm{c.m.}] = p_\deut^+ / M_{pn} = p_\deut^+ / (2 E)$, resulting in
\begin{subequations}
\begin{align}
s_n^+ \; &= \; \frac{p_\deut^+}{2E} \, (s_n^0 + s_n^3) [\textrm{c.m.}] ,
\\[2ex]
s_n^- \; &= \; \frac{2E}{p_\deut^+} \, (s_n^0 - s_n^3) [\textrm{c.m.}] ,
\\[2ex]
\bm{s}_{nT} \; &= \; \bm{s}_{nT} [\textrm{c.m.}] .
\end{align}
\end{subequations}

It is worth emphasizing that the deuteron polarization parameters appearing
in Eqs.~(\ref{eq:Pi_n_tensor}) and (\ref{eq:sn}) are the polarization 3-tensor
and 3-vector in the deuteron rest frame (see Sec.~\ref{subsec:deuteron_polarization}).
The c.m.\ frame expressions of the neutron density have been derived 
by importing the deuteron polarization from the rest frame, using the fact that both frames are
in the class of collinear frames, where boosts can be performed in a
simple manner (see Sec.~\ref{subsec:collinear_frames}) \cite{Cosyn:2020kwu}.
The representation in terms of c.m.\ frame variables thus describes the
LF spin structure very efficiently, expressing the neutron LF helicity distribution
in any collinear frame in terms of the deuteron rest frame polarization parameters.

\subsection{Neutron momentum distributions}
\label{subsec:neutron_distributions}
From the neutron spin density matrix one can derive the distributions of neutrons
with a given spin projection in the deuteron with proton tagging. These distributions
depend on the deuteron polarization, the tagged proton momentum, and the neutron
spin projections, and exhibit rich structure. They can be used to present the IA result
for the tagged cross section and its decomposition in structure functions in a
compact form, exhibiting the connection with the nucleon spin structure functions.

We define the spin-projected distribution of neutrons in proton tagging as
\begin{align}
&P (\alpha_p,\bm p_{pT} | \bm{S}_\deut, T_\deut)
\nonumber \\[1ex]
&\equiv \frac{1}{2 - \alpha_p} \sum_{\lambda_n,\lambda_n'} a(\lambda'_n, \lambda_n)
\rho_n(\alpha_p,\bm p_{pT} ; \lambda_n, \lambda'_n | \bm{S}_\deut, T_\deut)
\nonumber \\[1ex]
&= \frac{1}{2 - \alpha_p} \sum_{\lambda_n,\lambda_n'} a(\lambda'_n, \lambda_n)
\sum_{\lambda_\deut,\lambda_\deut'} \rho_\deut(\lambda_\deut,\lambda_\deut' | \bm{S}_\deut, T_\deut) 
\nonumber \\
&\times \sum_{\lambda_p}\Psi_\deut^\ast  (\alpha_p , \bm{p}_{pT}; \lambda_p, \lambda_n'| \lambda_\deut')
\Psi_\deut       (\alpha_p , \bm{p}_{pT}; \lambda_p, \lambda_n|  \lambda_\deut),
\label{distribution_sum}
\end{align}
where $\rho_n$ is the neutron spin density matrix Eq.~(\ref{rho_n_def}) and
$a(\lambda'_n, \lambda_n)$ is a matrix in the neutron LF helicities.
We have explicitly indicated the dependence on the deuteron rest-frame polarization
3-vector and tensor, $\bm{S}_\deut$ and $T_\deut$.
It is convenient to evaluate the neutron spin projection using the covariant representation of
the density matrix Eq.~(\ref{eq:effective_neutron_rho_cov}). Writing the matrix $a$
as a bilinear form in LF bispinors as in Eq.~(\ref{neutron_operator_bilinear})
\begin{align}
a(\lambda'_n, \lambda_n) \; &= \; \bar u(p_n, \lambdanp) \; 
\Gamma \; u(p_n, \lambda_n),
\end{align}
the average Eq.~(\ref{distribution_sum}) is computed using Eq.~(\ref{neutron_operator_trace})
\begin{align}
P (\alpha_p,\bm p_{pT}| \bm{S}_\deut, T_\deut) \; = \; 
\frac{\textrm{tr}[\Pi_n \Gamma]}{2 - \alpha_p} .
\label{distribution_general_trace}
\end{align}
%
%
\begin{table}
\renewcommand{\arraystretch}{1.2}
\[
\begin{array}{c|c|c|c}
a & \Gamma & \textrm{symbol} & \textrm{name}
\\
\hline
1 & \gamma^+ / p_n^+ & U & \textrm{unpolarized} \\
\sigma^3 & -\gamma^+\gamma_5 / p_n^+ & S_L & \textrm{longitud.\ polarized} \\
\sigma^i (i = 1,2) & i\sigma^{i+} \gamma_5 / p_n^+ & S_T & \textrm{transv.\ polarized}
\\
\hline
\frac{1}{2} (1 \pm \sigma^3) & & L\pm & \textrm{$\pm$1/2 longitud.}
\\
\frac{1}{2} (1 \pm \bm{n}_T\bm{\sigma}) & & T\pm & \textrm{$\pm$1/2 transv.\ $\bm{n}_T$}
\\
\hline
\end{array}
\]
\caption{Spin matrices used in the definition of the spin-projected neutron distributions,
Eqs.~(\ref{distribution_sum}) and (\ref{distribution_general_trace}).
Upper rows: Distributions of unpolarized, longitudinally polarized, and transversely polarized neutrons.
Lower rows: Distributions in pure spin states with projection $\pm 1/2$ along the longitudinal axis
or a transverse axis set by the vector $\bm{n}$.}
\label{tab:spin_structures}
\end{table}
The spin matrices used in the present study are summarized in
Table~\ref{tab:spin_structures}.\footnote{The form of the bispinor matrices $\Gamma$ is not unique.
They are contracted with
on-shell bispinors satisfying the Dirac equation and can be modified by terms that vanish
due to the Dirac equation. The form in Table~\ref{tab:spin_structures} are the standard
projectors used in partonic structure in QCD.}
They describe the unpolarized, longitudinally polarized, and transversely polarized distributions
of neutrons, as well as the the probabilistic distributions of neutrons in states with a defined
longitudinal or transverse spin projection $\pm 1/2$.
These distributions are analogous to the leading-twist quark parton distributions in a hadron
and provide a complete characterization of the one-body LF momentum distributions in a system
with spin-1/2 constituents. We use the notation
\begin{align}
&P_{[\text{neutron}]}(\alpha_p,\bm p_{pT}| \bm{S}_\deut, T_\deut)
\nonumber \\[1ex]
&\text{neutron} = U,S_L,S_T
\label{notation_distributions_neutron}
\end{align}
to denote the neutron spin projection in the distribution.
The neutron distributions here depend on the deuteron polarization parameters $\bm{S}_\deut$ and $T_\deut$
and refer to a general mixed polarization state; normalized distributions in particular deuteron spin
states will be defined in Secs.~\ref{subsec:unpolarized_distributions} and
\ref{subsec:polarized_distributions}.

The neutron distributions defined here are functions of the tagged proton LF momentum
variables $\alpha_p$ and $\bm{p}_{pT}$ are to be integrated as
\begin{align}
\int \frac{d\alpha_p}{\alpha_p} d^2 p_{pT} \; P (\alpha_p,\bm p_{pT}).
\end{align}
When the deuteron LF wave function is expressed in terms of c.m.\ wave function, the
neutron densities are obtained as functions of the c.m.\ momentum variable $\bm{k}$,
related to $\alpha_p$ and by Eq.~(\ref{CM_LF_variables}). In the c.m.\ frame one can
also consider a neutron distribution $N(\bm{k})$ defined as
\begin{align}
P (\alpha_p,\bm p_{pT}) = \frac{N(\bm{k})}{2 - \alpha_p},
\label{distribution_LF_CM}
\end{align}
which differs from the distribution $P$ by the flux factor $1/(2 - \alpha_p)$.
The distribution $N(\bm{k})$ corresponds to the 3-dimensional particle density of
the c.m.\ frame wave function and is to be integrated as [see Eq.~(\ref{integration_k})]
\begin{align}
\int \frac{d^3k}{E(k)} \; N (\bm{k}).
\end{align}
The unpolarized neutron projection of $N(\bm{k})$ exhibits 3-dimensional
rotational symmetry (see below); in the longitudinally and transversely polarized projections
the rotational symmetry is broken only by the Melosh rotations [see Eq.~(\ref{wf_3d})].
The neutron distribution in tagged DIS can be described either by LF distribution
$P$ or the c.m.\ momentum distribution $N$. The distribution $P$ is identical to the
the spectral function as defined in Refs.~\cite{Strikman:2017koc,Cosyn:2019hem}
and appears in tagged cross section differential in phase space element Eq.~(\ref{eq:recoil_phasespace}).
In following we quote the expressions for $P$; the expressions for
$N$ are obtained simply by stripping away the factor $1/(2 - \alpha_p)$,
see Eq.~(\ref{distribution_LF_CM}).

The functions introduced in Eqs.~(\ref{distribution_sum}) and (\ref{distribution_general_trace})
describe the neutron distribution in a general mixed polarization state of the deuteron,
characterized by the rest-frame polarization 3-vector and 3-tensor $\bm{S}_\deut$ and $T_\deut$.
It is useful to define normalized distributions corresponding to unit polarization along
given directions, i.e., distributions corresponding to unit values of the scalar parameters
$S_L, S_T$ and $T_{LL}, T_{LT}, T_{TT}$. These normalized distributions can then be
multiplied with the scalar parameters and combined to obtain the distributions in the
general polarization state.
This enables easy calculation of the tagged DIS cross section and produces simple expressions
of the tagged structure functions (see Sec.~\ref{sec:IA_strucfunc}). We use the notation
\begin{align}
&P_{[\text{deuteron, neutron}]} (\alpha_p,\bm p_{pT})
\nonumber \\[1ex]
&\text{neutron} = U,S_L,S_T
\nonumber\\[1ex]
&\text{deuteron} = U,S_L,S_T,T_{LL},T_{LT},T_{TT} 
\label{notation_distributions}
\end{align}
where ``neutron'' characterizes the neutron polarization as in Eq.~(\ref{notation_distributions_neutron})
and Table~\ref{tab:spin_structures}, and ``deuteron'' characterizes the deuteron polarization.
The normalized neutron distributions are analogous to the unpolarized and polarized parton densities
defined for hadron states with unit polarization.
In the following we compute the normalized neutron distributions needed for tagged DIS and and discuss
their properties.

\subsection{Unpolarized neutron distributions}
\label{subsec:unpolarized_distributions}
The distribution of unpolarized neutrons obtained from Eq.~(\ref{distribution_general_trace})
with the density matrix Eq.~(\ref{eq:Pi_n}) and the projector in Table~\ref{tab:spin_structures} is
\begin{align}
&P_{[U]}(\alpha_p,\bm p_{pT}| T_\deut)
\equiv \frac{\textrm{tr}[\Pi_n \gamma^+]}{(2 - \alpha_p)^2 \; p_\deut^+}
\nonumber \\[1ex]
&= \frac{f_0^2 + f_2^2}{2 - \alpha_p}
-\frac{3}{2 - \alpha_p} \; \frac{(\bm{k} T_\deut \bm{k})}{|\bm{k}|^2} 
\left( 2 f_0 + \frac{f_2}{\sqrt{2}} \right) \frac{f_2}{\sqrt{2}} .
\label{distribution_unpol_trace}
\end{align}
With respect to deuteron polarization, it contains a term independent of the deuteron
polarization and a term proportional to the polarization tensor $T_\deut$, but no term
involving the polarization vector $\bm{S}_\deut$.
For the unpolarized deuteron we define the normalized distribution in in the notation
of Eq.~(\ref{notation_distributions}) as
\begin{align}
P_{[U, U]}(\alpha_p,\bm p_{pT}) \equiv \frac{f_0^2 + f_2^2}{2 - \alpha_p}.
\label{spectral_U_U}
\end{align}
For the tensor-polarized deuteron we define normalized distributions 
corresponding to the 3-dimensional spherical tensors in the expansion of the rest-frame
polarization tensor, Eq.~(\ref{tensor_restframe}).
\begin{align}
&
\left.
\begin{array}{l}
P_{[T_{LL}, U]}
\\[1ex]
P_{[T_{LT}, U]}
\\[1ex]
P_{[T_{TT}, U]}
\end{array}
\right\} (\alpha_p,\bm p_{pT})
\; \equiv \;
\left\{
\begin{array}{l}
P_{[U]} (T_\deut = e_{LL})
\\[1ex]
P_{[U]} (T_\deut = e_{LT})
\\[1ex]
P_{[U]} (T_\deut = e_{TT})
\end{array}
\right\}
\nonumber \\[2ex]
&= -\frac{3}{2-\alpha_p} \left( 2 f_0 + \frac{f_2}{\sqrt{2}} \right)
\frac{f_2}{\sqrt{2}} \;
\left\{
\begin{array}{l}
\displaystyle \frac{(\bm{k} e_{LL} \bm{k})}{|\bm{k}|^2}
\\[2ex]
\displaystyle \frac{(\bm{k} e_{LT} \bm{k})}{|\bm{k}|^2}
\\[2ex]
\displaystyle \frac{(\bm{k} e_{TT} \bm{k})}{|\bm{k}|^2}
\end{array}
\right\} .
\label{distribution_tensor}
\end{align}
The three distributions have the same dependence on the radial wave functions,
as they arise from projections of the same rotationally covariant structure
in Eq.~(\ref{distribution_unpol_trace}). They are proportional to the $D$ wave
radial function $f_2$, showing that the deuteron tensor polarization can
influence the neutron distribution only through the $L=2$ state in the
orbital motion. The angular dependence is governed by the contraction of
the spherical basis tensors with the vector $\bm{k}$. Using the
explicit expressions in \firstpart, Eq.~(\ref{P1:basis_tensors}), we obtain
\begin{subequations}
\label{tensor_contractions}
\begin{align}
&\frac{(\bm{k} e_{LL} \bm{k})}{|\bm{k}|^2} 
= \frac{1}{\sqrt{6}} \left( \frac{3 (k^z)^2}{|\bm{k}|^2} - 1 \right)
= \frac{1}{\sqrt{6}} \left( 3 \cos^2\theta_k - 1 \right)
\nonumber
\\
&= \sqrt{\frac{2 \cdot 4\pi}{3\cdot 5}} Y_{20} (\theta_k, \phi_k = 0),
\\[2ex]
&\frac{(\bm{k} e_{LT} \bm{k})}{|\bm{k}|^2} 
= \sqrt{2} \, \frac{k^z |\bm{k}_T|}{|\bm{k}|^2}
= \sqrt{2} \cos\theta_k \sin\theta_k
\nonumber
\\
&= \sqrt{\frac{2 \cdot 4\pi}{3\cdot 5}} \frac{1}{\sqrt{2}} \left[ -Y_{21} + Y_{2-1} \right]
(\theta_k, \phi_k = 0),
\\[1ex]
&\frac{(\bm{k} e_{TT} \bm{k})}{|\bm{k}|^2} 
= \frac{1}{\sqrt{2}} \frac{|\bm{k}_T|^2}{|\bm{k}|^2}
= \frac{1}{\sqrt{2}} \sin^2\theta_k
\nonumber
\\
&= \sqrt{\frac{2 \cdot 4\pi}{3\cdot 5}} \frac{1}{\sqrt{2}} \left[ Y_{22} + Y_{2-2} \right]
(\theta_k, \phi_k = 0),
\end{align}
\end{subequations}
The polar angle of the vector $\bm{k}$ in the $x'y'z$ coordinate system is defined as
\begin{align}
\cos \theta_k \equiv \bm{e}_{z}\cdot\bm{k}, \hspace{2em} \sin \theta_k \equiv \bm{e}_{x'}\cdot\bm{k};
\end{align}
the azimuthal angle is zero, $\phi_k = 0$, because the $x'$-axis is chosen along the
direction of $\bm{k}_T \equiv \bm{p}_{pT}$; see
Eq.~(\ref{basis_vectors_prime}) and \firstpart, Fig.~\ref{P1:fig:kin}.
The contractions of the spherical tensors $e_{LT'}$ and $e_{TT'}$ with $\bm{k}$ are zero;
these components of the deuteron tensor polarization therefore do not influence the
neutron momentum distribution.

In terms of the normalized distributions, the unpolarized neutron distribution
in a deuteron with general tensor polarization, Eq.~(\ref{distribution_unpol_trace}),
can be expressed as
\begin{align}
&P_{[U]} (\alpha_p, \bm{p}_{pT} | T_\deut)
\nonumber \\[1ex]
&= P_{[U, U]}(\alpha_p, \bm{p}_{pT})
\nonumber \\[1ex]
&+ \sqrt{\frac{3}{2}} \, T_{LL} \, P_{[T_{LL}, U]}(\alpha_p, \bm{p}_{pT})
\nonumber \\[1ex]
&+ \sqrt{2} \, T_{LT} \, \cos (\phi_p - \phi_{T_L}) P_{[T_{LT}, U]}(\alpha_p, \bm{p}_{pT})
\nonumber \\[1ex]
&+ \frac{1}{\sqrt{2}} \, T_{TT} \, \cos (2 \phi_p - 2 \phi_{T_T}) P_{[T_{TT}, U]}(\alpha_p, \bm{p}_{pT}),
\label{unpolarized_from_normalized}
\end{align}
where we have used the expansion of $T_\deut$ in the basis tensors,
Eq.~(\ref{tensor_restframe}). Equation~(\ref{polarized_from_normalized}) shows the
explicit dependence of the neutron distributions on the deuteron tensor
polarization parameters and can be used in the calculation of tagged scattering observables.

The integral of the unpolarized neutron distribution Eq.~(\ref{spectral_U_U}) is
\begin{align}
& \int_0^2\frac{d\alpha_p}{\alpha_p} \, d^2 p_{pT} \; 
P_{[U, U]}(\alpha_p,\bm p_{pT})
\nonumber \\
&= \int \frac{d^3 k}{E} \; (f_0^2 + f_2^2)
= 1,
\label{integral_baryon_number}
\end{align}
because of the normalization condition of deuteron wave function Eq.~(\ref{normalization_radial}).
Furthermore,
\begin{align}
& \int_0^2\frac{d\alpha_p}{\alpha_p} \, d^2 p_{pT} \; (2 - \alpha_p) \; 
P_{[U, U]}(\alpha_p,\bm p_{pT})
\nonumber \\
&= \int \frac{d^3 k}{E} \; \left( 1 - \frac{k^z}{E} \right)(f_0^2 + f_2^2) = 1,
\label{integral_momentum}
\end{align}
because the term $\propto k^z$ integrates to due to rotational symmetry
(the relation can also be derived using the symmetry of the deuteron LF wave function
under $\alpha_p \rightarrow 2 - \alpha_p$).
The integrals of the tensor-polarized distributions Eq.~(\ref{distribution_tensor})
are zero,
\begin{subequations}
\label{integral_tensor}
\begin{align}
& \int_0^2\frac{d\alpha_p}{\alpha_p} \, d^2 p_{pT} \; 
P_{[T_{LL}, U]}(\alpha_p,\bm p_{pT}) = 0,
\\
& \int_0^2\frac{d\alpha_p}{\alpha_p} \, d^2 p_{pT} \; (2 - \alpha_p)
P_{[T_{LL}, U]}(\alpha_p,\bm p_{pT}) = 0
\\[1ex]
\nonumber
& \textrm{(same for $T_{LT}, T_{TT}$),}
\end{align}
\end{subequations}
because the polarization tensor is traceless ($L = 2$) and cannot produce any scalars.
Equations~(\ref{integral_baryon_number}) and (\ref{integral_momentum}) express the
baryon number and LF momentum sum rule at the nucleon level and play an essential role
in the applications to DIS. The fact that both sum rules are realized is an achievement
of LF quantization and the implementation of rotational symmetry through the c.m.\ frame
formulation. Equation~(\ref{integral_tensor}) guarantees that the sum rules hold
in a deuteron state with arbitrary tensor polarization.

%
%
\begin{figure}
\centering
\includegraphics[width=\linewidth]{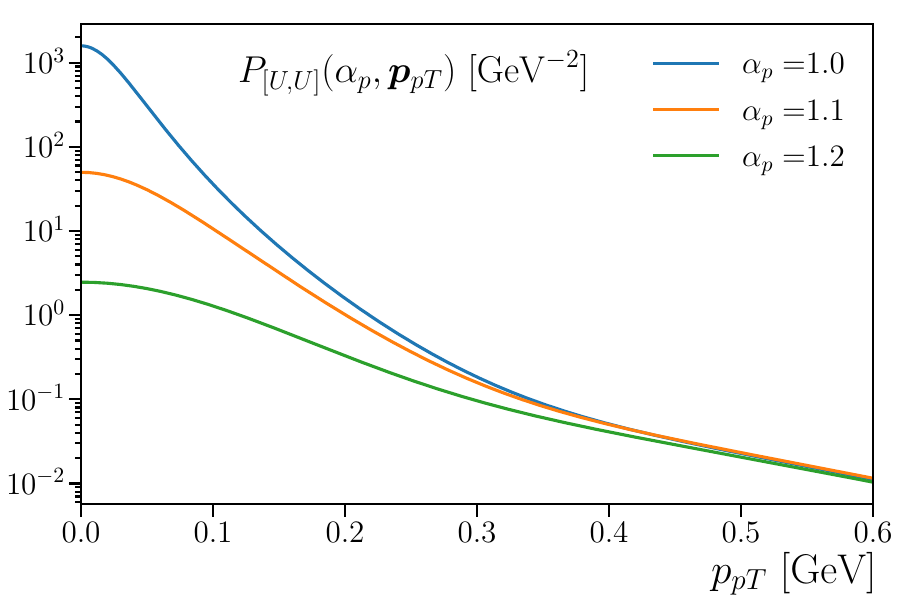}
\caption{The distribution of unpolarized neutrons in the unpolarized deuteron,
$P_{[U,U]}$, Eq.~(\ref{spectral_U_U}), as a function of the tagged proton transverse
momentum $p_{pT}$, for several values of the longitudinal momentum fraction $\alpha_p$.}
\label{fig:PUU_log}
\end{figure}
%
%
\begin{figure}
\centering
\includegraphics[width=\linewidth]{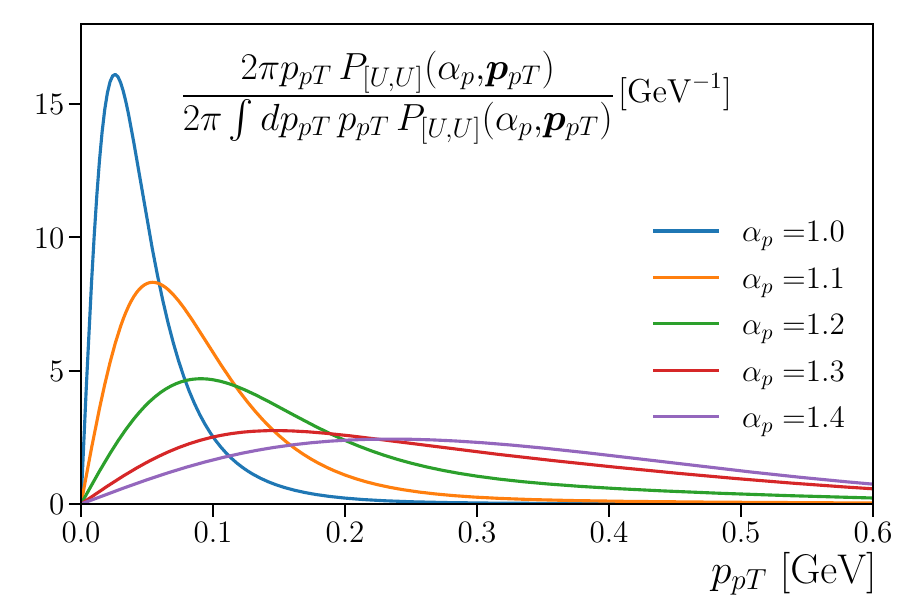}
\caption{Same as Fig.~\ref{fig:PUU_log}, but showing the distributions multiplied
by the phase space factor $2\pi p_{pT}$, and divided by the integral over $p_{pT}$
(normalized radial distributions in $p_{pT}$).}
\label{fig:PUU_normalized_pt}
\end{figure}
%
%
\begin{figure*}
\centering
\includegraphics[width=\linewidth]{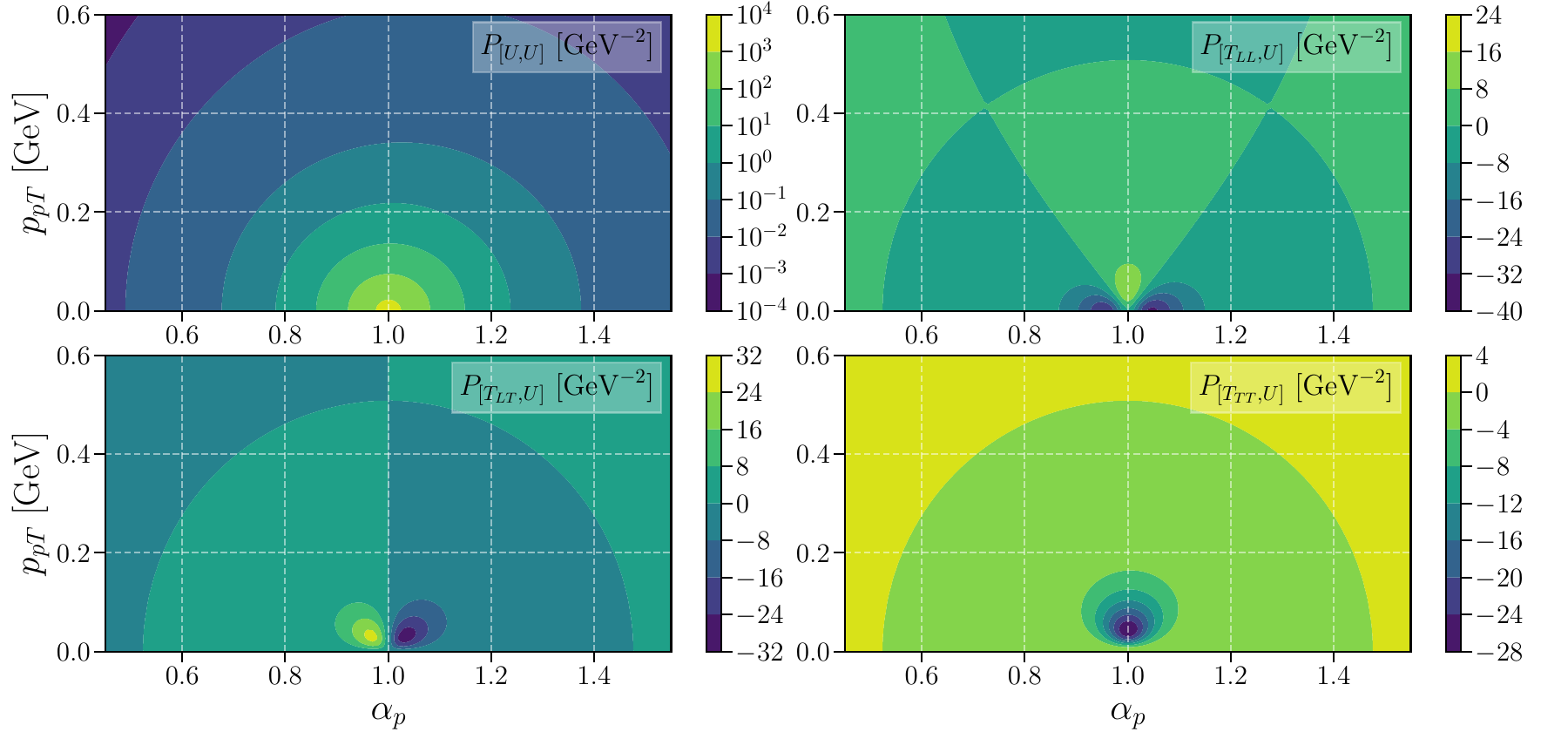}
\caption{The distribution of unpolarized neutrons in the unpolarized
and tensor-polarized deuteron, Eqs.~(\ref{spectral_U_U}) and (\ref{distribution_tensor}),
as functions of $\alpha_p$ and $p_{pT}$.
Upper left panel: Distribution in unpolarized deuteron, $P_{[U,U]}$, 
Upper right panel: $T_{LL}$ tensor-polarized deuteron, $P_{[T_{LL},U]}$. 
Lower left panel: $T_{LT}$ tensor-polarized deuteron, $P_{[T_{LT},U]}$. 
Lower right panel: $T_{TT}$ tensor-polarized deuteron, $P_{[T_{TT},U]}$.}
\label{fig:dist_tensor}
\end{figure*}

Numerical distributions are obtained by evaluating the expressions with models of the
radial wave functions $f_{0,2}(k)$. We use the nonrelativistic approximation
Eq.~(\ref{nonrel_approx}) and the nonrelativistic radial wave functions
obtained from empirical $NN$ interactions. All distributions shown in the
following are computed with the AV18 radial wave functions \cite{AV18};
the nuclear model dependence is investigated in Sec.~\ref{subsec:nuclear_model}.

Figure~\ref{fig:PUU_log} shows the distribution of unpolarized neutrons
in the unpolarized deuteron, $P_{[U,U]}$ as a function of $p_{pT} \equiv |\bm{p}_{pT}|$ for
several values of $\alpha_p$. The dependence on the variables is determined
by the way in which $p_{pT}$ and $\alpha_p$ combine to form the c.m.\ momentum $k$,
Eq.~(\ref{CM_LF_variables}), and by the behavior of the radial wave functions.
One observes: (i) $P_{[U,U]}$ is maximal at $\alpha_p = 1$ and $p_{pT} = 0$, where $k = 0$.
The distribution decreases with growing $|\alpha_p - 1|$ or $p_{pT}$.
(ii) The bulk of the strength is at values $|\alpha_p - 1|\leq 1.2; p_{pT} \leq 0.3~\text{GeV}$.
(iii) The effective dependence on $\alpha_p$ at fixed $p_{pT}$ changes with the value of $p_{pT}$,
and vice versa. At small $p_{pT} \leq$ 0.1 GeV, $P_{[U,U]}$ decreases roughly by an order of magnitude
if $|\alpha_p - 1|$ is increased by 0.1.  At large $p_{pT} \geq$ 0.5 GeV, $P_{[U,U]}$
decreases much more slowly in $|\alpha_p - 1|$. 

Figure~\ref{fig:PUU_normalized_pt} shows the radial distributions in transverse momentum,
$2\pi p_{pT} P_{[U,U]}$, as functions of $p_{pT}$ for fixed values of $\alpha_p$.
The functions are divided by the integral $2\pi \int dp_{pT} \, p_{pT} P_{[U,U]}
= \int d^2 p_{pT} \, P_{[U,U]}$ and represent the normalized $p_{pT}$ distributions
at the given value of $\alpha_p$. One observes that the shape of the normalized
$p_{pT}$ distributions strongly changes with $\alpha_p$. For $\alpha_p$ close to 1,
the distribution has a strong peak at a $p_{pT}$ value below 0.1 GeV and a long tail
at larger $p_{pT}$. For larger values of $\alpha_p$ the strength of the distribution
diffuses to larger values of $p_{pT}$, with a peak at larger values of $p_{pT}$
and a greatly increased width.

Figure~\ref{fig:dist_tensor} shows the distributions of unpolarized neutrons
in the unpolarized and tensor-polarized deuteron, $P_{[U,U]}$ and
$P_{[T_{LL},U]}, P_{[T_{LT},U]}, P_{[T_{TT},U]}$, as functions of $\alpha_p$ and $p_{pT}$.
The scales on the axes are such that the logitudinal and transverse 3-momentum of the proton
in the deuteron rest frame are represented in (approximately) the same graphical units,
such that the plot conveys an image of the shape of the distribution in physical 3-momentum
in the deuteron rest frame. One observes: (i) $P_{[U,U]}$ is positive and strongly centered
around $\alpha_p = 1$ and $p_{pT} = 0$ (see also Fig.~\ref{fig:PUU_log}). The distribution is
approximately spherical in the proton 3-momentum; the small deviations from spherical shape
are relativistic effects caused by the use of LF momentum variables.
(ii) $P_{[T_{LL},U]}, P_{[T_{LT},U]}$ and $P_{[T_{TT},U]}$ take positive and negative values.
They are numerically smaller than $P_{[U,U]}$ at $\alpha \approx 1$ and $p_{pT} \approx 0$
but become comparable at values $|\alpha - 1| \gtrsim 0.2$ and/or $p_{pT} \gtrsim$ 0.2 GeV, where the
$D$-wave becomes large. All three tensor polarized distributions have a node from the $(2f_0+f_2/\sqrt{2})$
factor, see Eq.~(\ref{distribution_tensor}). (iii) The angular dependence is governed by the polar angle
dependence of spherical harmonics in Eq.~(\ref{tensor_contractions}) and exhibits a
quadrupole pattern. These features are expressed in the tagged DIS observables
(see Sec.~\ref{sec:polarization_observables}) and can be tested experimentally.

\subsection{Polarized neutron distributions}
\label{subsec:polarized_distributions}
The longitudinally polarized neutron distribution obtained from Eq.~(\ref{distribution_general_trace})
with the density matrix Eq.~(\ref{eq:Pi_n}) and the projector in Table~\ref{tab:spin_structures} is
\begin{align}
&P_{[S_L]} (\alpha_p,\bm p_{pT} | \bm{S}_\deut)
\nonumber \\[1ex]
& \equiv \frac{\textrm{tr}[\Pi_n (-\gamma^+ \gamma^5)]}{(2 - \alpha_p)^2 \; p_\deut^+}
= \frac{2m s_n^+}{(2-\alpha_p)^2 p_\deut^+}.
\label{neutron_density_pol_trace}
\end{align}
Due to the presence of $\gamma^5$, only the term proportional to the polarization 4-vector $s_n$
in the density matrix Eq.~(\ref{eq:Pi_n}) contributes. Using the explicit expressions
of $s_n$ in the c.m. frame, Eq.~(\ref{eq:sn}), we obtain
\begin{subequations}
\label{distribution_helicity}
\begin{align}
&P_{[S_L]} (\alpha_p,\bm p_{pT} | \bm{S}_\deut)
= \frac{1}{2 - \alpha_p} \; 
\left( f_0 - \frac{f_2}{\sqrt{2}} \right)
\nonumber \\
& \times \left[ A \left( f_0 - \frac{f_2}{\sqrt{2}} \right) 
+ B \left( f_0 + \sqrt{2} f_2 \right)\right]\,,
\\[2ex]
A 
\, &= \, S_\deut^z - \frac{ E(E + k^z) |\bm{k}_T|^2}{(m^2 + |\bm{k}_T|^2) |\bm{k}|^2} \,  S_\deut^z
\nonumber \\[1ex]
&+ \; \frac{(E + k^z) [-|\bm{k}|^2 + E k^z]}{(m^2 + |\bm{k}_T|^2) |\bm{k}|^2} \, (\bm{S}_{\deut T} \bm{k}_T) ,
\\[2ex]
B 
\, &= \, \frac{ m(E + k^z) |\bm{k}_T|^2}{(m^2 + |\bm{k}_T|^2) |\bm{k}|^2} \,  S_\deut^z
\nonumber \\[1ex]
&- \; \frac{m k^z(E + k^z) }{(m^2 + |\bm{k}_T|^2) |\bm{k}|^2} \, (\bm{S}_{\deut T} \bm{k}_T)\,.\label{eq:C_AT}
\end{align}
\end{subequations}
At zero transverse nucleon momentum, 
\begin{align}
&A(\bm k_T=0)=S_\deut^z, &B(\bm k_T=0)=0\,.
\end{align}
The deviations from these values for $\bm k_T \neq 0$ are due to the
relativistic spin effects in LF quantization (Melosh rotations).

With respect to the deuteron polarization, the distribution Eq.~(\ref{distribution_helicity})
depends only on the polarization vector $\bm{S}_\deut$. As such it contains terms proportional
to the longitudinal and transverse vector polarization in the collinear frame.
We define normalized distributions in the longitudinally and transversely polarized deuteron
by choosing $\bm{S}_\deut$ as the unit vectors Eq.~(\ref{basis_vectors_prime})
(with $\bm{p}_{pT} \equiv \bm{k}_T$),
\begin{align}
\bm{e}_{z},
\hspace{1em}
\bm{e}_{x'} = \bm{k}_T / |\bm{k}_T|,
\hspace{1em}
\bm{e}_{y'} = \bm{e}_{z} \times \bm{e}_{x'}.
\label{unit_vectors_for_polarization}
\end{align}
We define
\begin{subequations}
\label{distribution_helicity_favored}
\begin{align}
& P_{[S_L, S_L]}(\alpha_p,\bm p_{pT} )
\equiv P_{[S_L]} (\bm{S}_\deut = \bm{e}_{z})
\nonumber \\[1ex]
&=\frac{1}{2-\alpha_p}\left( f_0-\frac{f_2}{\sqrt{2}}\right)
\nonumber \\
& \times
\left[A_{[=]} \left( f_0-\frac{f_2}{\sqrt{2}}\right)
+ B_{[=]} \left(f_0 + \sqrt{2}f_2\right)\right],
\\[2ex]
&A_{[=]}
\, \equiv \, 1 - \frac{ E(E + k^z) |\bm{k}_T|^2}{(m^2 + |\bm{k}_T|^2) |\bm{k}|^2} ,
\\[1ex]
&B_{[=]}
\, \equiv \,
\frac{ m(E + k^z) |\bm{k}_T|^2}{(m^2 + |\bm{k}_T|^2) |\bm{k}|^2},
\end{align}
\end{subequations}
and
\begin{subequations}
\label{distribution_helicity_unfavored}
\begin{align}
& P_{[S_T, S_L]}(\alpha_p,\bm p_{pT})
\equiv P_{[S_L]} (\bm{S}_\deut = \bm{e}_{x'}) 
\nonumber \\[1ex]
&=\frac{1}{2-\alpha_p}\left( f_0-\frac{f_2}{\sqrt{2}}\right)
\nonumber \\
&\times \left[A_{[\neq]} \left( f_0-\frac{f_2}{\sqrt{2}}\right)
- B_{[\neq]} \left(f_0 + \sqrt{2}f_2\right)\right],
\\[2ex]
&A_{[\neq]} \, \equiv \, \frac{(E + k^z) [-|\bm{k}|^2 + E k^z]|\bm k_T|}{(m^2 + |\bm{k}_T|^2) |\bm{k}|^2} ,
\\[1ex]
&B_{[\neq]} \, \equiv \, \frac{m k^z(E + k^z)|\bm k_T| }{(m^2 + |\bm{k}_T|^2) |\bm{k}|^2} .
\end{align}
\end{subequations}
At $\bm k_T=0$, the coefficients become
\begin{subequations}
\begin{align}
&A_{[=]}(\bm k_T=0) = 1, &B_{[=]}(\bm k_T=0)=0,
\\
&A_{[\neq]}(\bm k_T=0) = 0, &B_{[\neq]}(\bm k_T=0)=0.
\end{align}
\end{subequations}
Equation~(\ref{distribution_helicity_favored}) describes the distribution of longitudinally
polarized neutrons in a longitudinally polarized deuteron. This distribution
would be present also in a nonrelativistic system and is nonzero at $\bm{k}_T = 0$.
Equation~(\ref{distribution_helicity_unfavored}) describes the distribution of longitudinally polarized
neutrons in a transversely polarized deuteron. This distribution arises from relativistic
spin effects in the LF formulation and vanishes at $\bm{k}_T = 0$.
We refer to the two functions as the ``favored'' and ``unfavored'' distributions.
The terms will be explained further when comparing the longitudinally and transversely polarized distributions.

The distribution of transversely polarized neutrons obtained from Eq.~(\ref{distribution_general_trace})
with the density matrix Eq.~(\ref{eq:Pi_n}) and the projector in Table~\ref{tab:spin_structures} is
\begin{align}
&P^i_{[S_T]}(\alpha_p, \bm{p}_{pT}| \bm{S}_\deut) \;\equiv \; 
\frac{\text{tr}[\Pi_n (i\sigma^{+i} \gamma^5)]}{(2 - \alpha_p)^2 \; p_\deut^+}
\nonumber \\
&= \frac{2( p_n^+s_n^i-s_n^+p_n^i)}{(2-\alpha_p)^2 p_\deut^+}.
\label{distribution_transversity_trace}
\end{align}
As in the distribution of longitudinally polarized neutrons, only the term proportional to the
polarization 4-vector $s_n$ in the neutron density matrix contributes. With the explicit
expression of $s_n$ in the c.m.\ frame, Eq.~(\ref{eq:sn}), we obtain
\begin{subequations}
\label{distribution_transversity}
\begin{align}
& \bm{P}_{[S_T]} (\alpha_p, \bm{p}_{pT} | \bm{S}_\deut) \; = \; \frac{1}{2 - \alpha_p}
\left( f_0 - \frac{f_2}{\sqrt{2}} \right)
\nonumber \\
& \times \; \left[  \bm{A}_T  \left(f_0 +  \sqrt{2} f_2 \right)
+ \bm{B}_T  \left( f_0 - \frac{f_2}{\sqrt{2}} \right)  \right] ,
\end{align}
\begin{align}
\bm{A}_T &\equiv \bm S_{\deut T} - \frac{ E(E + k^z) \bm{k}_T}{(m^2 +
|\bm{k}_T|^2) |\bm{k}|^2} \, (\bm{S}_{\deut T} \bm{k}_T)
\nonumber \\[1ex]
&- \; \frac{(E + k^z) [-|\bm{k}|^2 + E k^z]\bm k_T}{(m^2 + |\bm{k}_T|^2) |\bm{k}|^2} \, S_\deut^z ,
\\
\bm{B}_T &\equiv \frac{ m(E + k^z) \bm{k}_T}{(m^2 + |\bm{k}_T|^2) |\bm{k}|^2} \,  
(\bm{S}_{\deut T} \bm{k}_T) \nonumber \\[1ex]
&+ \; \frac{m k^z(E + k^z)\bm k_T }{(m^2 + |\bm{k}_T|^2) |\bm{k}|^2} \, S_\deut^z .
\end{align}
\end{subequations}
At zero transverse nucleon momentum,
\begin{align}
&\bm{A}_T(\bm k_T=0) = \bm S_{\deut T}\,, &\bm{B}_T(\bm k_T=0)=0\,.
\end{align}

With respect to deuteron polarization, Eq.~(\ref{distribution_transversity})
contains terms proportional to the longitudinal and transverse vector polarization
in the collinear frame. The transverse neutron polarization described by the direction
of $\bm{P}_{[S_T]}$ has terms proportional to $\bm{k}_T$ and to $\bm{S}_{\deut T}$.
We define normalized distributions corresponding to deuteron and neutron polarizations
along the transverse unit vectors $\bm{e}_{x'}$ and $\bm{e}_{y'}$, Eq.~(\ref{unit_vectors_for_polarization}).
For transverse deuteron polarization,
\begin{subequations}
\label{transversity_favored}
\begin{align}
& P^\parallel_{[S_T, S_T]} \equiv \bm{e}_{x'} \bm{P}_{[S_T]} (\bm{S}_\deut = \bm{e}_{x'})
\nonumber \\[1ex]
&=\frac{1}{2-\alpha_p}\left( f_0-\frac{f_2}{\sqrt{2}}\right)
\nonumber \\
&\times 
\left[A_{[=]} \left(f_0 + \sqrt{2}f_2\right) + B_{[=]} \left( f_0-\frac{f_2}{\sqrt{2}}\right)\right],
\\[2ex]
& P^\perp_{[S_T, S_T]} \equiv \bm{e}_{y'} \bm{P}_{[S_T]} (\bm{S}_\deut = \bm{e}_{y'})
\nonumber \\[1ex]
&= \frac{1}{2-\alpha_p}\left( f_0-\frac{f_2}{\sqrt{2}}\right)\left(f_0 + \sqrt{2}f_2\right);
\end{align}
\end{subequations}
the other projections are zero
\begin{subequations}
\begin{align}
& \bm{e}_{y'} \bm{P}_{[S_T]} (\bm{S}_\deut = \bm{e}_{x'}) = 0,
\\[1ex]
& \bm{e}_{x'} \bm{P}_{[S_T]} (\bm{S}_\deut = \bm{e}_{y'}) = 0.
\end{align}
\end{subequations}
For longitudinal deuteron polarization
\begin{subequations}
\label{transversity_unfavored}
\begin{align}
&P_{[S_L, S_T ]}
\equiv \bm{e}_{x'} \bm{P}_{[S_T]} (\bm{S}_\deut = \bm{e}_{z})
\nonumber \\[1ex]
&=\frac{1}{2-\alpha_p}	\left( f_0-\frac{f_2}{\sqrt{2}}\right)
\nonumber \\
&\times \left[- A_{[\neq]}
\left(f_0 + \sqrt{2}f_2\right)
+ B_{[\neq]} \left( f_0-\frac{f_2}{\sqrt{2}}\right)\right],
\\[2ex]
&\bm{e}_{y'} \bm{P}_{[S_T]} (\bm{S}_\deut = \bm{e}_{z}) = 0.
\end{align}
\end{subequations}
The functions $A_{[=]}, B_{[=]}$ and $A_{[\neq]}, B_{[\neq]}$
in Eqs.~(\ref{transversity_favored}) and (\ref{transversity_unfavored})
are those defined in Eq.~(\ref{distribution_helicity_favored})
and (\ref{distribution_helicity_unfavored}) and thus the same as in the
helicity-dependent distributions.

In terms of the normalized distributions, the distributions of
longitudinally and transversely polarized neutrons
in a deuteron with general vector polarization,
Eqs.~(\ref{distribution_helicity})
and (\ref{distribution_transversity}), can be expressed as
\begin{subequations}
\label{polarized_from_normalized}
\begin{align}
&P_{[S_L]} (\alpha_p, \bm{p}_{pT} | \bm{S}_\deut)
\nonumber \\[1ex]
&= S_\deut^z P_{[S_L, S_L]}(\alpha_p, \bm{p}_{pT})
\nonumber \\[1ex]
&+ |\bm{S}_{\deut T}| \cos (\phi_p - \phi_S) P_{[S_T, S_L]}(\alpha_p, \bm{p}_{pT}),
\\[2ex]
&\bm{P}_{[S_T]} (\alpha_p, \bm{p}_{pT} | \bm{S}_\deut)
\nonumber \\[1ex]
&= |\bm{S}_{\deut T}| \left[
\cos (\phi_p - \phi_S) \bm{e}_{x'} P^\parallel_{[S_T, S_T]}(\alpha_p, \bm{p}_{pT})
\right.
\nonumber \\[1ex]
&+ \left. \sin (\phi_p - \phi_S) \bm{e}_{y'} P^\perp_{[S_T, S_T]}(\alpha_p, \bm{p}_{pT})
\right]
\nonumber \\[1ex]
&+ S_\deut^z \bm{e}_{x'} P_{[S_L, S_T]}(\alpha_p, \bm{p}_{pT}),
\end{align}
\end{subequations}
where we have used the expansion of $\bm{S}_\deut$ in the basis vectors,
Eq.~(\ref{vector_restframe}). Equations~(\ref{polarized_from_normalized}) show the
explicit dependence of the neutron distributions on the deuteron vector polarization
parameters and can be used in the calculation of tagged scattering observables.

The integrals of the polarized neutron distributions over the tagged proton momentum
describe the fraction of the deuteron spin carried by the neutron spin degrees of freedom
in LF quantization (``spin sum rules'').
The integrals can be computed as integrals over the c.m.\ momentum variable as in
Eqs.~(\ref{integral_baryon_number}) and (\ref{integral_momentum}), in which
rotational invariance is manifest and an expansion in $k/m$ can be performed
(see Appendix~\ref{app:spin_sum_rules}).
For the longitudinally polarized neutron distributions we obtain
\begin{subequations}
\label{sumrules_helicity}
\begin{align}
\int_0^2\frac{d\alpha_p}{\alpha_p} \, d^2 p_{pT} \; 
P_{[S_L, S_L]}(\alpha_p,\bm p_{pT}) &= 1 - \frac{3}{2} \omega_2,
\\
\int_0^2\frac{d\alpha_p}{\alpha_p} \, d^2 p_{pT} \; 
P_{[S_T, S_L]}(\alpha_p,\bm p_{pT}) &= \varepsilon_{[S_L]}.
\end{align}
\end{subequations}
Here $\omega_2$ is the $D$-state probability of the deuteron wave function in the c.m.\ frame,
given by the integral of $f_2^2$, Eq.~(\ref{dwave_probability}). $\varepsilon_{[S_L]}$ is
an integral of a quadratic form in $f_0$ and $f_2$, weighted by $k/m$, Eq.~(\ref{epsilon_sl_app})
(so-called relativistic correction).
For the transversely polarized neutron distributions we obtain
\begin{subequations}
\label{sumrules_transversity}
\begin{align}
\int_0^2\frac{d\alpha_p}{\alpha_p} \, d^2 p_{pT} \; 
P^\parallel_{[S_T, S_T]}(\alpha_p,\bm p_{pT}) &= 1 - \delta^\parallel,
\label{sumrule_transversity_parallel}
\\
\int_0^2\frac{d\alpha_p}{\alpha_p} \, d^2 p_{pT} \; 
P^\perp_{[S_T, S_T]}(\alpha_p,\bm p_{pT}) &= 1 - \delta^\perp,
\\
\int_0^2\frac{d\alpha_p}{\alpha_p} \, d^2 p_{pT} \; 
P_{[S_L, S_T]}(\alpha_p,\bm p_{pT}) &= \varepsilon_{[S_T]}.
\end{align}
\end{subequations}
Here $\delta^\parallel$ and $\delta^\perp$ are integrals of terms $f_0 f_2$ and $f_2^2$
(linear and quadratic in the $D$-wave), Eqs.~(\ref{delta_parallel_app}) and (\ref{delta_perp_app}).
$\varepsilon_{[S_T]}$ is the relativistic correction integral Eq.~(\ref{epsilon_st_app}),
of similar form as $\varepsilon_{[S_L]}$. The numerical values of the integrals
are summarized in Table~\ref{tab:sumrules}.

One observes: (i) The favored polarized neutron distributions (neutron polarization
and deuteron spin along same axis) integrate to unity up to depolarization
corrections resulting from the $D$-wave of the deuteron wave function, see
Eqs.~(\ref{sumrules_helicity}a), (\ref{sumrules_transversity}a), and (\ref{sumrules_transversity}b).
It shows the preservation of spin when the deuteron state is expanded in $pn$ states.
(ii)~The unfavored distributions (neutron polarization
and deuteron spin along different axes) integrate to zero,
up to relativistic corrections, see Eqs.~(\ref{sumrules_helicity}b) and Eq.~(\ref{sumrules_transversity}c).
It shows that some conversion of net transverse to longitudinal polarization (and vice versa)
happens because of the preferred direction in LF quantization.

The results for the integrals of the polarized neutron distributions are obtained thanks to
the manifest realization of 3-dimensional rotational invariance in the c.m.\ frame representation
of the LF wave function. They are essential for ensuring the sum rules for the tagged spin
structure functions in the IA (see Sec.~\ref{subsec:spin_sum_rules}).

%
%
\begin{figure}[t]
\centering
\includegraphics[width=\linewidth]{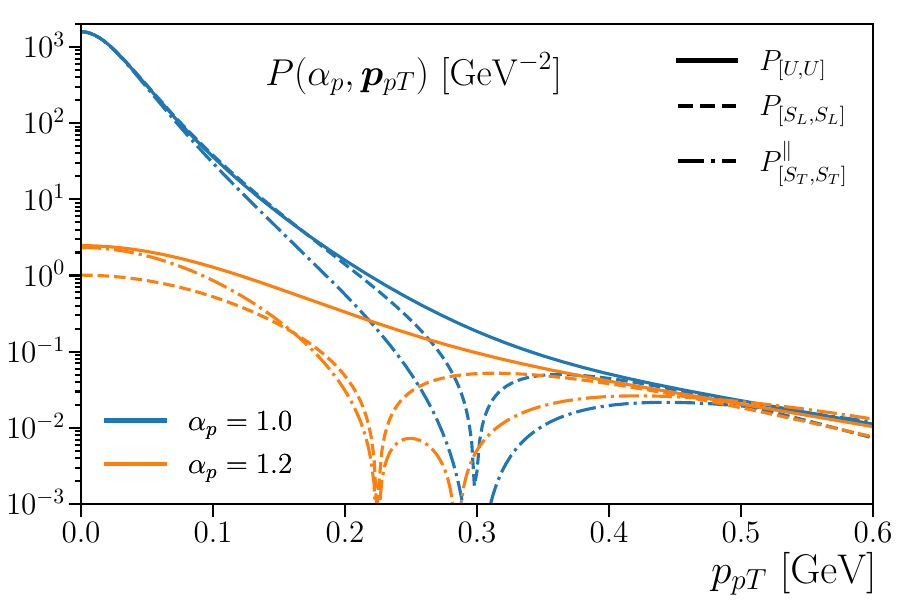}
\caption{Comparison of the distribution of unpolarized neutrons, $P_{[U,U]}$, Eq.~(\ref{spectral_U_U}),
and the ``favored'' distributions of longitudinally and transversely polarized neutrons,
$P_{[S_L,S_L]}$ and $P^\parallel_{[S_T,S_T]}$, 
Eqs.~(\ref{distribution_helicity_favored}) and (\ref{transversity_favored}).
The distributions are shown as functions of $p_{pT}$ for two values of $\alpha_p$.}
\label{fig:PUU_PSS}
\end{figure}
Numerical distributions of the polarized neutrons are obtained using the
same setup as in the unpolarized case. We want to compare the
unpolarized and polarized, and the favored and unfavored distributions.

Figure~\ref{fig:PUU_PSS} compares the unpolarized neutron distribution
$P_{[U,U]}$ with the favored distributions of longitudinally and
transversely polarized neutrons, $P_{[S_L,S_L]}$ and $P^\parallel_{[S_T,S_T]}$.
For $\alpha_p=1$ and $p_{pT}\leq 0.1~\text{GeV}$, the
deuteron $S$-wave dominates and there are very small depolarization
effects. Basically any deuteron polarization is transferred to the
nucleon, independent of the deuteron state, and all three
distributions consequently coincide.  For larger internal deuteron
momenta ($\alpha_p$ away from 1 or $p_{pT}
\sim~\text{few}~0.1~\text{GeV}$), the $D$-wave dominates and there is
significant depolarization.  In this region the favored
helicity-dependent distributions are suppressed relative to
$P_{[U,U]}$ and have a node for momenta where $f_0(k) =
f_2(k)/\sqrt{2}$, which is around $k\approx 0.3~\text{GeV}$.  

%
%
\begin{figure}[t]
\centering
\includegraphics[width=\linewidth]{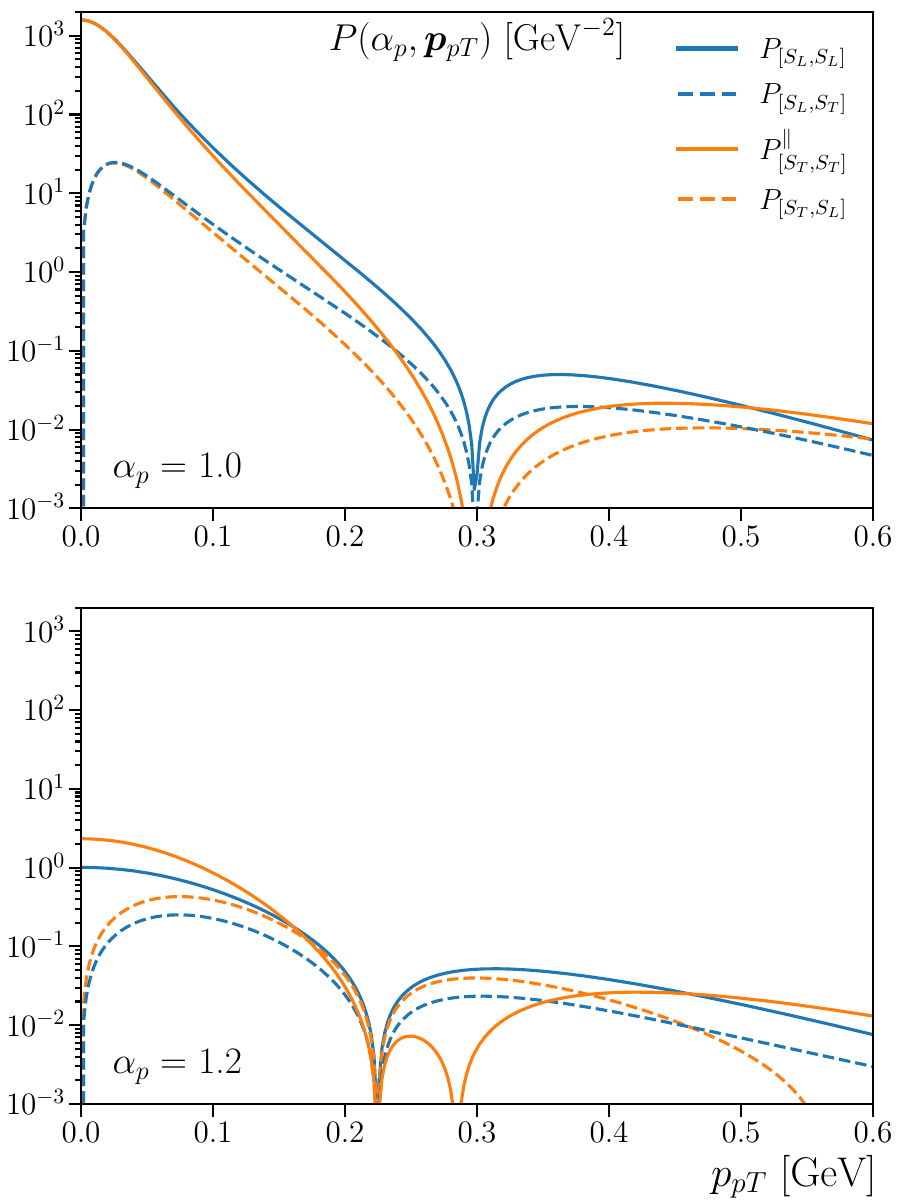}
\caption{Comparison of the favored and unfavored polarized neutron distributions
in the longitudinally polarized deuteron, $P_{[S_L, S_L]}$ and $P_{[S_L, S_T]}$,
and in the transversely polarized deuteron, $P_{[S_T, S_T]}^\parallel$ and $P_{[S_T, S_L]}$.
The distributions are plotted
as functions of $p_{pT}$ for fixed values of $\alpha_p$. Upper panel: $\alpha_p=1.0$.
Lower panel: $\alpha_p=1.2$. Both panels use the same vertical axis scale to
facilitate comparison of the magnitude of the functions.}
\label{fig:PSS_fav_unfav}
\end{figure}
Figure~\ref{fig:PSS_fav_unfav} compares the favored and unfavored polarized distributions
for all four combinations of deuteron and neutron polarization:
$P_{[S_L, S_L]}, P_{[S_L, S_T]}$ (deuteron longitudinal, neutron longitudinal/transverse),
and $P_{[S_T, S_T]}^\parallel, P_{[S_T, S_L]}$
(deuteron transverse, neutron transverse/longitudinal).
As observed in Fig.~\ref{fig:PUU_PSS}, all the polarized distributions have
a node for momenta where $f_0(k)=f_2(k)/\sqrt{2}$.  A second node can
appear (see examples in the bottom panel) for momenta where the factor
containing $A_{[=]},B_{[=]}$ (or $A_{[\neq]}, B_{[\neq]}$) goes to zero.
For all momenta away from those nodes, the unfavored distributions are
significantly suppressed relative to the favored distributions with
the same deuteron polarization.

Overall, the results of this section show a remarkable similarity of the longitudinally
and transversely polarized distributions of neutrons in the deuteron,
when comparing favored with favored and unfavored with unfavored distributions.
It appears even though the distributions refer to LF momentum and spin
variables and include relativistic effects and spin rotations.
The similarity is realized because: (i) the construction of the LF wave function in the c.m.\ frame
implements 3-dimensional rotational invariance; 
(ii) relativistic effects in the deuteron are moderate up to momenta $\sim$ few 100 MeV.

\subsection{Probabilistic distributions}
The neutron distributions discussed so far correspond to definite types of neutron polarization
(unpolarized, longitudinally polarized, transversely polarized)
and deuteron polarization (unpolarized, vector, tensor).
It is interesting to study the distributions describing the probability to find a neutron
with spin projection along a given axis in a deuteron state with given polarization state.
These probabilistic distributions are positive and bounded and have a simple interpretation.
They illustrate how the internal structure of deuteron ``polarizes'' or ``depolarizes''
the nucleons depending on the nuclear configuration selected by the spectator momentum.

The spin matrices defining the neutron distributions in states with spin projection
$\pm 1/2$ along the longitudinal or a transverse axis are given in
Table~\ref{tab:spin_structures} (second row). In the transverse case the spin
matrix depends on the transverse unit vector $\bm{n}_T$ specifying the quantization axis.
In the convention of Eq.~(\ref{notation_distributions}), we denote the
probabilistic distributions as
\begin{align}
&P_{[\text{neutron}]}(\alpha_p,\bm p_{pT}| \bm{S}_\deut, T_\deut)
\nonumber \\[1ex]
&\text{neutron} = L\pm, \, T\pm
\label{notation_distributions_probabilistic}
\end{align}
The probabilistic distributions are linear combinations of the unpolarized and longitudinally
or transversely polarized neutron distributions defined in
Eqs.~(\ref{distribution_unpol_trace}), (\ref{distribution_helicity}) and (\ref{distribution_transversity}),
\begin{subequations}
\label{distribution_probabilistic}
\begin{align}
& P_{[L\pm]}(\alpha_p,\bm p_{pT}| \bm{S}_\deut, T_\deut)
\nonumber \\
&= \tfrac{1}{2} \left[ P_{[U]} \pm P_{[S_L]} \right](\alpha_p,\bm p_{pT}| \bm{S}_\deut, T_\deut)
\\[1ex]
&P_{[T\pm]}(\alpha_p,\bm p_{pT}| \bm{S}_\deut, T_\deut)
\nonumber \\
&= \tfrac{1}{2} \left[ P_{[U]} \pm \bm{n}_T \bm{P}_{[S_T]} \right](\alpha_p,\bm p_{pT}| \bm{S}_\deut, T_\deut) .
\end{align}
\end{subequations}
Equations~(\ref{distribution_probabilistic}) describe the probabilistic neutron distributions
in a deuteron with a general (mixed) polarization state characterized by the
polarization parameters $\bm{S}_\deut$ and $T_\deut$.

For a pure deuteron spin state, with spin projection $\Lambda = \{ -1, 0, 1\}$ along an
axis described by the unit vector $\bm{N}$ in the deuteron rest frame, the
polarization parameters $\bm{S}_\deut$ and $T_\deut$ are given by Eqs.~(\ref{pure_state}) et seq.
Evaluating Eq.~(\ref{distribution_probabilistic}) with these expressions, we can compute the
probabilistic neutron distributions in a pure deuteron spin state. They depend on the
deuteron spin direction, the neutron spin direction, and the tagged proton momentum,
and exhibit a rich structure. We consider the following situations:
\begin{enumerate}[a)]
\item Neutron spin longitudinal or transverse along the
proton transverse momentum direction $(\bm{n}_T = \bm{e}_{x'})$, with spin projections $\pm 1/2$
\item Deuteron polarization longitudinal $(\bm{N} = \bm{e}_z)$ or transverse along the
proton transverse momentum direction $(\bm{N} = \bm{e}_{x'})$, with spin projections $\pm 1$
\end{enumerate}
We denote these distributions by
\begin{align}
&P_{[\text{deuteron, neutron}]} (\alpha_p,\bm p_{pT})
\nonumber \\[1ex]
&\text{neutron} = L\pm, \, T\pm
\nonumber\\[1ex]
&\text{deuteron} = L\pm, \, T\pm
\label{notation_distributions_probabilistic_pure}
\end{align}
We evaluate the distributions using the definitions Eq.~(\ref{distribution_probabilistic}),
the values of the deuteron polarization parameters in
Eqs.~(\ref{pure_state_longitudinal})--(\ref{pure_state_transverse_invariant}),
and the expressions of the distributions $P_{[U]}, P_{[S_L]}, \bm{P}_{[S_T]}$
in Eqs.~(\ref{unpolarized_from_normalized}) and (\ref{polarized_from_normalized}).
The result is expressed in terms of the normalized distributions introduced in
Sec.~\ref{subsec:unpolarized_distributions} and \ref{subsec:polarized_distributions}.

The probabilistic distributions of neutrons with longitudinal spin $\pm 1/2$
in a deuteron state with longitudinal spin $+1$, or with transverse spin $+1$
(along the proton transverse momentum $\bm{p}_{pT}$), are obtained as
\begin{subequations}
\label{distribution_pure_helicity}
\begin{align}
& P_{[L+, L\pm]}(\alpha_p, \bm{p}_{pT})
\nonumber \\
&= \tfrac{1}{2} \left[ P_{[U,U]} + \tfrac{1}{\sqrt{6}} P_{[T_{LL},U]} \pm P_{[S_L,S_L]} \right]
(\alpha_p, \bm{p}_{pT}),
\\[1ex]
& P_{[T+, L\pm]}(\alpha_p, \bm{p}_{pT})
\nonumber \\
&= \tfrac{1}{2} \left[ P_{[U,U]} -\tfrac{1}{2\sqrt{6}} P_{[T_{LL},U]}
+ \tfrac{1}{2\sqrt{2}}  P_{[T_{TT},U]} \right.
\nonumber \\
& \left. \phantom{\tfrac{0}{\sqrt{0}}} \pm P_{[S_T,S_L]} \right](\alpha_p, \bm{p}_{pT}).
\end{align}
\end{subequations}
The corresponding distributions in the deuteron states with longitudinal or transverse
spin projections $-1$ are given by
\begin{subequations}
\label{distribution_pure_helicity_reversed}
\begin{align}
& P_{[L-, L\pm]} = P_{[L+, L\mp]},
& P_{[T-, L\pm]} = P_{[T+, L\mp]},
\end{align}
\end{subequations}
as follows from fact that the neutron spin-dependent distributions Eq.~(\ref{distribution_helicity})
are proportional to the deuteron polarization vector and change sign when the deuteron polarization
is reversed.

The probabilistic distributions of neutrons with transverse spin $\pm 1/2$ in a deuteron state
with transverse spin $+1$ (both along the proton transverse momentum $\bm{p}_{pT}$),
or with longitudinal spin $+1$, are obtained as
\begin{subequations}
\label{distribution_pure_transversity}
\begin{align}
& P_{[T+, T\pm]}(\alpha_p, \bm{p}_{pT})
\nonumber \\
&= \tfrac{1}{2} \left[ P_{[U,U]}-\tfrac{1}{2\sqrt{6}} P_{[T_{LL}, U]}
+\tfrac{1}{2\sqrt{2}} P_{[T_{TT}, U]} \right.
\nonumber \\
&\left. \phantom{\tfrac{0}{\sqrt{0}}} \pm P^\parallel_{[S_T,S_T]} \right] (\alpha_p, \bm{p}_{pT}),
\\[1ex]
& P_{[L+, T\pm]}(\alpha_p, \bm{p}_{pT})
\nonumber \\
&= \tfrac{1}{2} \left[ P_{[U,U]}
+ \tfrac{1}{\sqrt{6}} P_{[U,T_{LL}]} \pm P_{[S_L,S_T]} \right](\alpha_p, \bm{p}_{pT}).
\end{align}
\end{subequations}
The corresponding distributions in the deuteron spin states with spin projections $-1$ are obtained
from relations analogous to Eq.~(\ref{distribution_pure_helicity_reversed}).

The distributions obtained in Eqs.~(\ref{distribution_pure_helicity}) and (\ref{distribution_pure_transversity})
are positive (as can be verified by explicit computation) and can be interpreted as the number densities
of neutrons with a given spin projection in a deuteron with given spin projection.
The distributions in which the directions of the neutron and deuteron spins are the same (both
longitudinal or both transverse), Eqs.~(\ref{distribution_pure_helicity}a) and
(\ref{distribution_pure_transversity}a), involve the favored polarized distributions
$P_{[S_L,S_L]}$ and $P_{[S_T,S_T]}$, 
Eqs.~(\ref{distribution_helicity_favored}) and (\ref{transversity_favored}).
The distributions in which the directions of the neutron and deuteron polarization are the
different (one longitudinal, one transverse), Eqs.~(\ref{distribution_pure_helicity}b) and
(\ref{distribution_pure_transversity}b), involve the unfavored polarized distributions
$P_{[S_T,S_L]}$ and $P_{[S_L,S_T]}$,
Eqs.~(\ref{distribution_helicity_unfavored}) and (\ref{transversity_unfavored}).

%
%
\begin{figure*}[t]
\includegraphics[width=.98\textwidth]{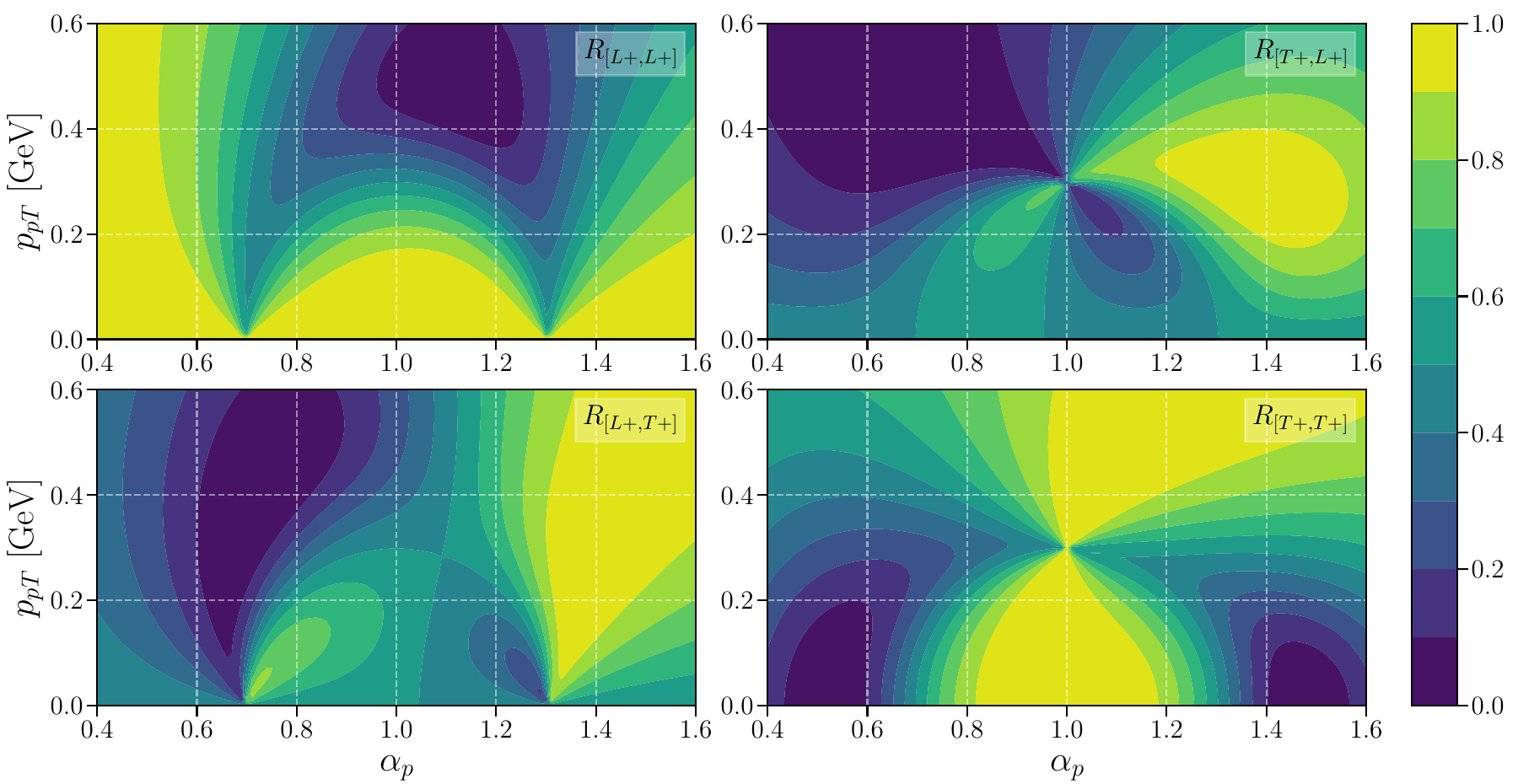}
\caption{The neutron polarization probabilities in spectator tagging with a polarized deuteron
(pure spin states with projections $\pm 1$, longitudinal or transverse),
Eqs.~(\ref{ratio_helicity}) and (\ref{ratio_transversity}),
as functions of the tagged proton LF momentum variables $\alpha_p$ and $p_{pT} \equiv |\bm{p}_{pT}|$.
Left/right column: Longitudinal/transverse deuteron polarization.
Upper/lower row: Longitudinal/transverse neutron spin.}
\label{fig:deut_lf_dist}
\end{figure*}
From the distributions Eqs.~(\ref{distribution_pure_helicity}) and (\ref{distribution_pure_transversity})
we can form the ratios
\begin{subequations}
\label{ratio_helicity}
\begin{align}
& R_{[L+, L+]} \equiv \frac{P_{[L+, L+]}}{P_{[L+, L+]} + P_{[L+, L-]}} ,
\\
& R_{[T+, L+]} \equiv \frac{P_{[T+, L+]}}{P_{[T+, L+]} + P_{[T+, L-]}} ,
\end{align}
\end{subequations}
and
\begin{subequations}
\label{ratio_transversity}
\begin{align}
& R_{[T+, T+]} \equiv \frac{P_{[T+, T+]}}{P_{[T+, T+]} + P_{[T+, T-]}} ,
\\
& R_{[L+, T+]} \equiv \frac{P_{[L+, T+]}}{P_{[L+, T+]} + P_{[L+, T-]}} .
\end{align}
\end{subequations}
The ratios in Eq.~(\ref{ratio_helicity}) describe the probability for the neutron to be in the
$+1/2$ longitudinal spin state in a deuteron polarized in the longitudinal or transverse direction,
as a function of the tagged proton LF momentum. The ones in Eq.~(\ref{ratio_helicity}) describe the
same probability for the neutron to be in the $+1/2$ transverse spin state in a deuteron polarized
in the transverse or longitudinal direction.  The ratios are bounded by $[0, 1]$ (because of the
positivity of the pure-state distributions) and can be interpreted as
probabilities. Eqs.~(\ref{ratio_helicity}a) and (\ref{ratio_transversity}a), where the direction of
polarization of the deuteron and neutron are the same, describe the polarization transfer from the
deuteron to the nucleon, as determined by the orbital motion and its angular momentum.
Eqs.~(\ref{ratio_helicity}b) and (\ref{ratio_transversity}b), where the direction of polarization of
the deuteron and neutron are different, describe the effective polarization perpendicular to the
deuteron spin as induced by the orbital motion. The corresponding ratios with the $-1/2$ neutron
spin state in the numerator are obtained as
\begin{align}
& R_{[L+, L-]} \equiv \frac{P_{[L+, L-]}}{P_{[L+, L+]} + P_{[L+, L-]}} =
1 - R_{[L+, L+]},
\end{align}
etc., and describe the complementary probability, for the neutron to be polarized in the ``opposite'' direction.

Note that in the ratios Eqs.~(\ref{ratio_helicity}) and (\ref{ratio_transversity}) the denominators are
defined as the sums of the neutron spin distributions in a deuteron with pure polarization state $+1$
(longitudinal or transverse). They are not the same as the distributions in the unpolarized deuteron,
as they contain some degree of tensor polarization, see Eqs.~(\ref{distribution_pure_helicity})
and (\ref{distribution_pure_transversity}).

Figure~\ref{fig:deut_lf_dist} shows the neutron polarization probabilities as functions of the tagged proton
LF momentum variables $\alpha_p$ and $p_{pT} \equiv |\bm{p}_{pT}|$. The left column shows the
neutron $L+$ polarization probabilities in the $L+$ and $T+$ polarized deuteron,
Eq.~(\ref{ratio_helicity}); the right column shows neutron $T+$ probabilities
in the $L+$ and $T+$ polarized deuteron. The probabilities exhibit a complex dependence
on the tagged proton momentum, which controls the $D/S$ wave ratio in the deuteron wave function
and thus the effective neutron polarization. One observes:

(i) At proton momenta with $0.8 < \alpha_p < 1.2$ and $p_{pT} <$ 0.2 GeV the neutron is almost completely
polarized along the deuteron spin direction, in both the longitudinally and transversely polarized deuteron.
Such LF momenta correspond to c.m.\ momenta $|\bm{k}| \lesssim$ 0.2 GeV, where the $S$-wave dominates.
The $[L+,L+]$ (upper left) and $[T+,T+]$ (lower right) probabilities show an effective
neutron polarization $\sim 1$. The $[L+,T+]$ (lower left) and $[T+,L+]$ (upper right) probabilities
have values $\sim 1/2$, showing the 50/50 probability for the neutron to be polarized in
either direction along the axis perpendicular to the deuteron spin direction.

(ii) At proton momenta with $\alpha_p < 0.7$ or $> 1.3$, or at $p_{pT} >$ 0.3 GeV, the $D$-wave becomes prominent
and causes significant depolarization of the neutron. The $[L+,L+]$ probability (upper left) becomes zero at
$p_{pT} \sim$ 0.5 GeV and $\alpha_p \approx 1$, showing that the neutron is polarized opposite to the
direction of the deuteron spin. These LF momenta correspond to c.m.\ momenta $|\bm{k}| \sim$ 0.5 GeV,
where the $S$-wave has a node and the $D$-wave dominates. Likewise, the $[T+,T+]$ probability (lower right)
becomes zero for $\alpha_p \sim 0.5$ or 1.5 and $p_{pT} \sim 0$. The $[L+,T+]$ (lower left) and
$[T+,L+]$ (upper right) probabilities again have values $\sim 1/2$ in these regions.

(iii) Some interesting effects of the LF spin structure can be seen in the probabilities at
zero transverse momentum, $p_{pT} = 0$. The $[L+,L+]$ probability (upper left) at $p_{pT} = 0$
remains exactly 1 for all values of $\alpha_p$. This happens because the LF spin structure causes
the $D$-wave to vanish at $p_{pT} = 0$. Using the
explicit expressions of the distributions in Eqs.~(\ref{distribution_pure_helicity}),
and Eqs.~(\ref{spectral_U_U}), (\ref{distribution_tensor}), and (\ref{distribution_helicity_favored}),
we obtain
\begin{align}
R_{[L+, L+]}(\alpha_p, \bm{p}_{pT} = 0) = 1 \hspace{2em} \textrm{($\alpha_p$ arbitrary).}
\end{align}
In contrast, the $[T+,T+]$ probability (lower right) at $p_{pT} = 0$ changes as $\alpha_p$ moves
away from unity. Using the expressions in Eqs.~(\ref{distribution_pure_transversity}),
and Eqs.~(\ref{spectral_U_U}), (\ref{distribution_tensor}), and (\ref{transversity_favored}),
we find
\begin{align}
&R_{[T+, T+]}(\alpha_p, \bm{p}_{pT} = 0)
\nonumber \\
&=1-\frac{\tfrac{9}{8}f_2^2}{(f_0+\frac{1}{2\sqrt{2}}f_2)^2+\tfrac{9}{8}f_2^2},
\end{align}
which shows that the probability decreases proportionally to the square of the $D$-wave as $\alpha_p$
moves away from unity.  Non-zero $(1-\alpha_p)$ implies $p_{pz}\neq 0$, where for a transversely polarized
deuteron the $D$-wave still contributes to the depolarization of the neutron. 
The $[L+,T+]$ (lower left) and $[T+,L+]$ (upper right) probabilities both attain values 1/2 at $p_{pT} = 0$,
\begin{subequations}
\begin{align}
R_{[T+, L+]}(\alpha_p, \bm{p}_{pT} = 0) = \frac{1}{2},
\\
R_{[L+, T+]}(\alpha_p, \bm{p}_{pT} = 0) = \frac{1}{2},
\end{align}
\end{subequations}
which shows that in these configurations the deuteron polarization causes equal neutron polarization
along either direction of the perpendicular axis.

(iv) All four plots show an approximately semicircular arc (centered at $\alpha_p=1, p_{pT}=0$,
with a radius of $p_{pT} \approx 0.3~\text{GeV}$ or $\Delta \alpha_p \approx 0.3$), where all
ratios take the value $R=1/2$, and there is no net nucleon polarization regardless
of the deuteron polarization state. This corresponds to the node in the spin-dependent
distributions caused by the vanishing of the factor $(f_0 - f_2/\sqrt{2})$ at $k \approx$ 0.3 GeV,
discussed in Sec.~\ref{subsec:polarized_distributions} and shown in
Figs.~\ref{fig:PUU_PSS} and \ref{fig:PSS_fav_unfav}.

(v) A striking spin-orbit effect is observed in the $[L+,T+]$ (lower left) and
$[T+,L+]$ (upper right) probabilities. Both probabilities attain values $\sim 1$
at $\alpha_p \gtrsim 1.3$ and $p_{pT} \gtrsim 0.3$ GeV, corresponding to complete
neutron polarization ``sideways'' to the deuteron spin. The effect is caused by the
entanglement of spin and orbital angular momentum in the deuteron wave function
and the selection of configurations by the tagged proton momentum.
In the particular kinematics here, the proton momentum has an angle $\sim 45^\circ$
relative to the $z$-axis and thus lies ``between'' the direction of the deuteron
spin and the neutron spin.

The spin and spin-orbit effects observed here can be observed in high-energy electron-deuteron
scattering with proton tagging, if the elementary electron-neutron scattering process
exhibits a dependence on the neutron spin. This can be realized in quasi-elastic electron scattering
(electric vs.\ magnetic neutron current) or in deep-inelastic scattering
(spin dependence of neutron structure functions).

The probabilistic distributions of neutrons in a pure deuteron spin state with projection $\Lambda = 0$
along a given axis can be studied in analogy to those in the $\Lambda = \pm 1$ states.
In the $\Lambda = 0$ state the vector polarization is absent, $\bm{S}_\deut = 0$, see Eqs.~(\ref{pure_state}),
and only the unpolarized neutron distribution $P_{[U]}$ in Eq.~(\ref{distribution_probabilistic}) is active.
The probabilistic distributions therefore do not depend on the neutron spin, longitudinal or transverse.
However, the distributions do depend on the angle of the proton momentum relative to
the deuteron spin quantization axis, because tensor polarization is present in the
$\Lambda = 0$ state, see Eqs.~(\ref{pure_state}).

\section{Tagged DIS structure functions}
\label{sec:IA_strucfunc}
\subsection{Unpolarized electron}
\label{sec:IA_strucfunc_unpol}
We now derive the explicit expressions of the structure functions of the polarized tagged DIS cross section.
The general decomposition of the cross section in independent structures and notation for the structure functions
are described in \firstpart, Sec.~\ref{P1:sec:cross_sec}. The particular kinematic variables and phase
space elements used in tagged DIS on the deuteron are described in Sec.~\ref{sec:process} of the present article.

The structure functions are derived from the master formula for the hadronic tensor in the IA,
Eq.~(\ref{w_impulse}), obtained in the LFQM formulation of the scattering process
in Sec.~\ref{subsec:quantum_mechanical}. The summation over the neutron LF helicities is performed using the
spin density matrix representation Eq.~(\ref{IA_density_matrix}) developed in
Sec.~\ref{subsec:nucleon_spin_density_matrix}.
The expressions of the structure functions are obtained by taking suitable components of the
IA hadronic tensor Eq.~(\ref{w_impulse}) ($++,TT,+T$) and matching them with the general
decomposition in terms of structure functions. The IA results for the tagged structure functions are
expressed in terms of the momentum- and spin-dependent neutron distributions introduced in
Secs.~\ref{subsec:neutron_distributions} -- \ref{subsec:polarized_distributions}, providing
a compact and transparent representation.

The calculations are performed in the DIS limit
\begin{align}
Q \gg \textrm{mass},
\label{dis_limit_ia}
\end{align}
where the generic mass scale includes the scale governing power corrections to
the nucleon structure functions in DIS,
\begin{align}
\textrm{mass} \sim x m ,
\end{align}
and the scales arising from the longitudinal and transverse motion of the nucleons in the deuteron 
\begin{align}
\textrm{mass} \sim (1 - \alpha_p) m, \, |\bm{p}_{pT}|.
\end{align}
Power corrections will be denoted simply as $\mathcal{O}(1/Q)$, without indicating the mass scale
governing the corrections; in most cases the origin of the mass scale will be clear; if necessary
it will be given explicitly.
In kinematic variables that have a finite value in the DIS limit, power corrections ${\mathcal O}(1/Q)$
to that value are neglected. The structure functions are computed to leading power accuracy.
In the structure functions that are $Q^2$-independent in the DIS limit
(``leading-twist'' in the terminology of QCD factorization),
power corrections ${\mathcal O}(1/Q)$ are neglected.
The power-suppressed structure functions (``higher-twist'') are evaluated to leading
nonzero power accuracy where possible.

The symmetric and antisymmetric parts of the hadronic tensor in Eq.~(\ref{IA_density_matrix})
\begin{align}
&W^{\mu\nu} = \tfrac{1}{2} W^{\{\mu\nu\}} + \tfrac{1}{2} W^{[\mu\nu]},
\\
&W^{\{\mu\nu\}}, W^{[\mu\nu]} \equiv W^{\mu\nu} \pm W^{\nu\mu},
\end{align}
give rise to the lepton-unpolarized and lepton-polarized structure functions,
see \firstpart, Sec.~\ref{P1:sec:cross_sec}.
We first compute the lepton-unpolarized structure functions. 
The symmetric part of the neutron hadronic tensor is diagonal in the neutron LF helicities
and independent of the value of the helicity,\footnote{We neglect the transverse spin 
dependence of the unpolarized nucleon tensor due to two-photon exchange effects
\cite{Christ:1966zz,Afanasev:2007ii}.} 
\begin{align}
W^{\{\mu\nu\}}_n(p_n, \widetilde{q}; \lambdanp, \lambda_n)
= \delta(\lambdanp, \lambda_n) \, W^{\{\mu\nu\}}_n(p_n, \widetilde{q}).
\end{align}
The decomposition of the symmetric neutron tensor is\footnote{Neutron and deuteron structure
functions are distinguished by adding square brackets and an $n,\deut$ index to the notation used in
\firstpart, e.g. $F_{[UU,T] n},F_{[UU,T] \deut}$ etc.}
\begin{align}
\tfrac{1}{2} W_n^{\{\mu\nu\}} &= \tfrac{1}{2} (e_L^\mu e_L^\nu - e_q^\mu e_q^\nu - g^{\mu\nu}) \, F_{[UU,T] n}
\nonumber
\\[1ex]
&+ \tfrac{1}{2} e_L^\mu e_L^\nu \, F_{[UU,L] n} ,
\label{eq:tensor_Wn}
\end{align}
where the basis vectors $e_L$ and $e_q$ are as defined in \firstpart, Eq.~(\ref{P1:basis_longit_set1}),
but constructed with the neutron 4-momentum $p_n$ and the effective 4-momentum transfer $\widetilde{q}$,
Eqs.~(\ref{momentum_neutron_alt}) and (\ref{qtilde_def}). [The decomposition in Eq.~(\ref{eq:tensor_Wn})
has the same form as the $\phi$-independent part of the unpolarized deuteron tensor
in \firstpart, Eq.~(\ref{P1:eq:41sf_first}).]
The neutron structure functions in Eq.~(\ref{eq:tensor_Wn}) depend on the invariants formed
with $p_n$ and $\widetilde{q}$. The effective invariant momentum transfer in the electron-neutron
scattering process is given by Eq.~(\ref{qtilde_squared}),
\begin{align}
-\widetilde{q}^{\,2} = Q^2 \left[1 + \mathcal{O}(1/Q^2)\right].
\label{qtilde_squared_dis}
\end{align}
The effective scaling variable is
\begin{align}
x_n \; \equiv \; \frac{-\widetilde{q}^2}{2 p_n \widetilde{q}}
\; = \; \frac{x}{2 - \alpha_p} \left[1 + \mathcal{O}(1/Q^2) \right];
\label{x_n}
\end{align}
its value is determined by the longitudinal LF momentum of the neutron, as fixed by the
tagged proton LF momentum. The neutron structure functions $F_{[UU,T] n}$ and $F_{[UU,L] n}$
in Eq.~(\ref{eq:tensor_Wn}) relate to the commonly used structure functions $F_{1n}$ and $F_{2n}$ as
\begin{subequations}
\begin{align}
F_{1n}  &=  \tfrac{1}{2}F_{[UU,T] n},
\\[1ex]
F_{2n}  &=  x_n (F_{[UU,T] n} + F_{[UU,L] n}),
\end{align}
\end{subequations}
where the coefficients are given up to terms $\mathcal{O}(1/Q^2)$.

The neutron tensor is averaged over the neutron LF helicities in Eq.~(\ref{IA_density_matrix}).
Because the symmetric neutron tensor is independent of the neutron LF helicity, the result for
the symmetric deuteron tensor involves only the unpolarized neutron distributions.
We obtain
\begin{align}
\tfrac{1}{2} \langle W^{\{\mu\nu\}}_\deut \rangle (p_\deut, q, p_p)
&= \frac{2 [2 (2\pi)^3]}{2 - \alpha_p} \; P_{[U]}(\alpha_p,\bm p_{pT}| T_\deut)
\nonumber \\
&\times \tfrac{1}{2} W^{\{\mu\nu\}}_n(p_n, \widetilde{q}),
\label{eq:tensor_ia_symm}
\end{align}
where $P_{[U]}$ is the unpolarized neutron distribution Eq.~(\ref{distribution_unpol_trace}).
With respect to deuteron polarization $P_{[U]}$ contains an unpolarized and a tensor-polarized part,
but no vector-polarized part, so for unpolarized electron scattering only unpolarized deuteron and
tensor-polarized deuteron structures appear. Expressions for the individual structure functions are
obtained by equating the IA result of Eqs.~(\ref{eq:tensor_ia_symm}) and (\ref{eq:tensor_Wn}) to the general
decomposition of the spin-1 semi-inclusive scattering tensor in \firstpart, Eq.~(\ref{P1:eq:41sf_first}),
and taking suitable components of the tensor equation. The calculation is straightforward; an explicit
demonstration of the steps is given in Ref.~\cite{Strikman:2017koc}, Appendix B.
For the unpolarized deuteron structure functions we obtain
\begin{subequations}
\label{eq:IA_FU} 
\begin{align}
& F_{[UU, T]\deut}(x, Q^2; \alpha_p, \bm{p}_{pT}) 
\nonumber \\[1ex]
&= \frac{2 \, [2(2\pi)^3]}{2 - \alpha_p} \, P_{[U,U]} (\alpha_p, \bm{p}_{pT})
\, F_{[UU, T]n} (x_n, Q^2),
\label{FT_ia}
\\[2ex]
& F_{[UU, L]\deut}(x, Q^2; \alpha_p, \bm{p}_{pT}) 
\nonumber \\[1ex]
&= \frac{2 \, [2(2\pi)^3]}{2 - \alpha_p} \, P_{[U,U]} (\alpha_p, \bm{p}_{pT})
\, F_{[UU, L]n} (x_n, Q^2),
\label{FL_ia}
\\[2ex]
& F_{[UU]\deut}^{\cos\phi_p}(x, Q^2; \alpha_p, \bm{p}_{pT}) 
\nonumber \\[1ex]
&= \frac{2 \, [2(2\pi)^3]}{2 - \alpha_p} \, P_{[U,U]} (\alpha_p, \bm{p}_{pT})
\, \frac{2x_n |\bm p_{pT}|}{Q}
\nonumber \\
& \hspace{2em} \times
\left[ F_{[UU, T]n} + F_{[UU, L]n} \right] (x_n, Q^2) 
\nonumber\\[1ex]
&= \frac{2 \, [2(2\pi)^3]}{2 - \alpha_p} \, P_{[U,U]} (\alpha_p, \bm{p}_{pT})
\, \frac{2|\bm p_{pT}|}{Q}
\nonumber \\
&\hspace{2em} \times F_{2n} (x_n, Q^2),
\\[2ex]
& F_{[UU]\deut}^{\cos2\phi_p}(x, Q^2; \alpha_p, \bm{p}_{pT}) 
\nonumber \\[1ex]
&= \frac{2 \, [2(2\pi)^3]}{2 - \alpha_p} \, P_{[U,U]} (\alpha_p, \bm{p}_{pT})
\, \frac{2x_n^2|\bm p_{pT}|^2}{Q^2}
\nonumber \\
& \hspace{2em} \times 
\left[ F_{[UU, T]n} + F_{[UU, L]n} \right] (x_n, Q^2)
\nonumber\\[1ex]
&= \frac{2 \, [2(2\pi)^3]}{2 - \alpha_p} \, P_{[U,U]} (\alpha_p, \bm{p}_{pT})
\, \frac{2x_n|\bm p_{pT}|^2}{Q^2}
\nonumber \\
&\hspace{2em} \times
F_{2n} (x_n, Q^2),
\end{align}
\end{subequations}
where $P_{[U,U]}$ is the unpolarized neutron distribution in the unpolarized distribution,
Eq.~(\ref{spectral_U_U}).  Equation~(\ref{eq:IA_FU}) represents the invariant structure functions of
tagged DIS on the deuteron in terms as products of the neutron LF momentum distribution in the
deuteron and the ordinary neutron DIS structure functions and embodies the ``factorization'' of
nuclear and nucleon structure. The factor
\begin{align}
2 \times 1/(2 - \alpha_p)
\end{align}
results from the relation between the neutron and deuteron tensors and ensures the proper sum rules and
compositeness relations for the nuclear structure functions (see below).

For the tensor polarized structure functions, we separate the contributions corresponding to
$T_{LL}, T_{LT}$ and $T_{TT}$ tensor polarization as described in Sec.~\ref{subsec:unpolarized_distributions},
using Eq.~(\ref{unpolarized_from_normalized}) and the normalized distributions.
For the $T_{LL}$ structure functions we obtain
\begin{subequations}
\label{eq:IA_FT_TLL}
\begin{align}
& F_{[UT_{LL}, T]\deut}(x, Q^2; \alpha_p, \bm{p}_{pT}) 
\nonumber \\[1ex]
&= \frac{2\, [2(2\pi)^3]}{2 - \alpha_p} \, \sqrt{\frac{3}{2}}P_{[T_{LL},U]} (\alpha_p, \bm{p}_{pT})
\, F_{[UU, T]n} (x_n, Q^2),
\label{FT_TLL_ia}
\\[0ex]
& F_{[UT_{LL}, L]\deut}(x, Q^2; \alpha_p, \bm{p}_{pT}) 
\nonumber \\[1ex]
&= \; \frac{2 \, [2(2\pi)^3]}{2 - \alpha_p} \; \sqrt{\frac{3}{2}}P_{[T_{LL},U]} (\alpha_p, \bm{p}_{pT})
\; F_{[UU, L]n} (x_n, Q^2),
\label{FL_TLL_ia}
\\[2ex]
& F_{[UT_{LL}]\deut}^{\cos\phi_p}(x, Q^2; \alpha_p, \bm{p}_{pT}) 
\nonumber \\[1ex]
&= \; \frac{2 \, [2(2\pi)^3]}{2 - \alpha_p} \; \sqrt{\frac{3}{2}}P_{[T_{LL},U]} (\alpha_p, \bm{p}_{pT})
\; \frac{2x_n|\bm p_{pT}|}{Q}
\nonumber \\
& \hspace{2em} \times
\left[ F_{[UU, T]n} + F_{[UU, L]n} \right] (x_n, Q^2),
\\[2ex]
& F_{[UT_{LL}]\deut}^{\cos2\phi_p}(x, Q^2; \alpha_p, \bm{p}_{pT}) 
\nonumber \\[1ex]
&= \; \frac{2 \, [2(2\pi)^3]}{2 - \alpha_p} \; \sqrt{\frac{3}{2}}P_{[T_{LL},U]} (\alpha_p, \bm{p}_{pT})
\; \frac{2x_n^2|\bm p_{pT}|^2}{Q^2}
\nonumber \\
& \hspace{2em} \times
\left[ F_{[UU, T]n} + F_{[UU, L]n} \right] (x_n, Q^2).
\end{align}
\end{subequations}
For the $T_{LT}$ structure functions we obtain
\begin{subequations}
\label{eq:IA_FT_TLT}
\begin{align}
& F_{[U T_{LT},T] \deut}^{\cos(\phi_p-\phi_{T_L})}(x, Q^2; \alpha_p, \bm{p}_{pT}) 
\nonumber \\[1ex]
&= \frac{2 \, [2(2\pi)^3]}{2 - \alpha_p} \; \sqrt{2} P_{[T_{LT},U]} (\alpha_p, \bm{p}_{pT})
\; F_{[UU, T]n} (x_n, Q^2),
\label{FT_TLT_ia}
\\[0ex]
& F_{[U T_{LT},L] \deut}^{\cos(\phi_p-\phi_{T_L})}(x, Q^2; \alpha_p, \bm{p}_{pT}) 
\nonumber \\[1ex]
&= \frac{2 \, [2(2\pi)^3]}{2 - \alpha_p} \; \sqrt{2} P_{[T_{LT},U]} (\alpha_p, \bm{p}_{pT})
\; F_{[UU, L]n} (x_n, Q^2),
\label{FL_TLT_ia}
\\[2ex]
& F_{[U T_{LT}] \deut}^{\cos\phi_{T_L}}(x, Q^2; \alpha_p, \bm{p}_{pT})
= F_{[U T_{LT}] \deut}^{\cos(2\phi_p-\phi_{T_L})}(...)
\nonumber \\[1ex]
&= \frac{2 \, [2(2\pi)^3]}{2 - \alpha_p} \; \frac{1}{\sqrt{2}} P_{[T_{LT},U]} (\alpha_p, \bm{p}_{pT})
\; \frac{2x_n|\bm p_{pT}|}{Q}
\nonumber \\
& \hspace{2em} \times
\left[ F_{[UU, T]n} + F_{[UU, L]n} \right] (x_n, Q^2),
\\[2ex]
& F_{[U T_{LT}]\deut}^{\cos(\phi_p+\phi_{T_L})}(x, Q^2; \alpha_p, \bm{p}_{pT})
= F_{[U T_{LT}]\deut}^{\cos(3\phi_p-\phi_{T_L})}(...)
\nonumber \\[1ex]
&= \frac{2 \, [2(2\pi)^3]}{2 - \alpha_p} \; \frac{1}{\sqrt{2}} P_{[T_{LT},U]} (\alpha_p, \bm{p}_{pT})
\; \frac{2x_n^2|\bm p_{pT}|^2}{Q^2}
\nonumber \\
& \hspace{2em} \times
\left[ F_{[UU, T]n} + F_{[UU, L]n} \right] (x_n, Q^2) .
\end{align}
\end{subequations}
For the $T_{TT}$ structure functions we obtain
\begin{subequations}
\label{eq:IA_FT_TTT}
\begin{align}
& F_{[U T_{TT},T]\deut}^{\cos(2\phi_p-2\phi_{T_T})}(x, Q^2; \alpha_p, \bm{p}_{pT}) 
\nonumber \\[1ex]
&= \frac{2 \, [2(2\pi)^3]}{2 - \alpha_p} \,  \frac{1}{\sqrt{2}}P_{[T_{TT},U]} (\alpha_p, \bm{p}_{pT})
\, F_{[UU, T]n} (x_n, Q^2),
\label{FT_TTT_ia}
\\[0ex]
& F_{[U T_{TT},L]\deut}^{\cos(2\phi_p-2\phi_{T_T})}(x, Q^2; \alpha_p, \bm{p}_{pT}) 
\nonumber \\[1ex]
&= \frac{2 \, [2(2\pi)^3]}{2 - \alpha_p} \,  \frac{1}{\sqrt{2}}P_{[T_{TT},U]} (\alpha_p, \bm{p}_{pT})
\, F_{[UU, L]n} (x_n, Q^2),
\label{FL_TTT_ia}
\\[2ex]
& F_{[U T_{TT}]\deut}^{\cos(\phi_p-2\phi_{T_T})}(x, Q^2; \alpha_p, \bm{p}_{pT})
= F_{[UT_{TT}]\deut}^{\cos(3\phi_p-\phi_{T_T})}(...)
\nonumber \\[1ex]
&= \frac{2 \, [2(2\pi)^3]}{2 - \alpha_p} \,  \frac{1}{2\sqrt{2}}P_{[T_{TT},U]} (\alpha_p, \bm{p}_{pT})
\, \frac{2x_n|\bm p_{pT}|}{Q}
\nonumber \\
& \hspace{2em} \times
\left[ F_{[UU, T]n} (x_n, Q^2)+ F_{[UU, L]n} (x_n, Q^2) \right],
\\[2ex]
& F_{[U T_{TT}]\deut}^{\cos2\phi_{T_T}}(x, Q^2; \alpha_p, \bm{p}_{pT})
= F_{[UT_{TT}]\deut}^{\cos(4\phi_p-2\phi_{T_T})}(...)
\nonumber \\[1ex]
&= \frac{2 \, [2(2\pi)^3]}{2 - \alpha_p} \,  \frac{1}{2\sqrt{2}}P_{[T_{TT},U]} (\alpha_p, \bm{p}_{pT})
\, \frac{2x_n^2|\bm p_{pT}|^2}{Q^2}
\nonumber \\
& \hspace{2em} \times
\left[ F_{[UU, T]n} (x_n, Q^2)+ F_{[UU, L]n} (x_n, Q^2) \right] .
\label{FU_TTT_ia}
\end{align}
\end{subequations}
Here $P_{[T_{LL},U]}, P_{[T_{LT},U]}$ and $P_{[T_{TT},U]}$ are the normalized neutron distributions
in the tensor-polarized deuteron defined in Eq.~(\ref{distribution_tensor}).

We observe that in the scattering of unpolarized electrons the following tagged structure functions are
leading-twist ($Q^2$-independent in the DIS limit): 
\begin{align}
& F_{[UU,T]\deut}, F_{[UU,L]\deut},
\nonumber \\[1ex]
& F_{[U T_{LL},T]\deut}, F_{[U T_{LT},T]\deut}^{\cos(\phi_p-\phi_{T_L})},
F_{[U T_{TT},T]\deut}^{\cos(2\phi_p-2\phi_{T_T})},
\nonumber\\[1ex]
& F_{[U T_{LL},L]\deut}, F_{[U T_{LT},L]\deut}^{\cos(\phi_p-\phi_{T_L})},
F_{[U T_{TT},L]\deut}^{\cos(2\phi_p-2\phi_{T_T})},\nonumber
\end{align}
for the unpolarized and tensor-polarized deuteron, respectively. All other structure functions are suppressed
by one or two powers of $|\bm{p}_{pT}|/Q$. Note that the leading-twist structure functions also contain
power-suppressed terms $|\bm{p}_{pT}|^2/Q^2$, which have been neglected here.

The general decomposition of the cross section of unpolarized electron scattering
in \firstpart, Eq.~(\ref{P1:eq:41sf_first}), also contains structures proportional to the
deuteron vector polarization. These structures are obtained as identically zero in the IA,
\begin{subequations}
\begin{align} 
& F_{[US_L]\deut}^{\sin\phi_p}, F_{[US_L]\deut}^{\sin2\phi_p}= 0,
\\
& F_{[US_T,L]\deut}^{\sin(\phi_p-\phi_S)}, F_{[US_T,T]\deut}^{\sin(\phi_p-\phi_S)}=0,
\\
& F_{[US_T]\deut}^{\sin(\phi_p+\phi_S)}, F_{[US_T]\deut}^{\sin(3\phi_p-\phi_S)},
F_{[US_T]\deut}^{\sin\phi_S}, F_{[US_T]\deut}^{\sin(2\phi_p-\phi_S)} = 0.
\end{align}
\end{subequations}
This result is specific to the dynamics of the IA, which does not generate spin-dependent phases
of the amplitudes (imaginary parts). FSI or other multi-step processes will
generate imaginary parts of the amplitudes and yield non-zero values for these
structure functions \cite{Strikman:2017koc}.

It should be noted that the IA expressions of the individual structure functions are
obtained from the $T$ component of the hadronic current in the deuteron hadronic tensor,
which is subject to interaction effects in LF quantization \cite{Frankfurt:1988nt}.
For the leading-twist structure functions, these interaction effects are generally suppressed
in the DIS limit, which is reflected in the fact that the structure functions satisfy the
momentum sum rule (see below). For the higher-twist structure functions, interaction effects
are generally not suppressed, and corrections to the IA approximations should be expected.

\subsection{Momentum sum rule}
The IA results for the tagged deuteron structure functions satisfy certain sum rules when integrated
over $x$ and over the tagged proton momentum. For the structure function
\begin{align}
\frac{x}{2} \left[ F_{[UU,T]\deut} + F_{[UU,L]\deut} \right](x, Q^2; \alpha_p, \bm{p}_{pT}),
\end{align}
corresponding to the conventional $F_{2\deut}$ deuteron structure function
with the scaling variable $x/2 = x_\deut \in [0,1]$, Eq.~(\ref{x_rescaled}),
we obtain \cite{Strikman:2017koc}
\begin{align}
& \int_0^2 dx \int d\Gamma_p \;
\frac{x}{2} \left[ F_{[UU,T]\deut} + F_{[UU,L]\deut} \right](x, Q^2; \alpha_p, \bm{p}_{pT}) 
\nonumber \\[2ex]
& = \int_0^2 dx \int_0^{2-x}\frac{d\alpha_p}{\alpha_p} d^2p_{pT}
\; P_{[U,U]} (\alpha_p, \bm{p}_{pT})
\nonumber \\
& \hspace{1em} \times \frac{x}{2 - \alpha_p}
\left[ F_{[UU, T]n} + F_{[UU, L]n} \right] (x_n, Q^2)
\nonumber \\[2ex]
& = \int_0^2 \frac{d\alpha_p}{\alpha_p} d^2p_{pT}
\; (2 - \alpha_p) \, P_{[U,U]} (\alpha_p, \bm{p}_{pT})
\nonumber \\
& \hspace{1em} \times \int_0^1 dx_n \; x_n
\left[ F_{[UU, T]n} + F_{[UU, L]n} \right] (x_n, Q^2)
\nonumber \\[2ex]
& = \int_0^1 dx_n \; x_n
\left[ F_{[UU, T]n} + F_{[UU, L]n} \right] (x_n, Q^2) .
\label{sumrule_structurefunction_unpolarized}
\end{align}
The integration bounds on $\alpha_p$ follow from the kinematic limit Eq.~(\ref{alphap_limit}).  In
the second expression we have substituted the IA results, Eqs.~(\ref{FT_ia}) and (\ref{FL_ia}).  In
the third expression we have changed the order of the integrations over $x$ and $\alpha_p$, and
changed the integration variable from $x$ to $x_n$ using Eq.~(\ref{x_n}).  In the last expression we
have used the momentum sum rule for the $P_{[U,U]}$ distribution, Eq.~(\ref{integral_momentum}). One
observes that the $x$-integral of the spectator-momentum-integrated deuteron structure function is
equal to the $x$-integral of the neutron structure function. This realizes the momentum sum rule for
the nuclear structure function and provides an important test of the consistency of the
approximations.  Similar integral relations can be derived for the tensor-polarized tagged structure
functions and evaluate to zero due to the angular integral over the tensor polarized densities; see
Eq.~(\ref{integral_tensor}).

\subsection{Polarized electron}
The antisymmetric part of the neutron hadronic tensor depends on the neutron LF helicities.
The dependence can be expressed in covariant form as
\begin{align}\label{eq:W_antisymm_nucleon}
& \tfrac{1}{2} W^{[\mu\nu]}_n(p_n, \widetilde{q}; \lambdanp, \lambda_n)
\nonumber \\[1ex]
&= A^{\mu\nu\rho} (p_n, \widetilde{q}) \;
\bar u (p_n, \lambdanp) \,
\left( \frac{-\gamma_\rho \gamma^5}{2m}\right) \, u (p_n, \lambda_n) .
\end{align}
The bilinear form in nucleon bispinors is the polarization 4-vector of the free neutron.
The rank-3 tensor is given by 
\begin{align}
\label{neutron_tensor_polarized}
A^{\mu\nu\rho} (p_n, \widetilde{q})
&= \frac{i}{2} \epsilon^{\mu\nu\sigma\tau} (e_q)_\sigma
\left\{ - (e_L)_\tau (e_{L\ast})^\rho \,  F_{[LS_L]n}
\right.
\nonumber \\
& + \; \left. \left[ (e_{L\ast})_\tau (e_{L\ast})^\rho + g_\tau^\rho \right] F_{[LS_T]n} \right\}.
\end{align}
Here the basis vectors $e_q, e_L$ and $e_{L\ast}$ are as defined in \firstpart, Eqs.~(\ref{P1:basis_longit_set1})
and (\ref{P1:basis_longit_set2}), but formed with the 4-momenta $p_n$ and $\widetilde{q}$. The parameter 
\begin{align}
\gamma_n &\equiv \frac{2x_n m}{\sqrt{-\widetilde q^2}}
= \frac{2 x m}{(2 - \alpha_p) Q} \left[1 + \mathcal{O}(1/Q^2)\right]
\end{align}
is as defined in \firstpart, Eq.~(\ref{P1:gamma_def}), but formed with the neutron DIS variables;
it is related to the parameter of the deuteron DIS process, Eq.~(\ref{gamma_approx}), by
\begin{align}
\gamma_n = \gamma / (2 - \alpha_p).
\end{align}
The neutron structure functions
in Eq.~(\ref{neutron_tensor_polarized}) depend on the invariant variables $\widetilde{q}^2$ and $x_n$, which
in the DIS limit are given by Eqs.~(\ref{qtilde_squared_dis}) and (\ref{x_n}). The structure functions
$F_{[LS_L]n}$ and $F_{[LS_T]n}$ in Eq.~(\ref{neutron_tensor_polarized}) relate to the
conventional neutron spin structure functions $g_{1n, 2n}$ as
\begin{subequations}
\label{neutron_polarized_conventional}
\begin{align}
F_{[LS_L]n} &= 2g_{1n},
\\
F_{[LS_T]n} &= -2\gamma_n (g_{1n}+g_{2n}).
\end{align}
\end{subequations}
Note that $F_{[LS_L]n}$ is leading-twist, while $F_{[LS_T]n}$ is higher-twist (power-suppressed).

When averaging the antisymmetric neutron tensor Eq.~(\ref{eq:W_antisymm_nucleon}) over the LF helicities
using Eq.~(\ref{neutron_operator_trace}), only the $\slashed{s}_n \gamma_5$ term in the covariant neutron spin
density matrix Eq.~(\ref{eq:Pi_n}) contributes. The average therefore takes a simple form: it replaces the
free neutron polarization 4-vector in Eq.~(\ref{eq:W_antisymm_nucleon}) by the effective polarization
vector provided by the density matrix Eq.~(\ref{eq:Pi_n}). We obtain
\begin{align}
& \tfrac{1}{2} \langle W^{[\mu\nu ]}_\deut(p_\deut, q, p_p) \rangle
\nonumber \\
&= \frac{2 \, [2 (2\pi)^3]}{(2 - \alpha_p)^2} \;
A^{\mu\nu\rho} (p_n, \widetilde{q}) \; s_{n, \rho} (\alpha_p, \bm{p}_{pT} | \bm{S}_\deut).
\label{tensor_ia_pol}
\end{align}
This result is achieved thanks to the covariant representation of the deuteron spin structure.
The effective polarization vector in Eq.~(\ref{tensor_ia_pol}) encodes all the information on
the deuteron polarization and the tagged proton momentum.

Expressions for the polarized structure functions are obtained by equating the IA
result of Eq.~(\ref{tensor_ia_pol}) with the general decomposition of the
spin-1 semi-inclusive scattering tensor in \firstpart, Eq.~(\ref{P1:eq:41sf_first}),
and taking suitable components of the tensor equation. 
The tensor equations involve the $+$ and $T$ components of the deuteron polarization vector,
and the structure functions can be expressed in terms of the longitudinally and transversely
polarized neutron distributions introduced in Sec.~\ref{subsec:polarized_distributions}.
The $\mu,\nu = 1, 2$ components of Eq.~(\ref{tensor_ia_pol}) give access to the structures
with longitudinal neutron polarization. We obtain
\begin{subequations}
\label{eq:F_IA_vector_hel}
\begin{align}
&F_{[LS_L]\deut}(x, Q^2; \alpha_p, \bm{p}_{pT}) 
\nonumber \\
&= \frac{2\, [2(2\pi)^3]}{(2-\alpha_p)} \, P_{[S_L,S_L]} (\alpha_p, \bm{p}_{pT}) \,
F_{[LS_L]n}(x_n,Q^2),
\\[2ex]
&F_{[LS_T]\deut}^{\cos(\phi_p-\phi_S)}(x, Q^2; \alpha_p, \bm{p}_{pT}) 
\nonumber \\
&=
\frac{2\, [2(2\pi)^3]}{(2-\alpha_p)} \, P_{[S_T,S_L]} (\alpha_p, \bm{p}_{pT}) \,
F_{[LS_L]n}(x_n,Q^2),
\end{align}
\end{subequations}
where $P_{[S_L,S_L]}$ and $P_{[S_T,S_L]}$ are given in Eqs.~(\ref{distribution_helicity_favored})
and (\ref{distribution_helicity_unfavored}). One observes that $F_{[LS_L]\deut}$ is proportional to the
favored distribution of longitudinally polarized neutrons in the longitudinally polarized deuteron,
$P_{[S_L,S_L]}$, while $F_{[LS_T]\deut}^{\cos(\phi_p-\phi_S)}$ is proportional to the unfavored
distribution in the transversely polarized deuteron, $P_{[S_T,S_L]}$;
see Sec.~\ref{subsec:polarized_distributions} for details.
Both structure functions are proportional to the neutron longitudinally polarized spin
structure function $F_{[LS_L]n}$ and are leading-twist, see Eq.~(\ref{neutron_polarized_conventional}).

The $\mu,\nu = +,T$ components of Eq.~(\ref{tensor_ia_pol}) give access to the structures
with transverse neutron polarization. We obtain
\begin{subequations}
\label{eq:F_IA_vector_transv}
\begin{align}
\label{eq:FLSL_IA}
&F_{[LS_L]\deut}^{\cos\phi_p}(x, Q^2; \alpha_p, \bm{p}_{pT}) 
\nonumber \\
&= \frac{2\, [2(2\pi)^3]}{(2-\alpha_p)} \, P_{[S_L,S_T]} (\alpha_p, \bm{p}_{pT}) \,
F_{[LS_T]n}(x_n,Q^2),
\\[2ex]
&\left[ F_{[LS_T]\deut}^{\cos\phi_S}+ F_{[LS_T]\deut}^{\cos(2\phi_p-\phi_S)} \right]
(x, Q^2; \alpha_p, \bm{p}_{pT}) 
\nonumber \\
&= \frac{2\, [2(2\pi)^3]}{(2-\alpha_p)} \, P^\parallel_{[S_T,S_T]} (\alpha_p, \bm{p}_{pT}) \,
F_{[LS_T]n}(x_n,Q^2),
\\[2ex]
&\left[ F_{[LS_T]\deut}^{\cos\phi_S}- F_{[LS_T]\deut}^{\cos(2\phi_p-\phi_S)}\right]
(x, Q^2; \alpha_p, \bm{p}_{pT}) 
\nonumber \\
&= \frac{2\, [2(2\pi)^3]}{(2-\alpha_p)} \, P^\perp_{[S_T,S_T]} (\alpha_p, \bm{p}_{pT}) \,
F_{[LS_T]n}(x_n,Q^2) ,
\\[2ex]
&F_{[LS_T]\deut}^{\cos\phi_S}(x, Q^2; \alpha_p, \bm{p}_{pT}) 
\nonumber \\
&= \frac{2\, [2(2\pi)^3]}{(2-\alpha_p)} \,
\frac{1}{2} \left[
P^\perp_{[S_T,S_T]} + P^\parallel_{[S_T,S_T]} \right] (\alpha_p, \bm{p}_{pT})
\nonumber \\
& \times F_{[LS_T]n}(x_n,Q^2) ,
\end{align}
\end{subequations}
where $P^\parallel_{[S_T,S_T]}, P^\perp_{[S_T,S_T]}$ and $P_{[S_L,S_T]}$ are given in
Eqs.~(\ref{transversity_favored}) and (\ref{transversity_unfavored}).
One observes that $F_{[LS_L]\deut}^{\cos\phi_p}$ is proportional to the unfavored distributions
of transversely polarized neutrons in the longitudinally polarized deuteron, $P_{[S_L,S_T]}$,
while $F_{[LS_T]\deut}^{\cos\phi_S}$ and $F_{[LS_T]\deut}^{\cos(2\phi_p-\phi_S)}$ are proportional to
the favored distributions in the transversely polarized deuteron, $P^\parallel_{[S_T,S_T]}$
and $P^\perp_{[S_T,S_T]}$; see Sec.~\ref{subsec:polarized_distributions} for details.
All three structure functions are proportional to the neutron transversely polarized spin structure function
$F_{[LS_T]n}$ and are consequently higher-twist, see Eq.~(\ref{neutron_polarized_conventional}).

The structure functions for polarized electron scattering on an unpolarized or tensor-polarized
deuteron are zero in the IA,
\begin{subequations}
\begin{align}
&F_{[LU]\deut}^{\sin\phi_p}=0,
\\
&
F_{[L T_{LL}]\deut}^{\sin\phi_p},
F_{[L T_{LT}]\deut}^{\sin(\phi_p+\phi_{T_\parallel})},
F_{[LT_{LT}]\deut}^{\sin\phi_{T_\parallel}},
F_{[LT_{LT}]\deut}^{\sin(2\phi_p-\phi_{T_\parallel})},
\nonumber \\
\hspace{1em}
&
F_{[LT_{TT}]\deut}^{\sin\phi_{T_\perp}},
F_{[LT_{TT}]\deut}^{\sin(\phi_p-\phi_{T_\perp})},
F_{[LT_{TT}]\deut}^{\sin(3\phi_p-\phi_{T_\perp})}=0\,.
\end{align}
\end{subequations}
This immediately follows from the fact that Eq.~(\ref{tensor_ia_pol}) is linear in the deuteron
vector polarization parameter $s_n \propto \bm{S}_\deut$ and does not give rise to unpolarized
or tensor-polarized structures.

\subsection{Spin sum rules}
\label{subsec:spin_sum_rules}
The IA results for the tagged deuteron structure functions in polarized electron scattering
satisfy certain sum rules when integrated over $x$ and over the tagged proton momentum,
similar to those in unpolarized electron scattering. For the principal leading-twist
spin structure function $F_{[L S_L]\deut}$, Eq.~(\ref{eq:FLSL_IA}), we obtain
\begin{align}
& \int_0^2 \frac{dx}{2} \int d\Gamma_p \; F_{[L S_L]\deut}(x, Q^2; \alpha_p, \bm{p}_{pT})
\nonumber \\[2ex]
& = \int_0^2 dx \int_0^{2-x}\frac{d\alpha_p}{\alpha_p (2-\alpha_p)} d^2p_{pT}
\; P_{[S_L,S_L]} (\alpha_p, \bm{p}_{pT})
\nonumber \\[1ex]
&\hspace{1em} \times F_{[LS_L]n} (x_n, Q^2)
\nonumber \\[2ex]
& = \int_0^2 \frac{d\alpha_p}{\alpha_p} d^2p_{pT} \; P_{[S_L,S_L]} (\alpha_p, \bm{p}_{pT})
\nonumber \\
& \hspace{1em} \times \int_0^1 dx_n \, F_{[LS_L]n} (x_n, Q^2)
\nonumber \\[2ex]
& = (1 - \tfrac{3}{2} \omega_2) 
\; \int_0^1 dx_n \, F_{[LS_L]n} (x_n, Q^2).
\label{sumrule_structurefunction_polarized}
\end{align}
The steps are the same as in the derivation of the momentum sum rule
Eq.~(\ref{sumrule_structurefunction_unpolarized}).
In the last step we have used the sum rule for the polarized neutron distribution,
Eq.~(\ref{sumrules_helicity}). Equation~(\ref{sumrule_structurefunction_polarized})
is the spin sum rule for the tagged DIS structure function. It shows that the
$x$-integral of the tagged spin structure function reduces to that of the neutron
spin structure function when integrating over the spectator momentum, and that the
only effect of nuclear binding is the neutron depolarization in the initial state.

Similar relations can be derived for the higher-twist tagged spin structure functions.
Here we demonstrate the Burkhardt-Cottingham (BC) sum rule \cite{Burkhardt:1970ti,Kodaira:1998jn}
for the tagged spin structure function
\begin{align}
g_{2D} = -\frac{1}{2} F_{[L,S_L]D} -\frac{1}{2\gamma} F_{[L,S_T]D}^{\cos\phi_S}
\label{g2_tagged}
\end{align}
by computing the integral
\begin{align}
\int_0^2 \frac{dx}{2} \int d\Gamma_p \; g_{2D}(x, Q^2; \alpha_p, \bm{p}_{pT}) .
\label{sumrule_bc}
\end{align}
The integral of the first term in Eq.~(\ref{g2_tagged}) is evaluated in
Eq.~(\ref{sumrule_structurefunction_polarized}); in terms of the neutron spin structure function $g_{1n}$
the result is (we omit the arguments of the functions for brevity)
\begin{align}
& \int_0^2 \frac{dx}{2} \int d\Gamma_p \; \left[ -\frac{1}{2} F_{[L S_L]\deut} \right]
\nonumber \\
& = - (1 - \tfrac{3}{2} \omega_2) \; \int_0^1 dx_n \, g_{1n}.
\label{sumrule_bc_g1}
\end{align}
The integral of the second term in Eq.~(\ref{g2_tagged}) is computed going through similar steps:
\begin{align}
&\int_0^2 \frac{dx}{2} \int d\Gamma_p \; \left[ -\frac{1}{2\gamma} F_{[L,S_T]D}^{\cos\phi_S} \right]
\nonumber\\[1ex]
&= \int_0^2 dx \int_0^{2 - x} \frac{d\alpha_p}{\alpha_p (2 - \alpha_p)} d^2 p_{pT}\;
\nonumber \\
& \qquad \times \frac{1}{2} \left[P^\parallel_{[S_T,S_T]}+P^\perp_{[S_T,S_T]} \right]
\frac{\gamma_n}{\gamma} (g_{1n}+g_{2n})
\nonumber\\[1ex]
&= \int_0^2 dx \int_0^{2 - x} \frac{d\alpha_p}{\alpha_p (2 - \alpha_p)^2} d^2 p_{pT}\;
\nonumber \\
& \qquad \times \frac{1}{2} \left[P^\parallel_{[S_T,S_T]}+P^\perp_{[S_T,S_T]} \right] (g_{1n}+g_{2n})
\nonumber\\[1ex]
&= \int_0^2 \frac{d\alpha_p}{\alpha_p (2 - \alpha_p)} d^2 p_{pT}
\; \frac{1}{2} \left[ P^\parallel_{[S_T,S_T]}+P^\perp_{[S_T,S_T]} \right]
\nonumber\\
&\qquad\times \int_0^1 dx_n (g_{1n}+g_{2n})
\nonumber\\
&= ( 1-\tfrac{3}{2}\omega_2) \; \int_0^1 dx_n (g_{1n}+g_{2n}).
\label{sumrule_bc_g12}
\end{align}
In the last step we used the sum rule Eq.~(\ref{sumrule_parallel_perp_inverse})
for the $\parallel$ and $\perp$ neutron distributions.
Combining Eqs.~(\ref{sumrule_bc_g1}) and (\ref{sumrule_bc_g12}), the terms involving
$g_{1n}$ cancel, and we obtain
\begin{align}
& \int_0^2 \frac{dx}{2} \int d\Gamma_p \, g_{2D}
\nonumber \\
&= ( 1 - \tfrac{3}{2}\omega_2) \, \int_0^1 dx_n \, g_{2n}(x_n,Q^2) = 0.
\end{align}
In the last step we have used the BC sum rule for the neutron structure function $g_{1n}$.
Thus the BC sum rule for the tagged deuteron structure function
is realized by combining the spin sum rule for the neutron LF momentum distributions
and the BC sum rule for the neutron structure function.

\section{Tagged polarization observables}
\label{sec:polarization_observables}

\subsection{Unpolarized deuteron}
The IA expresses the structure functions of DIS on the deuteron with proton tagging
in terms of calculable neutron momentum distributions and empirical neutron structure functions
(or vice versa, with neutron tagging and proton structure functions).
As such it offers a simple framework for estimating observables and analyzing the experiments.
The general form of the spin observables in polarized semi-inclusive scattering from a spin-1
target is discussed in \firstpart, Sec.~\ref{P1:sec:observables}.
We now compute the basic observables in polarized tagged DIS using the IA and
study the sensitivity to deuteron and neutron structure.

Experiments in polarized tagged DIS can be performed with various goals:
\begin{itemize}
\item validate the IA by testing the factorization of nuclear and nucleon structure,
testing the universality of the elements, quantifying deviations;
\item extract the free neutron spin structure functions;
\item extract the spin-dependent neutron LF momentum distributions in the deuteron,
including structures sensitive to the $D/S$ ratio, spin-orbit correlations, high-momentum components;
\item study effects beyond the IA, such as initial-state modifications
(EMC effect, antishadowing) or FSI ($T$-odd structures)
\end{itemize}
Each type has its specific requirements and uncertainties and merits a dedicated impact study.
Here we only provide an orientation and assess the basic feasibility of the measurements
in EIC kinematics. 

The following numerical estimates use the deuteron LF wave function obtained from the AV18
nonrelativistic deuteron wave function \cite{AV18} using the prescription of
Eq.~(\ref{nonrel_approx}).  The unpolarized neutron structure function $F_{2n}$ is evaluated using
the SLAC parametrization \cite{Bodek:1979rx}, and a constant value $R = F_L/F_T = 0.18$ is
assumed. The polarized neutron structure function $g_{1n}$ is evaluated using the DSSV09
parametrization \cite{deFlorian:2009vb}, and $g_{2n}$ is computed using the Wandzura-Wilczek
approximation \cite{Wandzura:1977qf,Kodaira:1998jn}. The simple input is chosen in order to make the
results reproducible; the numerical estimates could easily be refined using more elaborate
parametrizations.

The estimates are performed in the DIS limit, Eq.~(\ref{dis_limit_ia}).
The contributions of terms in the cross section to observables are ordered by 
(i) the scaling behavior of the kinematic factors, determined by their dependence on $\gamma$,
see \firstpart, Eq.~(\ref{P1:gamma_def}); (ii) the scaling behavior of structure functions, determined
by their dependence on $|\bm{p}_{pT}|/Q$ in the IA. The results are presented in leading nonvanishing
order of $1/Q$, without power corrections. 

Unpolarized tagged DIS cross section controls the reaction rates and determines feasibility of
measuring more complex observables such as spin asymmetries. Unpolarized measurements are also used
directly for the extraction of unpolarized neutron structure
\cite{Sargsian:2005rm,Strikman:2017koc,Jentsch:2021qdp} and studies of nuclear modifications
\cite{Melnitchouk:1996vp,Sargsian:2002wc}.

%
%
\begin{figure}[t]
\includegraphics[width=\linewidth]{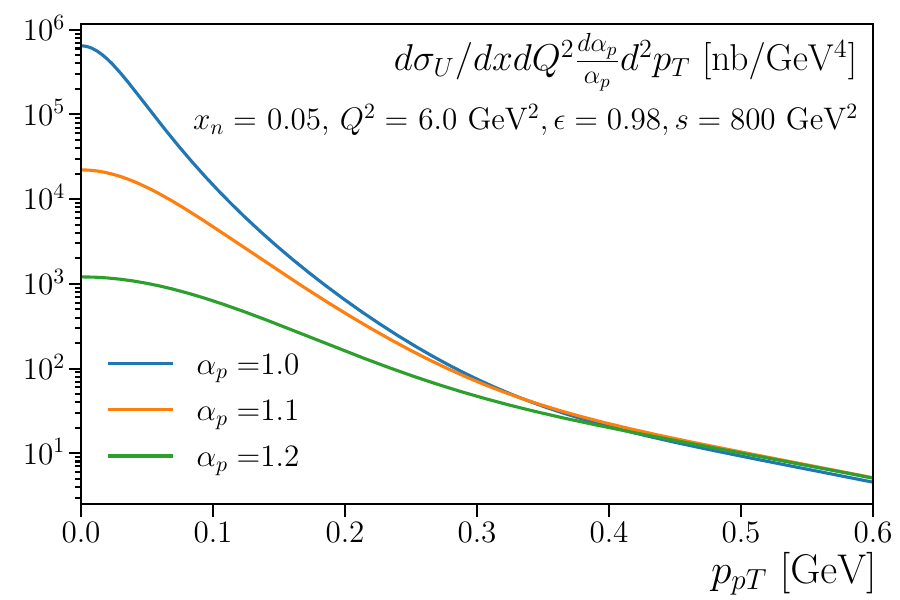}
\caption{Differential cross section of unpolarized tagged DIS, Eq.~(\ref{differential_unpol}),
as a function of the tagged proton transverse momentum $p_{pT}$,
for several values of the longitudinal momentum fraction $\alpha_p$ and fixed $x_n$
The cross section is averaged over the azimuthal angle $\phi_p$.}
\label{fig:sigma_u_log}
\end{figure}
Figure~\ref{fig:sigma_u_log} shows the magnitude and spectator momentum dependence of
the differential cross section of unpolarized tagged DIS in typical EIC kinematics.
The function plotted is
\begin{align}
d\sigma_U / dx dQ^2 (d\alpha_p / \alpha_p) d^2 p_{pT} ,
\label{differential_unpol}
\end{align}
where the differential cross section is given by Eq.~(\ref{cross_section_deuteron})
with $\mathcal{F}_U$ only ($\mathcal{F}_S, \mathcal{F}_T = 0$) and is averaged over the
tagged proton azimuthal angle $\phi_p$. This cross section depends only on the
leading-twist tagged structure functions $F_{[UU,T]}$ and $F_{[UU,L]}$.
It is plotted as a function of $p_{pT}$ for various values of $\alpha_p$,
and a fixed value of $x_n$, Eq.~(\ref{x_n}) (the value $x$ is changed when
$\alpha_p$ is varied such as to keep $x_n$ fixed).
One observes: (i)~The dependence on the tagged proton
momentum essentially tracks that of the unpolarized neutron distribution $P_{[U,U]}$
shown in Fig.~\ref{fig:PUU_log}. (ii)~Even at tagged proton momenta
corresponding to exceptional configurations in the deuteron
($|\alpha_p-1| \gtrsim 0.2; \, p_{pT} \gtrsim$ 0.2 GeV),
large event numbers can be collected at a projected EIC integrated
luminosity of 10 fb$^{-1}$ \cite{AbdulKhalek:2021gbh}. The rate for a given finite phase space element
in the tagged proton momentum is essentially the inclusive neutron DIS rate, multiplied by the
integral of the $P_{[U,U]}$ distribution over the phase space element.

%
%
\begin{figure}[t]
\includegraphics[width=\linewidth]{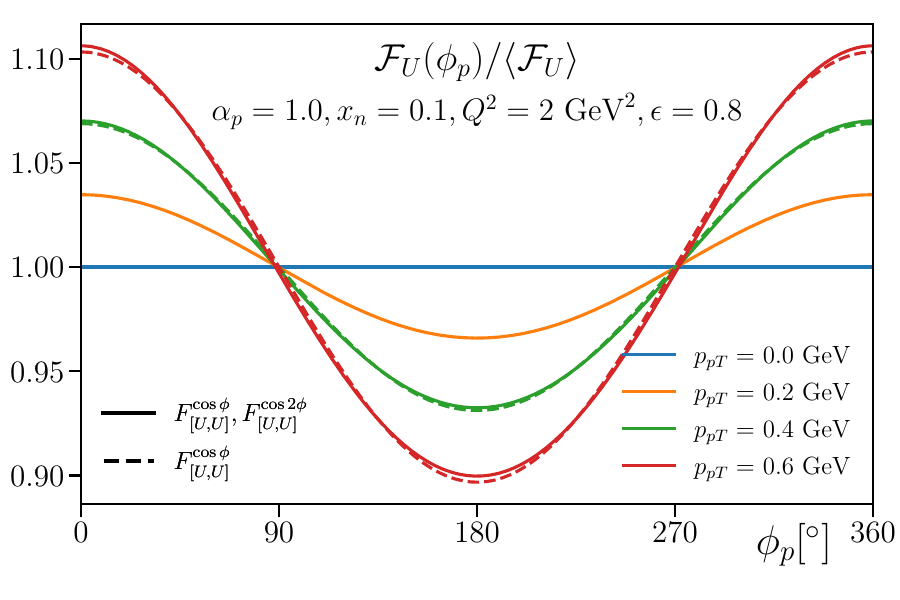}
\caption{Ratio of the $\phi_p$-dependent and $\phi_p$-averaged differential cross sections
in unpolarized tagged DIS, Eq.~(\ref{observable_unpol_ratio}),
as a function of the tagged proton azimuthal angle $\phi_p$.
\textit{Solid lines:} Ratio with all $\phi_p$-dependent structures in numerator.
\textit{Dashed lines:} Ratio with only $\cos \phi_p$ term in numerator.}
\label{fig:sigma_u_mod_relative}
\end{figure}
Figure~\ref{fig:sigma_u_mod_relative} shows the azimuthal modulation of the
unpolarized tagged DIS cross section. The function plotted is the ratio
of the $\phi_p$-dependent and $\phi_p$-averaged differential cross sections,
\begin{subequations}
\label{observable_unpol_ratio}
\begin{align}
\frac{\mathcal{F}_U(\phi_p)}{\langle \mathcal{F}_U \rangle}
&= 1
+ \sqrt{2\epsilon (1 + \epsilon)} \cos\phi_p \frac{F_{[U,U]}^{\cos\phi_p}}{F_{[UU,T]} + \epsilon F_{[UU,L]}}
\nonumber \\
&+ \epsilon \cos2 \phi_p \frac{F_{[U,U]}^{\cos 2\phi_p}}{F_{[UU,T]} + \epsilon F_{[UU,L]}},
\\[1ex]
\langle...\rangle &\equiv \int_0^{2\pi} \frac{d\phi_p}{2\pi} \; (...),
\end{align}
\end{subequations}
which includes the $\epsilon$-dependent factors accompanying the
angular structures in the cross section, see \firstpart, Eq.~(\ref{P1:cross_section_unpol}).
In the IA the two $\phi_p$-dependent structure functions are higher-twist,
being suppressed by factors $p_{pT}/Q$ (for $\cos\phi_p$) and
$(p_{pT}/Q)^2$ (for $\cos2\phi_p$).  This is reflected in the numerical size
of the azimuthal modulations, which are generally small and grow with $p_{pT}$.
The $\cos\phi_p$ harmonic reaches a magnitude of 10\% for $p_{pT} \sim$ 0.5 GeV
in the chosen kinematics. The $\cos2\phi_p$ modulations are $\lesssim$ 1\% even
for the largest $p_{pT}$ values shown here.

The extraction of the unpolarized neutron structure functions with spectator tagging
has been studied extensively for fixed-target experiments \cite{Sargsian:2005rm}
and for EIC \cite{Strikman:2017koc,Jentsch:2021qdp}. By extrapolating
to unphysical momenta $|\bm{p}_{pT}|^2 < 0$ one can reach the free nucleon pole
of the deuteron LF wave function, where the IA becomes exact and initial-state
modifications and FSI are absent \cite{Sargsian:2005rm,Strikman:2017koc};
in coordinate space it corresponds to configurations of infinite transverse size,
where the nucleons are free \cite{Cosyn:2020kwu}. The main challenge arises from the
strong kinematic variation of the cross section as a function of $|\bm{p}_{pT}|$
at values $\lesssim$ few 10 MeV, which places high demands on the momentum resolution.
Pole extrapolation in DIS with proton tagging appears feasible with the
EIC far-forward detectors \cite{Jentsch:2021qdp}.
Neutron tagging with the EIC zero-degree calorimeter is being explored \cite{Jentsch:2021qdp}.
It would allow one to extract the free proton structure functions from DIS on the deuteron
with neutron tagging, which can then be compared with direct measurements in DIS on the proton.

\subsection{Vector-polarized deuteron}
Tagged DIS on the vector-polarized deuteron can be used to extract the neutron spin structure functions
for both longitudinal and transverse polarization. It can also be used to determine the effective
neutron polarization as a function of the tagged proton momentum and validate the
$D/S$ wave ratio in the deuteron LF wave function.

The basic observable is the double spin asymmetry of the differential cross section with respect
to the electron helicity $\lambda_e = \pm 1/2$ and the deuteron spin projection $\Lambda = \pm 1$
along a given polarization axis; see \firstpart, Sec.~\ref{P1:subsec:spin_asymmetries}.
Here we compute the double spin asymmetry of the $\phi_p$-averaged differential cross section,
for deuteron polarization along an axis defined relative to the initial electron momentum direction
in the deuteron rest frame (or relative to the colliding beam direction in the collider),
as defined in \firstpart, Eq.~(\ref{P1:double_spin_asymmetry}).

The double spin asymmetry $A^V_{\parallel}$, for deuteron polarization parallel to the initial electron
momentum direction in the deuteron rest frame (or parallel to the deuteron beam direction in the collider),
is expressed in terms of the structure functions in \firstpart, Eq.~(\ref{P1:asymmetry_parallel_expanded}).
In the DIS limit, the kinematic factors are
\begin{subequations}
\label{depol_parallel_dis}
\begin{alignat}{2}
&D_{\parallel [S_L]} &&= \frac{y (1 - y/2)}{1 - y + y^2/2} + \mathcal{O}(1/Q),\label{depol_parallel_dis_SL}
\\
&D_{\parallel [S_T]} &&= \mathcal{O}(1/Q),
\\
&D_{\parallel [T_{LL}]} &&= \frac{1}{3} + \mathcal{O}(1/Q),
\\[.5ex]
&D_{\parallel [U T_{LT}]} &&= \mathcal{O}(1/Q),
\\[1ex]
&D_{\parallel [T_{TT}]} &&= \mathcal{O}(1/Q^2),
\end{alignat}
\end{subequations}
and, up to power corrections, the asymmetry becomes
\begin{subequations}
\label{asymmetry_parallel_expanded}
\begin{align}
A^V_{\parallel} &= D_{\parallel [S_L]} \frac{F_{[L S_L]\deut}}{\Sigma_\parallel F_\deut},
\\[1ex]
\Sigma_\parallel F_\deut &= F_{[UU,T]\deut} + \epsilon F_{[UU,L]\deut}
\nonumber \\
&+ \tfrac{1}{3} (F_{[UT_{LL},T]\deut} + \epsilon F_{[UT_{LL},L]\deut}).
\end{align}
\end{subequations}
The denominator of the structure function ratio includes the $T_{LL}$ tensor-polarized structure functions.
In the IA, the structure function ratio factorizes as
\begin{align}
\frac{F_{[L S_L]\deut}}{\Sigma_\parallel F_\deut}
&= \frac{P_{[S_L S_L]}}{P_{[U,U]} + \frac{1}{\sqrt{6}} P_{[T_{LL}, U]}}
\nonumber \\
& \times \frac{F_{[LS_L]n}}{F_{[UU,T]n} + \epsilon F_{[UU,L]n}} .
\label{asymmetry_parallel_ia}
\end{align}
The first factor is the ratio of the longitudinally polarized and unpolarized neutron momentum
distributions in deuteron; it describes the effective neutron polarization in the deuteron and
depends on the tagged proton momentum.  It can also be expressed as the ratio of the probabilistic
neutron distributions Eq.~(\ref{distribution_pure_helicity}a)
\begin{align}
\frac{P_{[L+,L+]} - P_{[L+,L-]}}{P_{[L+,L+]} + P_{[L+,L-]}},
\end{align}
which can be directly connected with the form of the spin asymmetry in \firstpart,
Eq.~(\ref{P1:double_spin_asymmetry}).  The second factor in Eq.~(\ref{asymmetry_parallel_ia}) is the
ratio of the longitudinally polarized and unpolarized neutron structure functions; it depends on the
subprocess DIS variables and is as would be measured in DIS on the free neutron in the subprocess
kinematics.

%
%
\begin{figure}[t]
\includegraphics[width=\linewidth]{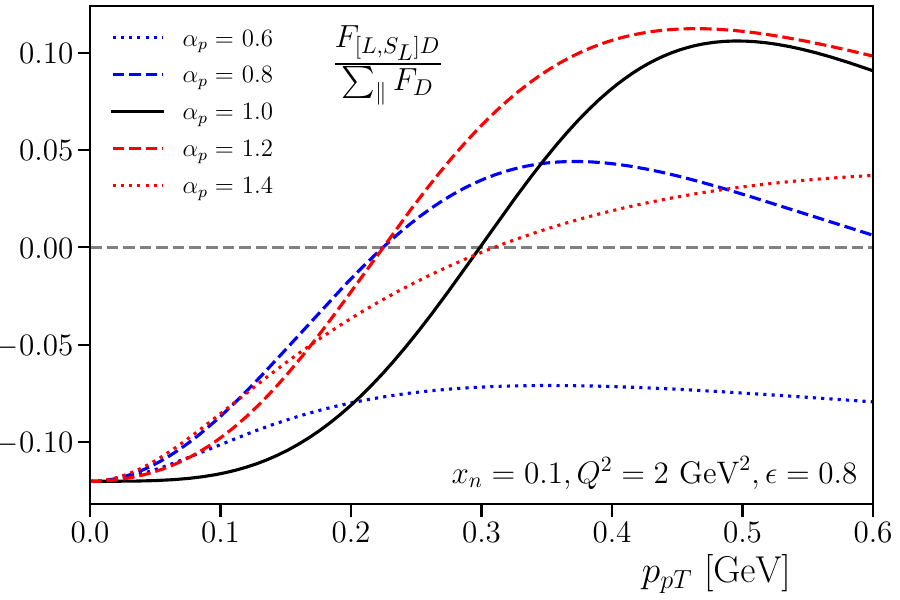}
\caption{The deuteron structure function ratio Eq.~(\ref{asymmetry_parallel_ia}), as a function of $p_{pT}$,
for several fixed values of $\alpha_p$. The ratio determines the double spin asymmetry for deuteron polarization
parallel to the electron direction in tagged DIS, $A^V_{\parallel}$, Eq.~(\ref{asymmetry_parallel_expanded}).
The observable asymmetry includes also the depolarization factor $D_{\parallel [S_L]}$,
Eq.~(\ref{depol_parallel_dis_SL}).}
\label{fig:AV_par}
\end{figure}
Figure~\ref{fig:AV_par} shows the deuteron structure function ratio in the parallel spin asymmetry,
Eq.~(\ref{asymmetry_parallel_ia}), as function of the tagged proton momentum. The ratio is plotted
for a fixed value of $x_n$ (the value of $x$ is changed as $\alpha_p$ is varied such as to keep
$x_n$ fixed), which is how the measurement would be performed in order to extract the neutron spin
structure function at fixed $x_n$.  One observes:
(i)~At $p_{pT} = 0$ the ratio has a negative value $\sim -0.1$ and
is independent of $\alpha_p$. At these proton momenta the $D$-wave is absent in the
deuteron LF wave function and the neutron is 100\% polarized along the
deuteron spin direction (see Fig.~\ref{fig:deut_lf_dist}, upper left panel).
The value of the deuteron structure function ratio is equal to
that of the neutron structure function ratio in Eq.~(\ref{asymmetry_parallel_ia}).
(ii)~As $p_{pT}$ increases to $\sim$ 0.1--0.2 GeV, the deuteron structure function ratio
decreases in magnitude. The depolarization effect is caused by the $D$-wave.
(iii)~Above $p_{pT} \sim$ 0.3 GeV the deuteron structure function ratio changes its sign.
For $\alpha_p \approx 1$ and $p_{pT} \approx$ 0.5 GeV, the ratio attains values $\sim +0.1$,
corresponding to complete reversal of the neutron polarization by the $D$-wave.
This illustrates how the tagged proton momentum controls the effective neutron polarization
in tagged DIS.

The double spin asymmetry $A^V_\parallel$ can be used to extract the neutron spin structure
function $F_{[LS_L]n} \propto g_1$. The pole extrapolation of the asymmetry is
discussed in Ref.~\cite{Cosyn:2019hem}. Note that Fig.~\ref{fig:AV_par} shows only the deuteron
structure function ratio in Eq.~(\ref{asymmetry_parallel_expanded}). The experimental asymmetry
includes also depolarization factor $D_{\parallel [S_L]}$,
Eq.~(\ref{depol_parallel_dis_SL}), whose values can be $\ll 1$ depending on $y$.

%
%
\begin{figure}[t]
\includegraphics[width=\linewidth]{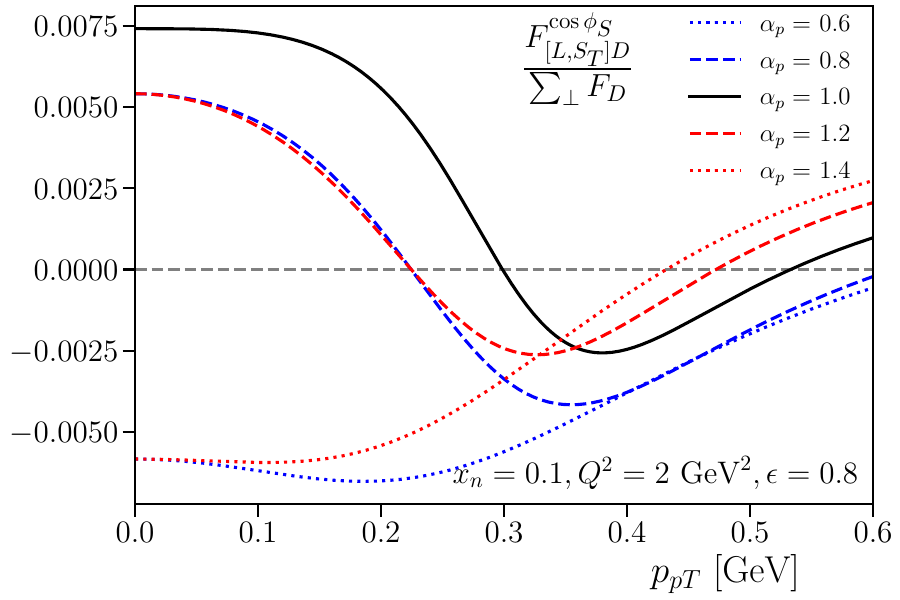}
\caption{The deuteron structure function ratio Eq.~(\ref{asymmetry_perp_ia}b), as a function of $p_{pT}$,
for several fixed values of $\alpha_p$. The ratio enters in the double spin asymmetry for perpendicular
deuteron polarization in tagged DIS, $A^V_\perp$, Eq.~(\ref{asymmetry_perp_expanded}), and controls
the contribution of the transversely polarized neutron structure function $F_{[LS_T]n}$ (or $g_{2n}$).
The observable asymmetry includes also the term with the longitudinally polarized neutron structure
function $F_{[LS_L]n}$ and the depolarization factors.}
\label{fig:AV_perp}
\end{figure}
The double spin asymmetry $A^V_{\perp}$, for deuteron polarization perpendicular to the initial electron
direction in the deuteron rest frame (or perpendicular to the deuteron beam direction in the collider),
is expressed in terms of the structure functions in \firstpart, Eq.~(\ref{P1:asymmetry_perp_expanded}).
In the DIS limit, the kinematic factors are
\begin{subequations}
\label{depol_perp_dis}
\begin{alignat}{2}
& D_{\perp [S_L]} && = \frac{\gamma y (1 - y/2) \sqrt{1-y}}{1 - y + y^2/2} \cos\psi_N
+ \mathcal{O}(1/Q^2),
\\[1ex]
& D_{\perp [S_T]} && = \frac{y \sqrt{1-y}}{1 - y + y^2/2} \cos\psi_N + \mathcal{O}(1/Q),
\\
& D_{\perp [T_{LL}]} && = -\frac{1}{6} + \mathcal{O}(1/Q),
\\[1ex]
& D_{\perp [U T_{LT}]} && = \mathcal{O}(1/Q),
\\[1.5ex]
& D_{\perp [T_{TT}]} && = \mathcal{O}(1),
\end{alignat}
\end{subequations}
where $\psi_N$ is the angle of the polarization axis relative to the lepton plane defined in
\firstpart, Sec.~\ref{P1:subsec:preparation} and Fig.~\ref{P1:fig:target_rest_frame}.
Taking into account the combined scaling behavior of the kinematic factors
the and structure functions, the asymmetry in the DIS limit becomes
\begin{subequations}
\label{asymmetry_perp_expanded}
\begin{align}
A^V_{\perp} &= D_{\perp [S_L]} \frac{F_{[L S_L]\deut}}{\Sigma_\perp F_\deut}
+ D_{\perp [S_T]} \frac{F_{[L S_T]\deut}^{\cos\phi_S}}{\Sigma_\perp F_\deut},
\\[1ex]
\Sigma_\perp F_\deut &\equiv F_{[UU,T]\deut} + \epsilon F_{[UU,L]\deut} 
\nonumber \\[1ex]
&- \tfrac{1}{6} (F_{[UT_{LL},T]\deut} + \epsilon F_{[UT_{LL},L]\deut}).
\end{align}
\end{subequations}
The two terms in Eq.~(\ref{asymmetry_perp_expanded}a) are both $\mathcal{O}(1/Q)$ and thus
of the same order in the DIS limit.
In the first term, $D_{\perp [S_L]} = \mathcal{O}(1/Q)$ and $F_{[L S_L]\deut} = \mathcal{O}(1)$;
in the second term $D_{\perp [S_T]} = \mathcal{O}(1)$ and $F_{[L S_T]\deut} = \mathcal{O}(1/Q)$;
see Eq.~(\ref{neutron_polarized_conventional}).
The denominator Eq.~(\ref{asymmetry_perp_expanded}b) again contains a
tensor-polarized contribution. Of the various tensor-polarized structures
in the original expression in {\firstpart} only the $T_{LL}$ structure survives in the DIS limit,
because $D_{\perp [T_{LL}]} = \mathcal{O}(1)$ and $F_{[UT_{LL},T]\deut}, F_{[UT_{LL},L]\deut}
= \mathcal{O}(1)$; the $T_{LT}$ structure is suppressed because $D_{\perp [U T_{LT}]} = \mathcal{O}(1/Q)$;
and $T_{TT}$ is suppressed because the structure function Eq.~(\ref{FU_TTT_ia}) is
\begin{align}
F_{[U T_{TT}]\deut}^{\cos 2\phi_{T_T}} = \mathcal{O}(1/Q^2).
\label{FU_TTT_order}
\end{align}

In the IA, the structure function ratios in Eq.~(\ref{asymmetry_perp_expanded}) factorize as
\begin{subequations}
\label{asymmetry_perp_ia}
\begin{align}
\frac{F_{[L S_L]\deut}}{\Sigma_\perp F_\deut} &= \frac{P_{[S_L S_L]}}{P_{[U,U]}
- \frac{1}{2 \sqrt{6}} P_{[T_{LL}, U]}}
\nonumber \\
& \times \frac{F_{[LS_L]n}}{F_{[UU,T]n} + \epsilon F_{[UU,L]n}},
\\[2ex]
\frac{F_{[L S_T]\deut}^{\cos\phi_S}}{\Sigma_\perp F_\deut}
&= \frac{\tfrac{1}{2} ( P_{[S_T S_T]}^\parallel + P_{[S_T S_T]}^\perp)}
{P_{[U,U]} - \frac{1}{2 \sqrt{6}} P_{[T_{LL}, U]}}
\nonumber \\
& \times \frac{F_{[LS_T]n}}{F_{[UU,T]n} + \epsilon F_{[UU,L]n}}.
\end{align}
\end{subequations}
The first ratio involves the leading-twist neutron structure function $F_{[LS_L]n}$;
the second involves the higher-twist structure function $F_{[LS_T]n}$.
Note that the contributions of the two structures to the asymmetry Eq.~(\ref{asymmetry_perp_expanded})
can be separated only using the $y$-dependence of the kinematic factors
(varying $y$ by changing the electron-deuteron collision energy), or
by performing an independent measurement of $F_{[LS_L]n}$ using the parallel spin asymmetry
Eq.~(\ref{asymmetry_parallel_expanded}).

Figure~\ref{fig:AV_perp} shows the deuteron structure function ratio Eq.~(\ref{asymmetry_perp_ia}b),
describing the contribution of the neutron structure function $F_{[LS_T]n}$ to the
perpendicular spin asymmetry, as a function of the tagged proton momentum.
One observes: (i)~The values of the transversely polarized structure function ratio
are an order of magnitude smaller than those of the longitudinally polarized ratio in Fig.~\ref{fig:AV_par},
because of the power suppression of the transversely polarized structure functions.
(ii)~At moderate values of the tagged proton momentum, $|1 - \alpha_p| \lesssim 0.2$ and
$p_{pT} \lesssim$ 200 MeV, the transversely polarized structure function ratio depends strongly
on $|1 - \alpha_p|$ but not on $p_{pT}$, in marked contrast to the longitudinally polarized
ratio in Fig.~\ref{fig:AV_par}. This happens because $\alpha_p \neq 1$ selects configurations with
nonzero longitudinal nucleon momentum and therefore transverse orbital angular momentum,
which causes $D$-wave depolarization of the transversely polarized neutron.
Note that the values at $p_{pT}=0$ depend only on $|1 - \alpha_p|$, the magnitude of the longitudinal momentum.

The extraction of the neutron structure functions from the perpendicular asymmetry is more complex than
in the parallel case, because the terms with $F_{[LS_L]n}$ and $F_{[LS_T]n}$ appear at the same order.
A future study should explore whether proton tagging could be used to separate $F_{[LS_L]n}$ and $F_{[LS_T]n}$,
or $g_{1n}$ and $g_{2n}$, in measurements at fixed $y$, using the different dependence of the terms
on the tagged proton momentum, including possibly the dependence on the azimuthal angle $\phi_p$.

\subsection{Tensor-polarized deuteron}
\label{subsec:tensor}
Tensor-polarized deuteron observables are particularly clean probes of the $D$-wave in the deuteron
LF wave function. In the tensor-polarized spin asymmetries the neutron structure functions cancel out in the IA,
so that the asymmetries depend only on deuteron structure. Tagging makes it possible to select
configurations where the $D$-wave is dominant, producing spin asymmetries of order unity.

The basic observable is the tensor-polarized asymmetry of the differential cross section averaged
over the electron helicity, computed with the deuteron spin states with projection
$\Lambda = \pm 1$ and $0$ along a given polarization axis;
see \firstpart, Sec.~\ref{P1:subsec:spin_asymmetries}.
Here we compute the tensor-polarized asymmetry of the $\phi_p$-averaged cross section
for deuteron polarization along an axis defined relative to the
initial electron momentum direction in deuteron rest frame (or relative to the
colliding beam direction in the collider), as defined in \firstpart, Eq.~(\ref{P1:eq:AT_gen}).
The steps are the same as in the study of the vector-polarized asymmetries above.

The tensor-polarized asymmetry for deuteron polarization parallel to the
initial electron direction, $A^T_{\parallel}$, is expressed
in terms of the deuteron structure functions in \firstpart, Eq.~(\ref{P1:asymmetry_tensor_parallel_expanded}).
In the DIS limit the kinematic factors are given by Eq.~(\ref{depol_parallel_dis}),
and, up to power corrections, the asymmetry becomes
\label{asymmetry_tensor_parallel_expanded}
\begin{align}
A^T_{\parallel} &=
\frac{2}{3} \frac{F_{[UT_{LL},T]\deut} + \epsilon F_{[UT_{LL},L]\deut}}
{F_{[UU,T]\deut} + \epsilon F_{[UU,L]\deut}}.
\end{align}
In the IA, substituting the factorized expressions of the deuteron structure functions,
Eqs.~(\ref{eq:IA_FT_TLL}) and (\ref{eq:IA_FU}), the neutron structure functions cancel, and we obtain
\begin{align}
A^T_\parallel &= \sqrt{\frac{2}{3}}\frac{P_{[T_{LL},U]}}{P_{[U,U]}}
\nonumber \\
&= \frac{\left(2f_0+\tfrac{1}{\sqrt{2}} f_2\right)
\tfrac{1}{\sqrt{2}} f_2}{f_0^2+f_2^2}\left(1-3\cos^2\theta_k \right).
\label{asymmetry_tensor_parallel_ia}
\end{align}
The asymmetry is given by a ratio of quadratic forms in the $S$- and $D$-wave radial wave functions in the
c.m.\ frame, multiplied by an angular factor. One can easily verify that the expression takes values in $[-2,1]$.
The ratio of quadratic forms attains a maximum value of $+1$ when
\begin{align}
f_2/f_0 = \sqrt{2},
\label{tensor_ratio_max}
\end{align}
(this is the location where the polarized neutron distributions have a node),
and a minimum value of $-1/2$ when
\begin{align}
f_2/f_0 = -1/\sqrt{2}.
\label{tensor_ratio_min}
\end{align}
The angular factor is proportional to the second-order Legendre polynomial $P_2(\cos\theta_k)$
and has a maximum value of $+1$ (at $\theta_k = \pi/2$) and minimum value of $-2$ (at $\theta_k = 0, \pi$).

%
%
\begin{table}[t]
\begin{tabular}{c|c|cc|cc|}
$A^T_\parallel$ & $f_2/f_0$ & \multicolumn{2}{c|}{c.m.\ variables} & \multicolumn{2}{c|}{LF variables} \\
& & $k$ & $\theta_k$ & $|\alpha_p - 1|$ & $p_{pT}$ \\
\hline
\hline
$-2$ & $\sqrt{2}$ & 0.3 GeV & 0 & 0.3 &0 \\
 $1$ & $\sqrt{2}$ & 0.3 GeV & $\pi/2$ & 0 & 0.3 GeV \\
 $1$ & $-1/\sqrt{2}$ & 1 GeV & 0 & 0.7 & 0 \\
\hline
\end{tabular}
\caption{Kinematic settings for which the tensor-polarized asymmetry
$A^T_\parallel$, Eq.~(\ref{asymmetry_tensor_parallel_ia}),
attains its minimal/maximal values. The spectator momentum
is given in terms of the c.m.\ and LF variables.
The numerical values are based on the AV18 radial wave functions.
(The high-momentum setting in the last line is not presumed
to be a realistic prediction and is listed only for completeness.)}
\label{tab:AT_extrema}
\end{table}
In inclusive measurements, where the numerator and denominator
in the asymmetry are separately averaged over the nucleon configurations in the deuteron,
most of the events come from momenta $k \sim$ few 10 MeV, where $|f_2| \ll |f_0|$, and the asymmetry
has values $\ll 1$. In tagged measurements, where the configuration is fixed by the spectator momentum,
one can select configurations where the asymmetry has values of order unity
or even reaches its extremal values $-2$ and $1$.

With the AV18 radial wave functions, the condition Eq.~(\ref{tensor_ratio_max}) is satisfied 
for $k =$ 0.30 GeV. At this value of the c.m.\ momentum, the minimal and maximal value of
$A^T_\parallel$ are attained with $\theta_k = 0$ and $\theta_k = \pi/2$, respectively.
In terms of the LF momentum variables, this corresponds to $|\alpha_p - 1| \approx 0.3,  p_{pT} = 0$ and
$|\alpha_p - 1| = 0, p_{pT} =$ 0.3 GeV, respectively. These momentum values are in the
domain where the present treatment of deuteron structure is well applicable and the results
are robust (see below). Our findings therefore imply that the
extremal values $A^T_\parallel =$ 1 and $-2$ can be achieved in experiments
with these kinematic settings.

The condition Eq.~(\ref{tensor_ratio_min}) is satisfied only at $k \approx$ 1 GeV
with the AV18 radial wave functions. At such large momenta the treatment of the deuteron
as an $NN$ bound state becomes questionable and the results are model-dependent.
While formally $A^T_\parallel$ attains extremal values at this c.m.\ momentum,
we cannot suggest this as a realistic prediction.
Table~\ref{tab:AT_extrema} summarizes the kinematic settings
in which $A^T_\parallel$ reaches its extremal values in the c.m.\ variables $k$ and $\theta_k$
and the corresponding LF variables $\alpha_p$ and $p_{pT}$.

%
%
\begin{figure}[t]
\includegraphics[width=0.9\linewidth]{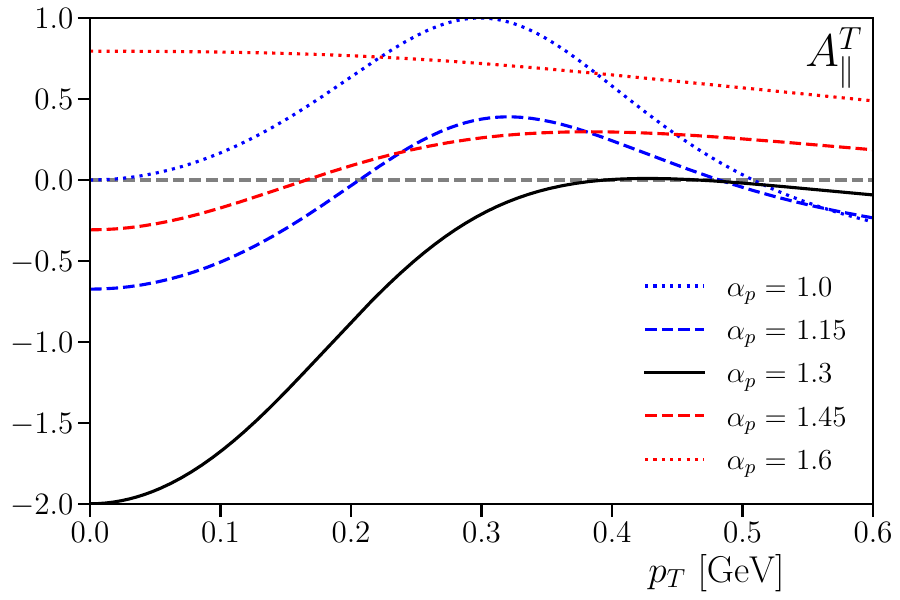}
\includegraphics[width=0.9\linewidth]{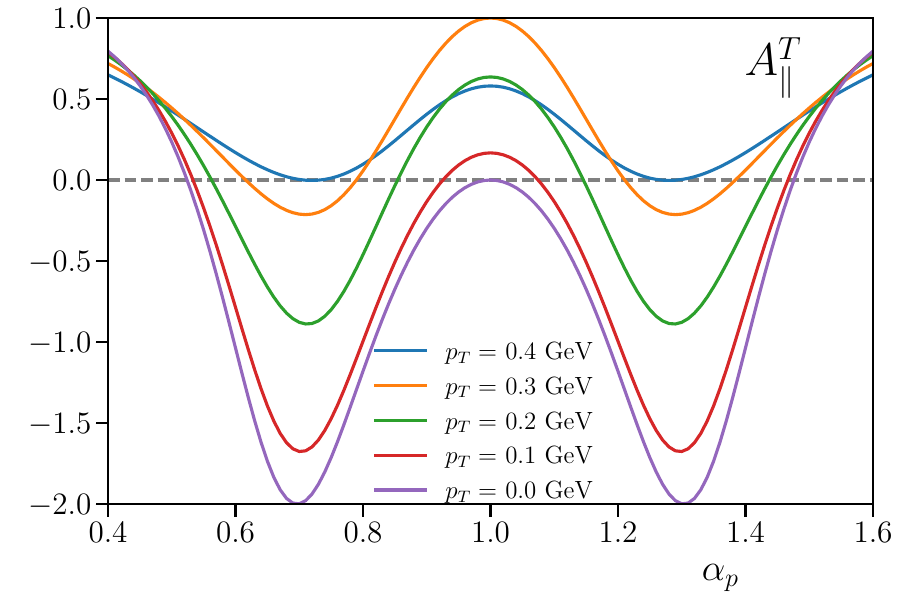}
\caption{The tensor-polarized asymmetry $A^T_\parallel$ in $\phi_p$-averaged
tagged DIS in the IA, Eq.~(\ref{asymmetry_tensor_parallel_ia}),
as a function of the tagged proton momentum variables $\alpha_p$ and $p_{pT}$.
The deuteron is polarized parallel to the electron direction. 
\textit{Upper panel:} Asymmetry as a function of $p_{pT}$, for fixed $\alpha_p$.
\textit{Lower panel:} As a function of $\alpha_p$, for fixed $p_{pT}$.
The extremal values correspond to the settings listed in Table~\ref{tab:AT_extrema}.}
\label{fig:Atensor}
\end{figure}
Figure~\ref{fig:Atensor} shows the dependence of $A^T_\parallel$ on the tagged proton
momentum variables $\alpha_p$ and $p_{pT}$. The extremal
values in the first two lines of Table~\ref{tab:AT_extrema} can be
identified in the graphs. The functional dependence reflects the
interplay of the two factors in Eq.~(\ref{asymmetry_tensor_parallel_ia}).
In the lower panel one sees that, as a function of $\alpha_p$,
$A^T_\parallel$ always peaks at $\alpha_p=1$ (maximum of the angular
factor at $\theta_k=\pi/2$), and always has a minimum at
$|\alpha_p-1|\approx 0.3$ (maximum of the ratio of quadratic forms in the
radial wave function).

The tensor-polarized asymmetry for deuteron polarization along an axis perpendicular to
the electron momentum is expressed in terms of the structure functions
in \firstpart, Eq.~(\ref{P1:asymmetry_tensor_perp_expanded}).
In the DIS limit the kinematic factors are given by Eq.~(\ref{depol_perp_dis}).
Taking into account the combined scaling behavior of the kinematic factors and
the structure functions, neglecting power corrections, the asymmetry becomes
\begin{align}
A^T_{\perp} &= -\frac{1}{3}
\frac{F_{[UT_{LL},T]\deut} + \epsilon F_{[UT_{LL},L]\deut}}
{F_{[UU,T]\deut} + \epsilon F_{[UU,L]\deut}};
\label{asymmetry_tensor_perp}
\end{align}
the other structures in the original expression in {\firstpart} are power-suppressed
either in the kinematic factors or in the structure functions,
see Eq.~(\ref{FU_TTT_order}). The asymmetry is independent of the angle
of the perpendicular polarization axis, $\psi_N$. Therefore, in the DIS limit 
\begin{align}
&A^T_\perp = -\frac{1}{2} A^T_\parallel	= -\sqrt{\frac{1}{6}} \frac{P_{[T_{LL},U]}}{P_{[U,U]}}.
\label{asymmetry_tensor_perp_ia}
\end{align}
The perpendicular asymmetry is $-1/2$ times the parallel asymmetry
in the IA, Eq.~(\ref{asymmetry_tensor_parallel_ia}), and thus contains no new information.
An experimental test of the relation Eq.~(\ref{asymmetry_tensor_perp_ia}) would test
the accuracy of the IA predictions and quantify possible deviations.

So far we have considered the tensor-polarized asymmetries of the $\phi_p$-averaged differential cross section.
We can also compute the tensor-polarized asymmetry of the $\phi_p$-dependent differential cross section,
defined as in \firstpart, Eq.~(\ref{P1:eq:AT_gen}), but with $d\sigma$ now the full
$\phi_p$-dependent differential cross section. These asymmetries involve the $\phi_p$-dependent
structures of the cross section in both numerator and denominator and exhibit a rich structure.
In particular, certain $\phi_p$-dependent structures corresponding to a given tensor polarization
$(T_{LL}, T_{LT}, T_{TT})$ can be $\mathcal{O}(1)$ in cases where the $\phi_p$-averaged
structures are suppressed, implying a change in the counting of structures.

%
%
\begin{figure}[t]
\includegraphics[width=0.9\linewidth]{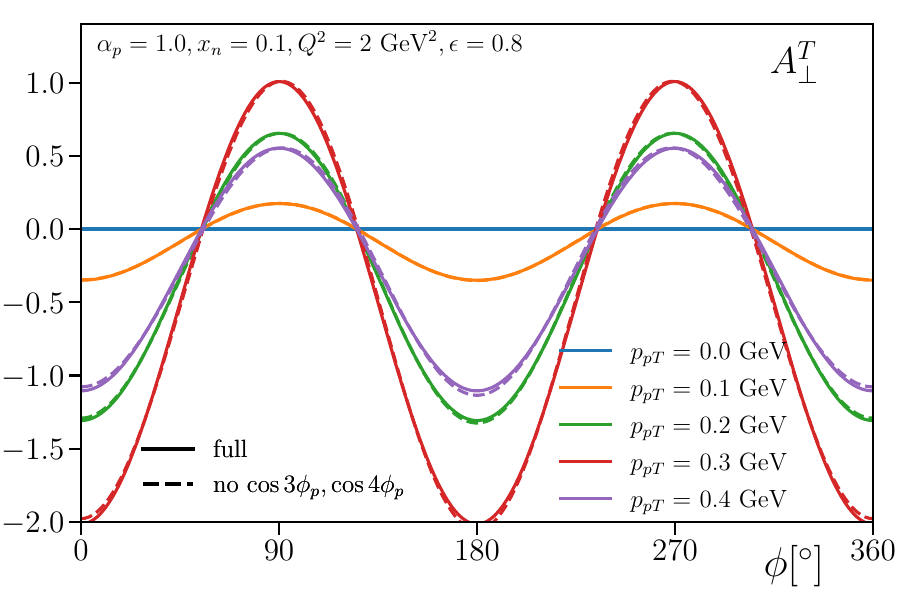}
\caption{The tensor-polarized asymmetry $A^T_\perp$ in $\phi_p$-dependent tagged DIS,
as a function of $\phi_p$, for $\alpha_p = 1$ and several values of $p_{pT}$.
Here the deuteron is polarized perpendicular to the initial electron direction in the rest frame,
at the same angle as the final electron direction (see text).
The denominator is the full $\phi_p$-dependent unpolarized cross section.}
\label{fig:Atensor_phi}
\end{figure}
Here we illustrate the behavior of the $\phi_p$-dependent tensor polarized asymmetry for deuteron
polarization perpendicular to the initial electron momentum in the rest frame,
at the same angle as the final electron direction ($\psi_N = 0$).
For this polarization direction the $\phi_p$-averaged asymmetry is
given by \firstpart, Eq.~(\ref{P1:asymmetry_tensor_perp_expanded}). In the $\phi_p$-dependent case,
the numerator includes the leading-twist $T_{TT}$ structure functions of Eq.~(\ref{eq:IA_FT_TTT}),
\begin{align}
F_{[U T_{TT},T]\deut}^{\cos(2\phi_p-2\phi_{T_T})}, \hspace{1em}
F_{[U T_{TT},L]\deut}^{\cos(2\phi_p-2\phi_{T_T})},
\end{align}
which qualitatively changes the form compared to Eq.~(\ref{asymmetry_tensor_perp}).
The denominator remains unchanged at the leading-twist level, as the $\phi_p$-dependent
structure functions of the $U$ cross section in Eq.~(\ref{eq:IA_FU}) are $\mathcal{O}(|\bm{p}_{pT}|/Q)$.
Figure~\ref{fig:Atensor_phi} shows the $\phi_p$-dependent $A^T_\perp$,
as a function of $\phi_p$, for $\alpha_p=1$ and several values of $p_{pT}$.
One observes: (i)~The asymmetry shows a clear $\cos2\phi_p$ modulation,
whose amplitude grows with increasing $p_{pT}$. (ii)~The higher-order modulations
in the numerator, which are associated with power-suppressed structure functions,
are tiny in the kinematics shown.  The same is the case for the $\cos\phi_p$ and
$\cos2\phi_p$ modulations in the denominator, which could otherwise
distort the clear $2\phi_p$ modulation in the curves. (iii)~The asymmetry
respects the bounds $[-2,1]$ and is able to attain its extremal values
for $p_{pT} \approx$ 0.3 GeV, which again corresponds to c.m.\ momenta
where Eq.~(\ref{tensor_ratio_max}) is satisfied.

The tensor-polarized asymmetries can be used to extract the $D/S$ wave ratio of the deuteron
radial wave functions, which is an important characteristic of short-range deuteron structure.
Tensor-polarized asymmetries of order unity can be achieved in spectator tagging, qualitatively
changing the magnitude compared to inclusive measurements.
The asymmetries for parallel and perpendicular polarization contain the same information
in different form and can be compared for validation.

The tensor-polarized asymmetries are in principle sensitive to FSI, because the deuteron spin
states with different spin projections have different spatial distributions of the nucleons
and therefore different probabilities for rescattering in the final state.
The effect of FSI on the tensor-polarized asymmetries should be investigated in a dedicated study.

\subsection{Nuclear structure model dependence}
\label{subsec:nuclear_model}
%
%
\begin{figure}[t]
\centering
\includegraphics[width=\linewidth]{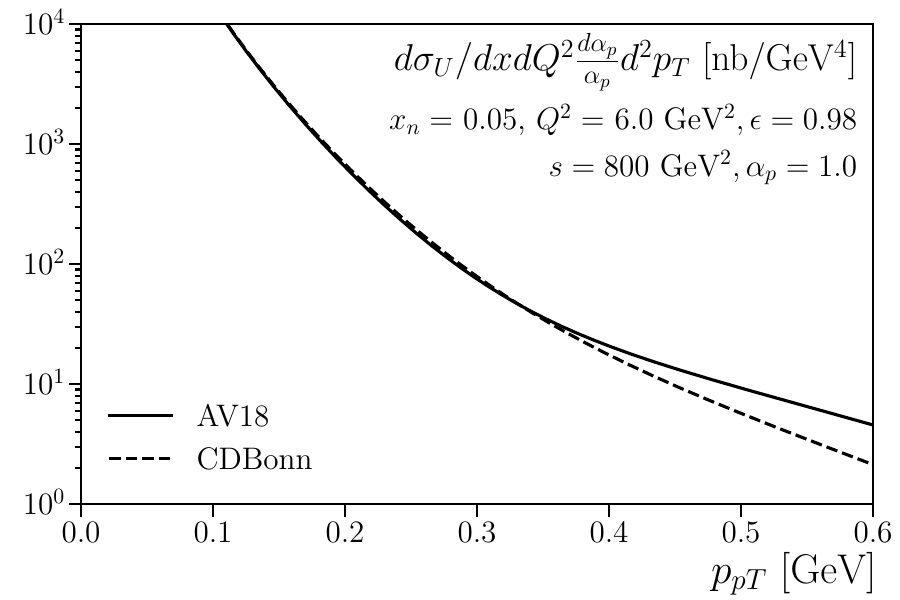}
\caption{Same as Fig.~\ref{fig:sigma_u_log}, but now comparing results
obtained with the AV18 and CDBonn deuteron wave functions.}
\label{fig:sigma_unpol_ndep}
\end{figure}
The numerical estimates of tagged DIS observables in the IA rely on the
deuteron LF wave function as a dynamical input.
In our approach the deuteron LF wave function is constructed from the
nonrelativistic wave function in the approximation described in
Sec.~\ref{subsec:spin_structure}, see Eq.~(\ref{nonrel_approx}).
The deuteron non-relativistic wave function is obtained from empirical $NN$ potentials
constrained by $NN$ scattering and nuclear bound state data;
see Ref.~\cite{Machleidt:2001rw} for a review.
At small nucleon momenta $|\bm{k}| \lesssim$ 100 MeV the $NN$ potentials
and the deuteron wave function are well determined in this scheme;
at larger momenta $|\bm{k}| \gtrsim$ 300 MeV there are substantial differences
between the various parametrizations. We now estimate the numerical uncertainty
of the IA predictions resulting from $NN$ interaction model.

The observables in this section have been computed with the deuteron wave function from the
AV18 potential \cite{AV18}, which is considered a ``hard'' $NN$ interaction (strong high-momentum components).
To estimate the uncertainty, we compare the results with those computed with the wave function
from the CDBonn potential \cite{Machleidt:2000ge}, which is a ``softer'' interaction.

Figure~\ref{fig:sigma_unpol_ndep} compares the unpolarized tagged DIS cross sections
for two different deuteron wave functions at $\alpha_p = 1$ (see Fig.~\ref{fig:sigma_u_log}).
One observes that at $p_{pT} \lesssim$ 300 MeV the results agree at few percent level,
but that at $\sim$ 500 MeV the CDBonn result is only $\sim$ 60\% of the AV18 one.

\begin{figure}[t]
\centering
\includegraphics[width=\linewidth]{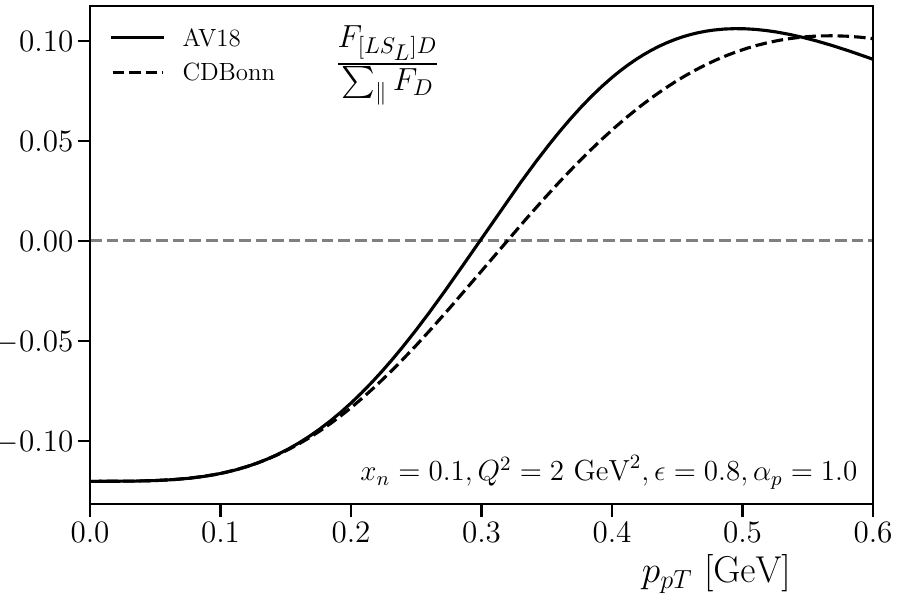}
\caption{Same as Fig.~\ref{fig:AV_par}, but now comparing results
obtained with the AV18 and CDBonn deuteron wave functions.}
\label{fig:ALL_ndep}
\end{figure}

Figure~\ref{fig:ALL_ndep} compares the deuteron structure function ratios in the
vector-polarized asymmetry (see Fig.~\ref{fig:AV_par}). One observes the same pattern: for small $p_{pT}$
the results are virtually the same, but discrepancies appear at larger values $p_{pT} \gtrsim$ 300 MeV.

\begin{figure}[t]
\centering
\includegraphics[width=\linewidth]{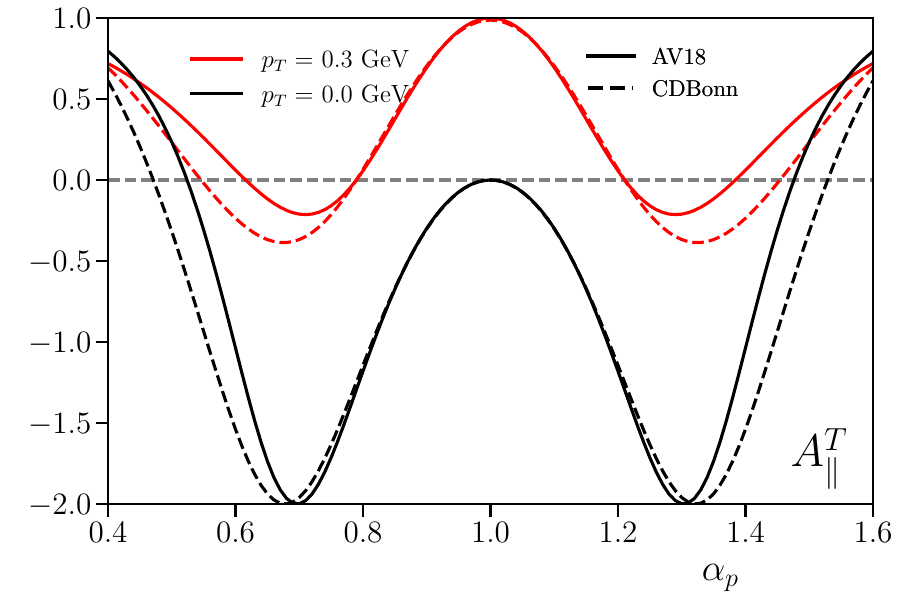}
\caption{Same as the lower panel in Fig.~\ref{fig:Atensor}, but now comparing results
obtained with the AV18 and CDBonn deuteron wave functions.}
\label{fig:Atensor_ndep}
\end{figure}

Figure~\ref{fig:Atensor_ndep} compares the predictions for the tensor polarized asymmetry
$A^T_\parallel$ (see Fig.~\ref{fig:Atensor}, lower panel). Here an interesting situation arises.
While the maximum value occurs at $\alpha_p=1$ and shows no dependence on the deuteron wave function,
the minimum value occurs at values $\alpha_p \neq 1$ and shows a substantial model dependence,
especially for $p_{pT} \gtrsim$ 300 MeV. The tensor polarized asymmetry can therefore
be used to discriminate between deuteron wave function models, especially since it is independent of
neutron structure in the IA (see the comments regarding FSI in Sec.~\ref{subsec:tensor}).
%
%
%
%

It is worth noting that the nuclear structure uncertainty estimated here is not the only
uncertainty affecting the tagged DIS cross section, and may not even be the dominant one
for tagged spectator momenta $\sim$ few 100 MeV. Other uncertainties arise from the implementation
of the LF description in the IA (the restriction to $NN$ states, the matching with the
nonrelativistic wave function) and from FSI neglected in the IA. In the differential cross section, FSI
effects are expected to have a strong dependence on the angle of the spectator momentum
($\theta_k$ in the c.m.\ frame) and reach a magnitude comparable to the IA for momenta
$\gtrsim$ 300 MeV \cite{Strikman:2017koc}.

\section{Conclusions and extensions}
\label{sec:conclusions}
In this work we have computed the cross section and spin observables of DIS on the
polarized deuteron with spectator nucleon tagging, combining methods of LF quantization,
the IA, and dynamical input from nonrelativistic nuclear structure.
The main results can be summarized as follows:

(i)~The LFQM formulation enables an efficient description of deuteron LF structure including
the spin degrees of freedom. The collinear boost invariance transports the deuteron
polarization parameters from the deuteron rest frame (where they are given experimentally)
to the $pn$ c.m.\ frame (where the spin wave function is constructed) and to any collinear
frame (where the scattering process happens). The resulting formulas have quasi 3-dimensional
appearance yet embody the full LF spin structure.

(ii)~The tagged DIS structure functions can be represented as products of the active neutron structure
functions and the neutron LF momentum distributions in the deuteron, which depend on the
tagged proton momentum and the neutron and deuteron spin variables (or vice versa for
active proton and tagged neutron).
This form expresses the factorization of hadronic and nuclear structure and provides
a compact and transparent representation of the structure functions (effective neutron polarization,
subprocesses kinematics).

(iii)~The spin-dependent neutron (or proton) LF momentum distributions can be used to quantify and
visualize relativistic deuteron structure and are interesting objects of study in themselves.
They have the same structural properties as the TMD parton distributions in QCD \cite{Boussarie:2023izj}
(spin-orbit couplings, $T$-even and odd structures) but are governed by nuclear dynamics and can be computed
from empirical interactions. Within the IA, the nucleon distributions are process-independent and
can be extracted and validated using various nucleon-level scattering processes,
e.g.\ quasi-elastic scattering.

(iv)~Spectator tagging is a special case of target fragmentation in DIS.
The treatment of nuclear breakup by soft nuclear interactions
is consistent with the factorization theorem and the twist expansion of the structure functions.
The tagged structure functions computed here can be connected with the fracture functions in the
conventions of Refs.~\cite{Trentadue:1993ka,Collins:1997sr,Anselmino:2011ss}.
The dynamical modeling of target fragmentation involves matrix elements of twist-2 QCD
operators and is simpler than current factorization involving TMD operators.

(v)~Spectator tagging enables the selection of specific nuclear configurations during the DIS process.
The spectator momentum controls the $D/S$ wave ratio in the $pn$ configuration. This produces striking
effects in spin observables. The vector-polarized spin asymmetry changes sign between low and
high spectator momenta, as the effective neutron polarization changes from along the
deuteron spin ($S$-wave) to opposite to it ($D$-wave).

(vi)~The tensor-polarized spin asymmetry in tagged DIS is generally of order unity and attains its
minimal/maximal values of $-2$ and $1$ in configurations where the $D/S$ ratio is $f_2/f_0 = \sqrt{2}$,
which occurs at spectator momenta $\sim$ 300 MeV and certain angles. This is very different from
untagged (inclusive) DIS, where the tensor-polarized asymmetry is $\ll 1$ because
the scattering is dominated by $S$-wave configurations. 

The main limitations of the present study are:

(i) The present treatment of the scattering process is limited to the IA and should be
extended to include FSI. FSI has a large effect on the unpolarized tagged DIS differential
cross section at spectator momenta $\gtrsim$ 300 MeV (the rescattering process changes the
tagged nucleon momentum and thus the functional dependence of the cross section), and shows
a strong dependence on the spectator momentum angle \cite{Strikman:2017koc}.
The effect on the polarized tagged DIS
differential cross section and the spin asymmetries should be studied.
Interesting questions are: (a) Can a possible spin dependence of the rescattering process
enable interference of scattering amplitudes with $S$- and $D$-wave in the initial state?
(b) What is the effect of FSI on tensor-polarized asymmetries, where the polarization state
controls the spatial configuration of the nucleons? (c) What is the size of $T$-odd
structure functions, which are zero in the IA and provide a sensitive test of FSI?

(ii)~The expansion of the deuteron state in the LFQM calculation is restricted to the $pn$ component.
While this approximation is adequate for c.m.\ momenta $\lesssim$ 300 MeV
\cite{Frankfurt:1981mk}, for higher momenta nonnucleonic components ($NN\pi, \Delta\Delta$)
should be included in the expansion \cite{Frankfurt:1992ny}. These components then participate
in the DIS process, bringing in new hadronic structure functions and interference effects
at the amplitude level. A proper treatment of these components in tagged DIS remains a major challenge.
A study of quasi-elastic scattering, where the final state is simpler and limits the interference
of amplitudes, would be a good first step.

The results of the present study can be applied and extended in several other ways:

The structure function models can be used to simulate polarized tagged DIS with the EIC,
using the far-forward detectors to capture spectator protons or neutrons.
Measurements of the unpolarized $\phi_p$-integrated tagged cross sections have been
simulated in Ref.~\cite{Jentsch:2021qdp}. The predictions for the $\phi_p$-dependence
obtained in this work can be used to control the systematic uncertainties
in measurements with incomplete azimuthal coverage. They can also be used to simulate the
extraction of $\phi_p$-harmonics, valuable observables which provide critical tests of the
IA and possible FSI effects. The predictions for the spin dependence can be used to simulate
measurements of polarized tagged DIS with possible polarized deuteron
beams at EIC \cite{Huang:2020uui,Huang:2021gbi,Huang:2025gqx}. Important questions are:
(a) What magnitude and uncertainty of the deuteron polarization in the various spin states
are required to measure vector- and tensor-polarized asymmetries?
(b) What are the luminosity requirements for measuring spin asymmetries at
spectator momenta $\gtrsim$ 100 MeV, which correspond to rare configurations in the deuteron?
(c) What are the trade-offs in measuring the tensor-polarized asymmetry at spectator momenta
$\sim$ few 100 MeV, where it takes values of order unity (low rates vs.\ large spin asymmetry)?

The treatment can be extended to exclusive processes in electron scattering on the deuteron
with spectator tagging, $e + \deut \rightarrow e' + M + n + p$, where $M = $ meson, photon.
For such exclusive processes the dynamical calculations are performed at the amplitude level
(in inclusive scattering they are performed at the cross section level). The nucleon-level process
is now described by the exclusive amplitude $e + N \rightarrow e' + M + N'$.
In the regime of QCD factorization, this amplitude can be expressed in terms of quark/gluon
scattering amplitudes and the nucleon GPDs. A new aspect in the nuclear breakup calculations
(compared to inclusive scattering) is that the active nucleon appears intact in the final state,
which creates the possibility of the interference of amplitudes with different active nucleons.
Exclusive processes on the deuteron with spectator tagging would provide valuable new
information on the neutron GPDs.

The technique of spectator nucleon tagging and nuclear breakup detection can be extended to
scattering on $A > 2$ nuclei, e.g. $^3$He. While this greatly expands the range of possibilities,
the theoretical challenges are very considerable. The implementation of the LF structure of
the 3-body system is much more complex than for the 2-body system. The realization of
rotational invariance using c.m.\ frame (see Sec.~\ref{subsec:spin_structure})
is more complicated, because each 2-body subsystem has its own c.m.\ frame, which is
boosted relative to the nuclear rest frame in a way that depends on the
3rd particle \cite{Sokolov:1977im,Bakker:1979eg,Lev:1993pfz,Alessandro:2021cbg}.
Because the ``spectator'' system now
consists of multiple nucleons subject to nuclear interactions, the amplitudes of the
nuclear breakup into defined channels are determined by wave function overlap integrals
that show large model uncertainties. Also, FSI are much more complex than in $A = 2$ nuclei,
because of the many possible configurations of the remnant system and the resulting
interaction trajectories. Substantial further efforts are needed to extend the
present methods to $A > 2$ nuclei.
\appendix
\section{Sum rules for polarized neutron distributions}
\label{app:spin_sum_rules}
\subsection{Longitudinally polarized neutron}
In this appendix we compute the integrals of the longitudinally and transversely polarized
neutron distributions over the tagged proton momentum. The integrals describe the contribution of the
nucleon spin to the total spin of the deuteron, in analogy to the spin sum rules for the quarks and
gluons in the nucleon in QCD. The exercise demonstrates the realization of rotational invariance
in the LF description and the role of relativistic effects in deuteron structure.

The integral of the longitudinally polarized neutron distribution Eq.~(\ref{distribution_helicity})
over the tagged proton momentum can be expressed as \cite{Cosyn:2020kwu}
\begin{align}
& \int_0^2\frac{d\alpha_p}{\alpha_p} \, d^2 p_{pT} \; 
P_{[S_L]}(\alpha_p,\bm p_{pT}| \bm{S}_\deut)
= I_{[S_L]} S_\deut^z .
\label{integral_longitudinal_def}
\end{align}
That it is proportional to $S_\deut^z$, and independent of $\bm{S}_{\deut T}$, is required by
rotational invariance and confirmed by the explicit expressions below.
The integral is evaluated as an integral over the c.m.\ momentum using Eq.~(\ref{integration_k}),
\begin{align}
& \int_0^2\frac{d\alpha_p}{\alpha_p} \, d^2 p_{pT} \; 
P_{[S_L]}(\alpha_p,\bm p_{pT}| \bm{S}_\deut)
\nonumber \\
&= \int \frac{d^3 k}{E} \;
\left( f_0-\frac{f_2}{\sqrt{2}}\right)
\nonumber \\
& \hspace{1em} \times 
\left[A \left( f_0-\frac{f_2}{\sqrt{2}}\right)
+ B \left(f_0 + \sqrt{2}f_2\right)\right],
\nonumber \\
&= 4\pi \int \frac{dk \, k^2}{E} \;
\left( f_0-\frac{f_2}{\sqrt{2}}\right)
\nonumber \\
& \hspace{1em} \times 
\left[\bar A \left( f_0-\frac{f_2}{\sqrt{2}}\right)
+ \bar B \left(f_0 + \sqrt{2}f_2\right)\right],
\label{integral_helicity}
\end{align}
where
\begin{align}
&\bar A \equiv \int\frac{d\Omega_k}{4\pi} A(\bm{k}),
\hspace{2em}
\bar B \equiv \int\frac{d\Omega_k}{4\pi} B(\bm{k})
\label{angular_average}
\end{align}
are the angular averages of the coefficients given
in Eqs.~(\ref{distribution_helicity}b) and (\ref{distribution_helicity}c).
The angular averages can be computed exactly. Here we evaluate them by expanding
the coefficients in $k/m$, which produces simple analytic expressions,
explains the connection with nonrelativistic nuclear structure, and provides
an excellent numerical approximation to the exact results. 
Expanding in $k/m$, the coefficients become
\begin{subequations}
\begin{align}
A(\bm{k}) &= \cos^2 \theta_k \, S_\deut^z
- \sin^2 \theta_k \cos\theta_k \frac{|\bm k|}{m} S_\deut^z
\nonumber\\ 
& \quad + \left(\cos\theta_k - \sin^2 \theta_k \frac{|\bm k|}{m}\right)
\frac{(\bm S_{\deut T}\bm k_T)}{|\bm k|}
\nonumber\\
&\quad  + \mathcal{O} \left(\frac{k^2}{m^2}\right),
\\
B(\bm{k}) &= \sin^2 \theta_k \, S_\deut^z + \sin^2\theta_k \cos\theta_k \frac{|\bm k|}{m} S_\deut^z
\nonumber\\ 
& \quad - \cos\theta_k\left(1+\cos\theta_k\frac{|\bm k|}{m}\right)
\frac{(\bm S_{\deut T}\bm k_T)}{|\bm k|}
\nonumber\\
&\quad + \mathcal{O} \left(\frac{k^2}{m^2}\right).
\end{align}
\end{subequations}
Only the first terms $\propto S_\deut^z$ survive in the angular averaging; the terms
$\propto \bm{S}_{\deut T}$ average to zero because of rotational invariance. We obtain
\begin{align}
&\bar A = \frac{1}{3} S_\deut^z, \hspace{2em} \bar B = \frac{2}{3} S_\deut^z.
\end{align}
We can thus identify the scalar integral in Eq.~(\ref{integral_longitudinal_def}) as
\begin{align}
I_{[S_L]} &= 4\pi \int \frac{dk \, k^2}{E} \;
\left( f_0-\frac{f_2}{\sqrt{2}}\right)
\left( f_0 + \frac{f_2}{\sqrt{2}}\right)
\nonumber \\
&= 4\pi \int \frac{dk \, k^2}{E} \;
\left( f_0^2 - \frac{f_2^2}{2} \right)
\nonumber \\
&= 4\pi \int \frac{dk \, k^2}{E} \;
\left( f_0^2 + f_2^2 - \frac{3}{2} f_2^2 \right)
\nonumber \\
&= 1 - \frac{3}{2} \omega_2 .
\label{sumrule_helicity_steps}
\end{align}
In the last step we have used the normalization condition of the radial wave functions in the c.m.\ frame,
Eq.~(\ref{normalization_radial}), and substituted the expression of the $D$-state probability
of the deuteron wave function in the c.m. frame,
\begin{align}
\omega_2 \; \equiv \;
4\pi \int \frac{dk \, k^2}{E(k)} \; [f_2(k)]^2 .
\label{dwave_probability}
\end{align}
The longitudinally polarized neutron distribution satisfies a spin sum rule,
with a depolarization correction proportional to the $D$-state probability.

The integrals of the normalized longitudinally polarized distributions
Eqs.~(\ref{distribution_helicity_favored}) and (\ref{distribution_helicity_unfavored})
are computed in the same way. They are given by expressions similar to
Eq.~(\ref{integral_helicity}), in which the coefficients $A, B$ are replaced by
$A_{[=]}, B_{[=]}$ and $A_{[\neq]}, B_{[\neq]}$, respectively. Expanding
the coefficients in $k/m$,
\begin{subequations}
\begin{align}
A_{[=]}(\bm{k}) &= \cos^2 \theta_k - \sin^2 \theta_k \cos\theta_k \frac{k}{m}
+ \mathcal{O} \left(\frac{k^2}{m^2}\right),
\\
B_{[=]}(\bm{k}) &= \sin^2 \theta_k + \sin^2\theta_k \cos\theta_k \frac{k}{m}
+ \mathcal{O} \left(\frac{k^2}{m^2}\right),
\\
A_{[\neq]}(\bm{k}) &= \sin\theta_k \cos \theta_k - \sin^3 \theta_k \frac{k}{m}
+ \mathcal{O} \left(\frac{k^2}{m^2}\right),
\\
B_{[\neq]}(\bm{k}) &= \sin\theta_k \cos \theta_k + \cos^2 \theta_k \sin \theta_k \frac{k}{m}
\nonumber \\
& \qquad + \mathcal{O} \left(\frac{k^2}{m^2}\right),
\end{align}
\end{subequations}
we obtain the angular averages Eq.~(\ref{angular_average}) as
\begin{subequations}
\label{angular_averages_favored_unfavored}
\begin{align}
&\bar A_{[=]} = \frac{1}{3}, &\bar B_{[=]} = \frac{2}{3},
\\
&\bar A_{[\neq]} = -\frac{3\pi}{16}\frac{k}{m}, &\bar B_{[\neq]} = \frac{\pi}{16}\frac{k}{m}.
\end{align}
\end{subequations}
Going through the same steps as in Eq.~(\ref{sumrule_helicity_steps}) we obtain
for the favored distribution
\begin{align}
\int_0^2\frac{d\alpha_p}{\alpha_p} \, d^2 p_{pT} \; 
P_{[S_L, S_L]}(\alpha_p,\bm p_{pT}) &= 1 - \frac{3}{2} \omega_2;
\end{align}
for the unfavored distribution
\begin{align}
\int_0^2\frac{d\alpha_p}{\alpha_p} \, d^2 p_{pT} \; 
P_{[S_T, S_L]}(\alpha_p,\bm p_{pT}) &= \varepsilon_{[S_L]},
\label{integral_longitudinal_unfavored}
\end{align}
where
\begin{align}
\varepsilon_{[S_L]} = \frac{\pi^2}{4} \int \frac{d k k^2 }{E}\, \frac{k}{m}\,
\left( f_0-\frac{f_2}{\sqrt{2}}\right)
\left( -4 f_0 + \frac{f_2}{\sqrt{2}}\right).
\label{epsilon_sl_app}
\end{align}
For the favored distribution, the depolarization correction to the integral is not suppressed
in $k/m$ (present in the nonrelativistic limit) and quadratic in the $D$-wave.
For the unfavored distribution the correction is of first order in $k/m$
(relativistic correction) and quadratic in the $S$-wave.
These findings are naturally explained by the angular momentum addition implied by
rotational invariance in the c.m.\ frame.

Table~\ref{tab:sumrules} gives the numerical values of the depolarization corrections, computed with
the AV18 deuteron radial wave functions \cite{AV18}. The $D$-state probability $\omega_2$ is a basic
prediction of the $NN$ potential parametrization. The integral $I_{[S_L]}$ is given as computed
exactly using Eq.~(\ref{integral_longitudinal_def}), and approximately by expanding in $k/m$ using
Eq.~(\ref{sumrule_helicity_steps}); the difference between the two values is seen to be of order
$\sim$ 10$^{-3}$. The relativistic correction $\epsilon_{[S_L]}$ is of the order of a few percent.
%
%
\newcommand\mc[1]{\multicolumn{1}{c}{#1}}
\begin{table}[t]
\centering
\begin{tabular}{ld{2.4}d{2.4}l}
integral & \mc{exact} & \mc{expanded} & ref \\
\hline
\hline
$\omega_2$ & 0.0576 & & (\ref{dwave_probability}) \\
$I_{[S_L]}$ & 0.911 & 0.913 & (\ref{integral_longitudinal_def}) \\
$I_{[S_T]}$ & 0.912 & 0.913 & (\ref{integral_transverse_def}) \\
$\delta^\parallel$ & 0.125 & 0.122 & (\ref{integral_favored_parallel}) \\
$\delta^\perp$ & 0.0512 & & (\ref{integral_favored_perp}) \\
$\varepsilon_{[S_L]}$ & -0.0568 & -0.0579 & (\ref{integral_longitudinal_unfavored}) \\
$\varepsilon_{[S_T]}$ & 0.0550 & 0.0552 & (\ref{integral_transverse_unfavored}) \\
\hline
\end{tabular}
\caption{Values of the integrals appearing in the sum rules
of the polarized neutron distributions, evaluated with the AV18
deuteron wave function. \textit{Second column:} Results obtained with exact
numerical integration. \textit{Third column:} Results obtained
with integrand expanded in $k/m$, neglecting $\mathcal{O}(k^2/m^2)$ (see text).
\textit{Fourth column:} References in text.}
\label{tab:sumrules}
\end{table}
\subsection{Transversely polarized neutron}
The integral of the distribution of transversely polarized neutrons Eq.~(\ref{distribution_transversity})
is computed in the same way as for the longitudinally polarized neutrons. The integral is
expressed as 
\begin{align}
&\int_0^2\frac{d\alpha_p}{\alpha_p} \, d^2 p_{pT} \;
\bm{P}_{[S_T]} (\alpha_p, \bm{p}_{pT}| \bm{S}_\deut)
= I_{[S_T]} \bm{S}_{\deut T}.
\label{integral_transverse_def}
\end{align}
The integral is again evaluated in the c.m.\ momentum variables,
\begin{align}
& \int_0^2\frac{d\alpha_p}{\alpha_p} \, d^2 p_{pT} \; 
\bm{P}_{[S_T]}(\alpha_p, \bm{p}_{pT}| \bm{S}_\deut)
\nonumber \\
&= \int \frac{d^3 k}{E} \;
\left( f_0 - \frac{f_2}{\sqrt{2}} \right)
\nonumber \\
& \hspace{1em} \times \left[  \bm{A}_T  \left(f_0 +  \sqrt{2} f_2 \right)
  + \bm{B}_T  \left( f_0 - \frac{f_2}{\sqrt{2}} \right)  \right]
\nonumber \\
&= 4\pi \int \frac{dk \, k^2}{E} \;
\left( f_0-\frac{f_2}{\sqrt{2}}\right)
\nonumber \\
& \hspace{1em} \times
\left[  \bar{\bm{A}}_T  \left(f_0 +  \sqrt{2} f_2 \right)
  + \bar{\bm{B}}_T  \left( f_0 - \frac{f_2}{\sqrt{2}} \right)  \right],
\label{integral_transversity}
\end{align}
where now
\begin{align}
&\bar{\bm{A}}_T \equiv \int\frac{d\Omega_k}{4\pi} \bm{A}_T(\bm{k}),
\hspace{2em}
\bar{\bm{B}}_T \equiv \int\frac{d\Omega_k}{4\pi} \bm{B}_T(\bm{k}),
\label{angular_average_transversity}
\end{align}
are the angular averages of the transverse vector-valued coefficients Eqs.~(\ref{distribution_transversity}b)
and (\ref{distribution_transversity}c). Expanding in $k/m$, the coefficients become
\begin{subequations}
\label{coefficients_expanded_transversity} 
\begin{align}
\bm{A}_T &= \bm{S}_{\deut T}
- \left( 1+ \cos\theta_k \frac{|\bm k|}{m}\right) \frac{\bm{k}_T (\bm{S}_{\deut T} \bm{k}_T)}{|\bm{k}|^2}
\nonumber\\
& \quad - \left(\cos\theta_k - \sin^2\theta_k \frac{|\bm k|}{m} \right)\frac{\bm k_T}{|\bm{k}|} \, S_\deut^z
+ \mathcal{O}\left(\frac{k^2}{m^2}\right),
\\
\bm{B}_T &=
\left(1 + \cos\theta_k \frac{|\bm k|}{m} \right)\frac{\bm{k}_T (\bm{S}_{\deut T} \bm{k}_T)}{|\bm{k}|^2}
\nonumber\\ &\quad 
+ \cos\theta_k\left(1+\cos\theta_k \frac{|\bm k|}{m}  \right)\frac{\bm k_T}{|\bm{k}|} \, S_\deut^z
+ \mathcal{O} \left(\frac{k^2}{m^2}\right).
\end{align}
\end{subequations}
Now the terms $\propto S_\deut^z$ average to zero and the terms $\propto \bm{S}_{\deut T}$ survive.
We obtain
\begin{align}
&\bar{\bm{A}}_T = \frac{2}{3} \bm{S}_{\deut T} , & \bar{\bm{B}}_T = \frac{1}{3} \bm{S}_{\deut T}.
\end{align}
We thus identify the scalar integral in Eq.~(\ref{integral_transverse_def}) as
\begin{align}
I_{[S_T]} &= 4\pi \int \frac{dk \, k^2}{E} \;
\left( f_0-\frac{f_2}{\sqrt{2}}\right)
\left( f_0 + \frac{f_2}{\sqrt{2}}\right)
\nonumber \\
&= I_{[S_L]} = 1 - \frac{3}{2} \omega_2 .
\label{sumrule_transversity_scalar}
\end{align}
The scalar integral in the sum rule for the transversely polarized neutron distribution
is the same as that in the longitudinally polarized distribution Eq.~(\ref{integral_longitudinal_def}).
This result would be expected for a nonrelativistic system.
In our LF description it is obtained including terms of the order $\bm{k}/m$, neglecting only
terms of order $k^2/m^2$, and is thus a non-trivial consequence of the realization of
rotational invariance by the c.m.\ frame representation of the deuteron LF wave function.

The integrals of the normalized transversely polarized distributions
Eqs.~(\ref{transversity_favored}) and (\ref{transversity_unfavored})
can be computed in a similar way. Here we have to take into account that the
vectors used for the projections of the transverse spins,
Eq.~(\ref{unit_vectors_for_polarization}), depend on the transverse momentum $\bm{k}_T$ and thus
are affected by the angular averaging in the integrals.
For the $\parallel$ favored distribution Eq.~(\ref{transversity_favored}a),
\begin{align}
& \int_0^2\frac{d\alpha_p}{\alpha_p} \, d^2 p_{pT} \; 
P^\parallel_{[S_T, S_T]}(\alpha_p,\bm p_{pT}| \bm{S}_\deut)
\nonumber \\
&= \int \frac{d^3 k}{E} \;
\left( f_0-\frac{f_2}{\sqrt{2}}\right)
\nonumber \\
& \hspace{1em} \times 
\left[ A_{[=]} \left(f_0 + \sqrt{2}f_2\right)
+ B_{[=]} \left( f_0-\frac{f_2}{\sqrt{2}}\right) \right]
\nonumber \\
&= 4\pi \int \frac{dk \, k^2}{E} \;
\left( f_0-\frac{f_2}{\sqrt{2}}\right)
\nonumber \\
& \hspace{1em} \times 
\left[ \bar A_{[=]} \left(f_0 + \sqrt{2}f_2\right)
+ \bar B_{[=]} \left( f_0-\frac{f_2}{\sqrt{2}}\right) \right],
\end{align}
and using the angular averages Eq.~(\ref{angular_averages_favored_unfavored}) we obtain
\begin{align}
& \int_0^2\frac{d\alpha_p}{\alpha_p} \, d^2 p_{pT} \; 
P^\parallel_{[S_T, S_T]}(\alpha_p,\bm p_{pT}| \bm{S}_\deut)
\nonumber \\
&= 4\pi \int \frac{dk \, k^2}{E} \;
\left( f_0-\frac{f_2}{\sqrt{2}}\right) f_0
\nonumber \\
&= 4\pi \int \frac{dk \, k^2}{E} \;
\left[ f_0^2 + f_2^2 - \left( f_0 + \sqrt{2} f_2 \right) \frac{f_2}{\sqrt{2}} \right]
\nonumber \\
&= 1 - \delta^\parallel , 
\label{integral_favored_parallel}
\end{align}
where
\begin{align}
\delta^\parallel \equiv
4\pi \int \frac{dk \, k^2}{E} \;
\left( f_0 + \sqrt{2} f_2 \right) \frac{f_2}{\sqrt{2}} .
\label{delta_parallel_app}
\end{align}
For the $\perp$ favored distribution Eq.~(\ref{transversity_favored}b) we obtain
\begin{align}
& \int_0^2\frac{d\alpha_p}{\alpha_p} \, d^2 p_{pT} \; 
P^\perp_{[S_T, S_T]}(\alpha_p,\bm p_{pT}| \bm{S}_\deut)
\nonumber \\
&= \int \frac{d^3 k}{E} \;
\left( f_0-\frac{f_2}{\sqrt{2}}\right) \left(f_0 + \sqrt{2}f_2\right)
\nonumber \\
&= 1 - \delta^\perp ,
\label{integral_favored_perp}
\end{align}
where
\begin{align}
\delta^\perp \equiv
4\pi \int \frac{dk \, k^2}{E} \;
\left( -f_0 + 2 \sqrt{2} f_2 \right) \frac{f_2}{\sqrt{2}} .
\label{delta_perp_app}
\end{align}
In both the $\parallel$ and $\perp$ distribution the depolarization effects are caused by the $D$-wave;
however, the corrections are not simply quadratic in the $D$-wave radial wave function as in the sum rule
for the unprojected distribution, Eqs.~(\ref{dwave_probability}) and (\ref{sumrule_transversity_scalar}),
but involve the product of the $D$- and $S$-wave functions.
These differences are a consequence of the angular momentum addition implied by the projection
of the distribution on the angular-dependent transverse vectors Eq.~(\ref{unit_vectors_for_polarization}).
The numerical values of $\delta^\parallel$ and $\delta^\perp$ are given in Table~\ref{tab:sumrules}.
Comparing the integrals in Eqs.~(\ref{dwave_probability}), (\ref{delta_parallel_app}),
and (\ref{delta_perp_app}), we observe that
\begin{align}
\frac{1}{2} (\delta^\parallel + \delta^\perp) = \frac{3}{2} \omega_2,
\label{delta_parallel_perp}
\end{align}
so that
\begin{align}
& \int_0^2\frac{d\alpha_p}{\alpha_p} \, d^2 p_{pT} \; 
\frac{1}{2} \left[ P^\parallel_{[S_T, S_T]} + P^\perp_{[S_T, S_T]} \right]
= 1 - \frac{3}{2} \omega_2 .
\label{sumrule_parallel_perp}
\end{align}
In the sum of the $\parallel$ and $\perp$ projected distributions, the mixed $D$- and $S$-wave terms
in the depolarization correction cancel, and the correction is proportional to the square of
the $D$-wave, as it should be.

In the sum rule for the tagged spin structure function $g_{2D}$ in Sec.~\ref{subsec:spin_sum_rules}
we encounter the integral Eq.~(\ref{sumrule_parallel_perp}) with an additional factor
\begin{align}
\frac{1}{2 - \alpha_p}
= 1 + \cos\theta_k \frac{k}{m} + \mathcal{O}\left(\frac{k^2}{m^2}\right)
\end{align}
in the integrand. This integral can be computed in the same manner as above
and is equal to Eq.~(\ref{sumrule_parallel_perp}) up to corrections $\mathcal{O}(k^2/m^2)$,
\begin{align}
& \int_0^2\frac{d\alpha_p}{\alpha_p (2 - \alpha_p)} \, d^2 p_{pT} \;
\frac{1}{2} \left[ P^\parallel_{[S_T, S_T]} + P^\perp_{[S_T, S_T]} \right]
\nonumber \\
&= 1 - \frac{3}{2} \omega_2 .
\label{sumrule_parallel_perp_inverse}
\end{align}
This result is again a consequence of rotational invariance in the c.m.\ frame.

For the unfavored distribution of transversely polarized neutrons, Eq.~(\ref{transversity_unfavored}),
we obtain
\begin{align}
& \int_0^2\frac{d\alpha_p}{\alpha_p} \, d^2 p_{pT} \; 
P_{[S_L, S_T ]}(\alpha_p,\bm p_{pT}| \bm{S}_\deut) = \varepsilon_{[S_T]},
\label{integral_transverse_unfavored}
\end{align}
where
\begin{align}
& \varepsilon_{[S_T]} = \frac{\pi^2}{4} \int \frac{d k k^2 }{E}\, \frac{k}{m}\,
\left( f_0-\frac{f_2}{\sqrt{2}}\right)
\left( 4 f_0 + 5\frac{f_2}{\sqrt{2}}\right).
\label{epsilon_st_app}
\end{align}
The angular averages of the coefficients $A_{[\neq]}, B_{[\neq]}$ were used,
see Eq.~(\ref{angular_averages_favored_unfavored}). The result
Eq.~(\ref{integral_transverse_unfavored}) is of the same form as that for
the unfavored distribution of longitudinally polarized neutrons, Eq.~(\ref{integral_longitudinal_unfavored}),
but with a different coefficient. Namely,
\begin{align}
\varepsilon_{[S_T]} \approx -\varepsilon_{[S_L]} ,
\end{align}
as can be seen by comparing the sign of the squared $S$-wave term in the integrals.
The numerical value of $\varepsilon_{[S_T]}$ is given in Table~\ref{tab:sumrules}.
The sum rules for the unfavored distributions show that there is some net
longitudinal neutron polarization induced by transverse deuteron polarization,
and vice versa. The effect is of the order $k/m$ (relativistic correction) and approximately
of the same size in either direction.

\acknowledgments
This study greatly benefited from earlier work and exchanges with D.~Higinbotham, Ch.~Hyde, A.~Jentsch,
S.~Kumano, P.~Nadel-Turonski, M.~Sargsian, M.~Strikman, and Zhoudunming Tu.

This material is based upon work supported by the U.S.~Department of Energy, Office of Science,
Office of Nuclear Physics under contract DE-AC05-06OR23177, and by the U.S.~National Science Foundation
under awards PHY-2111442 and PHY-2239274.

\bibliography{spin1}

\end{document}